\documentclass[nofootinbib,onecolumn ,aps,longbibliography,a4paper,superscriptaddress,tightenlines,10pt]{revtex4-2}
\RequirePackage{silence}
\RequirePackage{float}
\usepackage[utf8]{inputenc}
\usepackage[USenglish]{babel}
\usepackage[T1]{fontenc}
\usepackage{array,multirow}
\usepackage{algorithm}
\usepackage{algorithmic}
\usepackage{graphicx,epsf}
\usepackage[toc,page,header]{appendix}
\usepackage{minitoc}
\usepackage{tikz}
\usetikzlibrary{quantikz}
\usepackage[bookmarksnumbered,hypertexnames=false,colorlinks,colorlinks=true,linkcolor=blue,urlcolor=black,citecolor=blue,anchorcolor=green]{hyperref}
\usepackage{bbm,braket,microtype,mathrsfs,amsmath,amssymb,color,amsthm,mathtools,fullpage,enumitem,ae,aecompl,bm,thm-restate}
\usepackage{tikz}
\usepackage[justification=justified]{caption}
\usepackage{subcaption}
\usepackage[capitalize]{cleveref}
\usepackage{silence}
\usepackage{newpxtext}
\usepackage{mathtools}

\usepackage[margin=0.75in]{geometry}
\usepackage[euler-digits,euler-hat-accent]{eulervm}
\allowdisplaybreaks
\newcommand{\be}{\begin{equation}}
\newcommand{\ee}{\end{equation}}
\newcommand{\ba}{\begin{eqnarray}}
\newcommand{\ea}{\end{eqnarray}}

\newcommand{\ignore}[1]{}

\newcommand{\parent}[1]{\left( {#1} \right)}

\def\CC{{\rm\kern.24em \vrule width.04em height1.46ex depth-.07ex
    \kern-.29em C}}
\def\P{{\rm I\kern-.25em P}}
\def\RR{{\rm
         \vrule width.04em height1.58ex depth-.0ex
         \kern-.04em R}}

\def\bbbc{{\mathchoice {\setbox0=\hbox{$\displaystyle\rm C$}\hbox{\hbox
to0pt{\kern0.4\wd0\vrule height0.9\ht0\hss}\box0}}
{\setbox0=\hbox{$\textstyle\rm C$}\hbox{\hbox
to0pt{\kern0.4\wd0\vrule height0.9\ht0\hss}\box0}}
{\setbox0=\hbox{$\scriptstyle\rm C$}\hbox{\hbox
to0pt{\kern0.4\wd0\vrule height0.9\ht0\hss}\box0}}
{\setbox0=\hbox{$\scriptscriptstyle\rm C$}\hbox{\hbox
to0pt{\kern0.4\wd0\vrule height0.9\ht0\hss}\box0}}}}
\def\bbbz{{\mathchoice {\hbox{$\sf\textstyle Z\kern-0.4em Z$}}
{\hbox{$\sf\textstyle Z\kern-0.4em Z$}}
{\hbox{$\sf\scriptstyle Z\kern-0.3em Z$}}
{\hbox{$\sf\scriptscriptstyle Z\kern-0.2em Z$}}}}

\usepackage{hyperref}

\begin{document}
\setcounter{secnumdepth}{3}

\title{Measuring magic on a quantum processor}
\author{Salvatore F.E. Oliviero}\thanks{}\email{s.oliviero001@umb.edu}
\affiliation{Physics Department,  University of Massachusetts Boston, Boston, MA, USA}
\author{Lorenzo Leone}\email{lorenzo.leone001@umb.edu}
\affiliation{Physics Department,  University of Massachusetts Boston, Boston, MA, USA}
\author{Alioscia Hamma}
\affiliation{Physics Department,  University of Massachusetts Boston, Boston, MA, USA}
\affiliation{Dipartimento di Fisica Ettore Pancini, Università degli Studi di Napoli Federico II, Via Cinthia, Fuorigrotta, 80126, Napoli, NA, Italy}
\affiliation{INFN, Sezione di Napoli, Naples, Italy}
\author{Seth Lloyd}
\affiliation{Department of Mechanical Engineering, Massachusetts Institute of Technology, Cambridge, MA, USA}
\affiliation{Turing Inc., Brooklyn, NY, USA}
\renewcommand{\thefootnote}{\fnsymbol{footnote}}
\footnotetext[1]{$^{\dag}$ The two authors contributed  equally to this paper.}
\begin{abstract}
Magic states are the resource that allows quantum computers to attain an advantage over classical computers. This resource consists in the deviation from a property called stabilizerness which in turn implies that stabilizer circuits can be efficiently simulated on a classical computer. Without magic, no quantum computer can do anything that a classical computer cannot do. Given the importance of magic for quantum computation, it would be useful to have a method for measuring the amount of magic in a quantum state. In this work, we propose and experimentally demonstrate a protocol for measuring magic based on randomized measurements. Our experiments are carried out on two IBM Quantum Falcon processors. This protocol can provide a characterization of the effectiveness of a quantum hardware in producing states that cannot be effectively simulated on a classical computer. We show how from these measurements one can construct realistic noise models affecting the hardware.
\end{abstract}
\maketitle
\section*{Introduction}
In the era of Noisy Intermediate Scale Quantum Computers (NISQs)\cite{Preskill2018quantumcomputing} it is of paramount importance to be able to characterize the proposed quantum hardware in order to check how good these machines are in performing quantum computation with the purpose of attaining an advantage over classical computers. 
This paper shows how to perform accurate and robust measurements of the stabilizer R\'enyi entropy, which in turn is known to quantify the resource known as `magic'\cite{leone2021renyi}
It is well known that the preparation of stabilizer states, the implementation of Clifford gates and measurements in the computational basis can be made fault tolerant\cite{campbell2010bound,Campbell2011Catalysis,campbell2012magic,Campbell2017fault,campbell2014enhanced,campbell2017unified,campbell2017unifying}. However, computers based on the Clifford resources can be efficiently simulated on classical computers\cite{Knill1996codes,Gottesman:1998hu,gottesmann1998fault,aaronson2004improved}, similarly to what happens for matchgate circuits(MGCs). This means that the power of quantum advance requires resources beyond the Clifford group, like the Phase $\pi/8$ gate (T gate) or the Toffoli gate and non-Gaussian states for the MCGs\cite{Hebenstreit2019gaussian,Hebenstreit2020match}. The precious resource that makes quantum computers special is colloquially dubbed as `magic' and a resource theory of magic has been developed in recent years\cite{campbell2010bound,Campbell2011Catalysis,Veitch2014resource,howard2017application,Ahmadi2018magic,Wang2019magic,seddon2019magic,liu2020many,leone2021renyi,seddon2021quantifying,White2021cft,qassim2021improved,Kouko2021magic,Hahn2021magic,Saxena2022stab}.

It is a striking fact that these resources are difficult to implement\cite{campbell2010bound,campbell2012magic,anwar2012qutrit,Campbell2012dist,Bravyi2012magic,Dawkins2015dist,bravyi2016resources,Hastings2018dist}. The very reason why these resources are powerful makes them fragile. Moreover, the amount of these resources that needs to be used in a computation must be calibrated accurately: just like entanglement\cite{Gross2009ent}, too much magic is not useful for quantum computation (see Supplementary Note $1$), see also\cite{liu2020many}. Moreover, decoherence is not magic preserving, and it can  both increase or decrease the amount of magic in a system, as we will show in the experiments. To the extent that decoherence is spoiling quantum computation, then one needs the amount of magic created and manipulated throughout the computation to be accurate: in this paper, we prove  that an excess amount of unwanted magic makes the task of distinguishing the state $\psi$ from a random state an exponentially difficult task, see Supplementary Note $1$ for the proof. Moreover, 
 since inaccurate Clifford gates can produce magic, the presence of excess magic is in fact a signal of noise. We exploit this fact to show that the measurement of magic can be used to quantify and characterize the noise in the quantum circuit. 
 It is thus important to be able to quantify this resource and measure it to characterize the fitness of real quantum hardware.  Unfortunately - until recently - proposed measures of magic\cite{Campbell2011Catalysis,howard2017application,beverland2019lower,seddon2021quantifying} have been based on extremization procedures and no experimental measurement scheme has been proposed.

In this work, we propose and experimentally demonstrate a protocol based on randomized measurements\cite{Enk2012measure,Tran2016Correlations,Elben2018Entropies,Elben2019probing,Vermersch2019meas,Brydges2019randomized,Ketterer2019ranmeas,Elben2020randomized,Knips2020entanglement,Ketterer2020randmeas,Zhou2020Estimation,Cian2021chern,Satoya2021ranmeas,Rath2021sampling}  to measure magic in a quantum system with $n$ qubits and to characterize quantum hardware. We adopt the magic measure called stabilizer $2$-R\'enyi entropy defined as\cite{leone2021renyi}
\begin{equation}\label{Mdef}
M_{2}(\mid\!\psi\rangle):=-\log_{2} W(\psi)-S_2(\psi)-\log d
\end{equation}
where $W(\psi):=\operatorname{tr}(Q\psi^{\otimes 4})$, $Q:={d^{-2}}\sum_{P}P^{\otimes 4}$ and $d=2^n$, where the sum is taken over all 
multi-qubit strings of Pauli operators, applied to four copies of the state, and $S_2(\psi)=-\log_{2}\operatorname{tr}{\psi^2}$ is the $2$-R\'enyi entropy.  In order to measure $M_2$ we propose an improved protocol compared to the one presented in Ref.~\cite{leone2021renyi} as it only involves randomized one-qubit measurements instead of global multi-qubit measurements, with obvious advantages in terms of errors and quantum control.

 As $M_2$ depends on the state $\psi$, a direct evaluation of $M_2$ would be possible by knowing all the expectation values $\operatorname{tr} (P\psi)$ of multi-qubit Pauli strings in the state $\psi$. This, of course, is equivalent to tomography and it is very expensive as it involves the evaluation of $d^2$ expectation values for a total cost in resources scaling as $\mathcal{O}(d^3)$. Here, we employ a protocol based on randomized measurements which does not rely on tomographic techniques. Remarkably, randomized measurement protocols are highly favorable compared to state tomography\cite{Elben2018Entropies,Elben2019probing,Brydges2019randomized,Elben2020randomized}. As we shall see, we will employ a number of resources scaling as $\mathcal{O}(\epsilon^{-2}d^2)$ for an estimate with error $\epsilon$.  

 \subsection*{Results}
\subsubsection*{The protocol}
The protocol consists in first drawing a string of random one-qubit Clifford operations, namely $C=\bigotimes_{i=1}^{n}c_{i}$
and applying it to four copies of the state of interest. The protocol extracts correlations between these copies. Indeed, the quantity of interest in the first term of Eq.~\eqref{Mdef} can then be computed as 
\begin{equation}\label{trQ}
-\log_{2}\operatorname{tr}(Q\psi^{\otimes 4})=-\log_{2}\sum_{\vec{\mathbf{s}}}(-2)^{-\|\vec{\mathbf{s}}\|}\mathbb{E}_{C} P(\mathbf{s}_1\!\mid\! C) P(\mathbf{s}_2\!\mid\! C) P(\mathbf{s}_3 \!\mid\! C) P(\mathbf{s}_4\!\mid\! C)
\end{equation}
The formula above features  the expectation value over the randomized measurements of the Clifford operator $C$ on states of the computational basis $\mathbf{s}_a$ and  the Hamming weight $\|\vec{\mathbf{s}}\|$  of the string $\mathbf{s}_1\oplus \mathbf{s}_2\oplus \mathbf{s}_3\oplus \mathbf{s}_4$. The quantity $P(\mathbf{s}_a\!\mid\! C)= \operatorname{tr}({C\psi C^\dagger}\mathbf{s}_a)$ represents the probability of finding the computational basis state $\mathbf{s}_a$ when measuring the state $C\psi C^\dagger$. The second term in Eq.~\eqref{Mdef} is the usual $2$-R\'enyi entropy and can be measured by randomized measurements using the techniques of Ref.~\cite{Brydges2019randomized}. {An important feature of our protocol is the fact that it only needs randomized operations over the Clifford group  instead of the full unitary group as in Ref.~\cite{Elben2018Entropies}. In fact, by collecting the occupation probabilities $P(C\psi C^{\dagger}\!\mid\! \mathbf{s}_a)$ one can estimate both $W(\psi)$ and the purity $P(\psi)$ together thanks to the fact that the Clifford group forms a $2$-design.} 
See Methods.   The operational meaning of the protocol is the following: 
randomized measurement protocols are usually conducted on a (Haar) random basis. Here we select a (local) stabilizer basis. Clifford rotations constitute the free resources for magic state resource theory.  General unitaries would result in a change in quantum magic. Clifford orbits of a given quantum state instead are filled out by \textit{iso-magic} states. A Clifford randomized measurement protocol measures the magic of the entire Clifford orbit, rather than of a single quantum state.


The experiments have been conducted on two IBM Quantum Falcon processors: a $5$ qubit system, \textit{ibmq$\_$quito} and a $7$ qubit system \textit{ibmq$\_$casablanca}\cite{IBMquantum}.

\begin{figure}[H]
    \centering
    \includegraphics[scale=0.3]{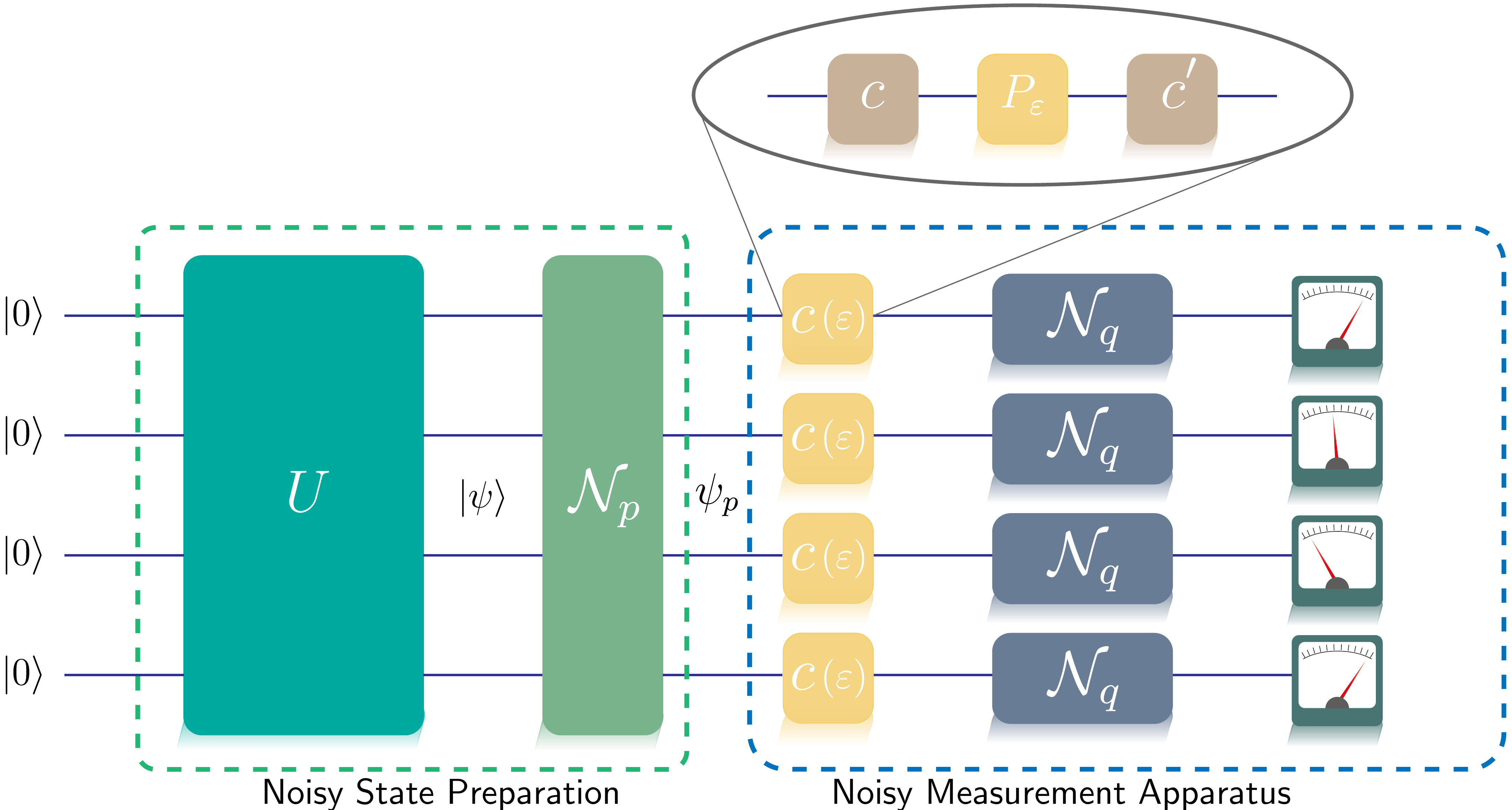}
\caption{{\bf Schematic of the implementation of the experiment for measuring magic on a quantum processor.} From left to right: Initialization of the system in the state $\!\mid\!\!0\rangle^{\otimes n}$; preparation of the target state $\mid\!\psi\rangle$ by a unitary quantum circuit $U_t$ containing a number $t$ of non-Clifford gates;  intervention of the noise $\mathcal N_p$ affecting the system effectively prepares the (mixed) state $\psi_p$; measurement. The measurement apparatus is composed of $n$ local Clifford operators $C=\bigotimes_{i=1}^{n}c_i$  randomly sampled from the single qubit Clifford group $c_{i}\in\mathcal{C}_{1}$, followed by $n$ measurements in the computational basis $\{\mid\!\mathbf{s}\rangle\}$ which are performed to estimate the occupation probabilities $P(C\psi C^{\dag}\!\mid\! \mathbf{s})$. The gate imperfection in the application of the Clifford operators is denoted by $c(\epsilon)$. }
\label{Fig: noisyexperiment}
\end{figure}

The experiment can be schematized as follows (see Fig.~\ref{Fig: noisyexperiment}).  Starting with a $n$-qubit state initialized in the $\mid \!0\rangle^{\otimes n}$ state, we pass it through a unitary quantum circuit $U$ resulting in the state preparation $\mid\!\!\psi\rangle$. We want to characterize the fitness of such a circuit in providing a state with the promised magic. At this point, one extracts $n$ one-qubit Clifford operations $c_i$, applies them to the state $\!\mid\!\psi\rangle$, and measures the state in the computational basis. 
At this point, we want to analyze the scaling of the cost of necessary resources, both analytically and numerically.
The experiment is repeated $N_M$ times for every string $C=\bigotimes_{i=1}^{n}c_{i}$ in order to collect statistics to compute the occupation probabilities $P(C\psi C^{\dagger}\!\mid\! \mathbf{s}_a)$. Then, in order to compute the expectation value over the whole Clifford group $\mathbb{E}_{C}$, one samples the Clifford group with $N_U$ elements. In order to sample the Clifford group properly and to build sufficient statistics we simulate numerically the total number of measurements needed for $M_{2}$, i.e. $N_{TOT} = N_M\times N_U$. By evaluating the variance of the estimator for $W$, through the use of standard statistical analysis (Bernstein inequality), one can bound the probability of making an error $\epsilon$ as a function of the total resources $N_U\times N_M$ employed. In Methods, we prove that by  employing a total number of resources $\mathcal{O}(\epsilon^{-2}d^{2})$ the randomized measurement protocol is able to estimate the purity within an error $\le \epsilon$ and the stabilizer purity within an error $\epsilon d^{-1}$. These theoretical bounds can be optimized by numerical analysis.  The optimal number of unitaries $N_U$ and of measurements $N_M$ is found by numerical simulations imposing that the relative error on the theoretical value of stabilizer purity to be below $12\%$ and an average value of the purity greater than $0.88$, thus imposing a relative error of $12\%$ on the purity as well. An important remark is that both $N_U$ and $N_M$ depend on the state $\psi$. Remarkably, low-magic states (like the states in the computational basis - which have exactly zero magic) require a higher $N_U\times N_M$ compared to states with high magic, see Supplementary Table I in Supplementary Note $4$. 


In order to characterize the fitness of a quantum processor in producing resources beyond stabilizer states, we adopt the model of a $t$-doped Clifford circuit\cite{zhou2020single,leone2021quantum,oliviero2021transitions}. This circuit consists of a block of Clifford gates in which $t$ non-Clifford gates are injected.  The non-Clifford gates we inject are $P_\vartheta= \mid\!\!0\rangle\langle 0\!\!\mid+e^{i\vartheta}\mid\!\!1\rangle\langle 1\!\!\mid$ gates: these constitute the resources that are injecting magic in the system, while the Clifford circuits are free resources. For $\vartheta=\pi/2$ one obtains the phase flip gate that still belongs to the Clifford group and thus is a free resource. The value $\vartheta=\pi/4$ instead, the so-called $T$ gate, yields the maximal amount of magic achievable for a $P_\vartheta$ gate. The $T$-gates will be called the ``magic seeds'' of the quantum circuit.  These circuits are efficient in entangling so the output state of the circuit is in general not a trivial product state but a state that is both entangled and possesses magic. 

\subsubsection*{Measuring magic}
We start with the characterization of the quantum processor on single-qubit states, and thus without entanglement. The single-qubit magic states are obtained by applying $P_\vartheta$ on the states $\mid\!\!+\rangle=\frac{1}{\sqrt{2}}(\mid\!\!\!\!\, 0\rangle +\!\mid\!\!\!\!\, 1\rangle)$ obtaining $\mid\!\!P_{\vartheta}\rangle\equiv P_{\vartheta}\!\!\mid\!\!+\rangle=\frac{1}{\sqrt{2}}(\mid\!\! 0\rangle+e^{i\vartheta}\!\mid\!\! 1\rangle)$ whose stabilizer $2$-R\'enyi entropy reads $M_{2}(\mid\!\!P_{\vartheta}\rangle)=-\log_{2}\left(\frac{7+\cos (4\vartheta)}{8}\right)$, achieving its maximum for $M_{2}(\mid\!\!P_{\pi/4}\rangle)=1-\log_{2}3/2$ and its minimum for $M_{2}(\mid\!\!P_{\pi/2}\rangle)=0$.

\begin{figure}[H]
    \centering
    \includegraphics[scale=0.3]{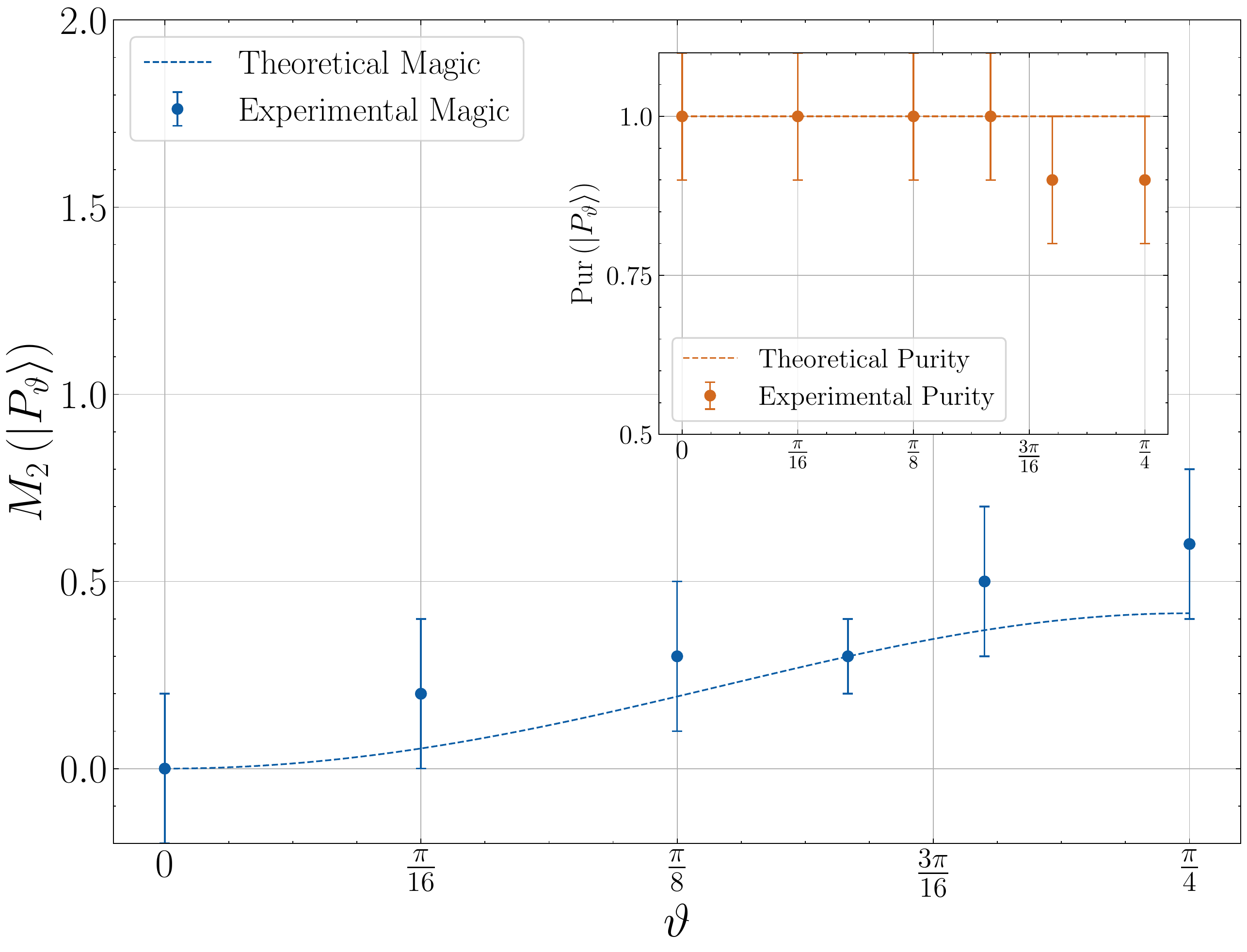}
\caption{{\bf Stabilizer $2$-R\'enyi entropy for $\mid\!\! P_{\vartheta}\rangle$.} Plot of the magic of the single qubit $\mid\!\! P_{\vartheta}\rangle$-states, for $\theta=0,\frac{\pi}{16},\frac{\pi}{8},\frac{\pi}{6},\frac{\pi}{5},\frac{\pi}{4}$. The data displayed (blue dots) are obtained from the quantum processor ibmq$\_$quito. The blue dashed curve represents the theoretical value of the magic for $\mid\!\! P_{\vartheta}\rangle$-states, i.e. $M_{2}(\mid\!\! P_{\vartheta}\rangle)=3-\log_{2}\left(7+\cos (4\vartheta)\right)$. Additionally, a plot of the purity for these states is displayed in the upper right corner: as the data show, the purity is $1$ within the experimental errors, showing that the decoherence affecting the system is negligible for $n=1$ and also the experimental values of magic are in perfect agreement with the theoretical ones. See Supplementary Table II in the Supplementary Note $4$ for the data. }
\label{Exp: singlequbiexp} 
\end{figure}

The results of the experiment on the ibmq$\_$quito are shown in Fig. \ref{Exp: singlequbiexp}. As we can see, the experimental data are in very good accordance with the theoretical prediction for the target state, showing the fitness of ibmq$\_$quito in preparing single-qubit magic states. Decoherence effects are also very low, as we can see from the purity, see Fig.~\ref{Exp: singlequbiexp}.

\begin{figure}[H]
    \centering
    \includegraphics[scale=0.24]{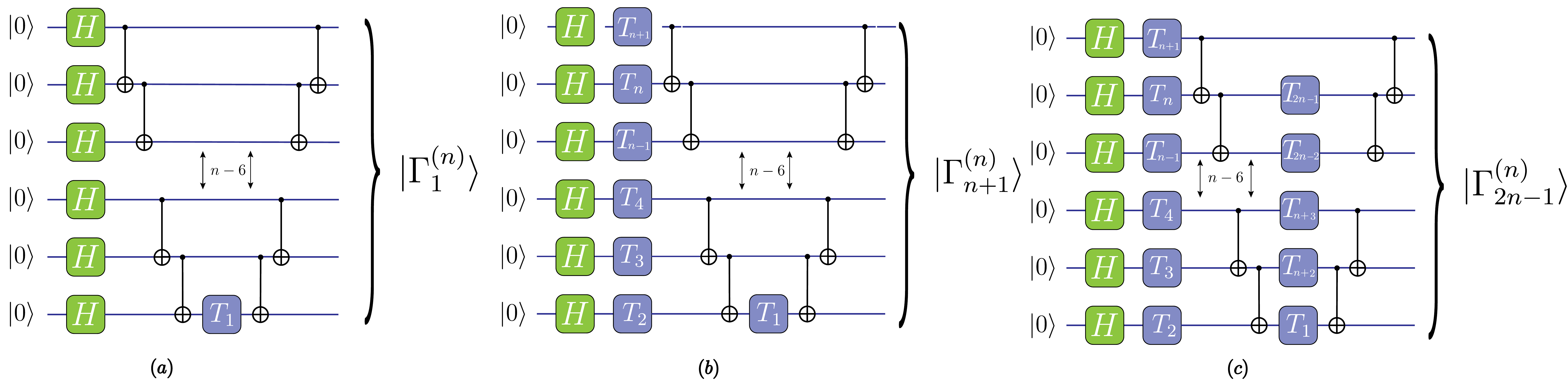}
\caption{{\bf Preparation of $\mid\!\!\Gamma_{t}^{(n)}\rangle$ states.} The magic seeds ($T$-gates) are placed either on the first layer (immediately after the Hadamard gates $H$), or in the second layer (immediately after the first layer of $C-NOT$ gates). We start with a $T$-gate in the second layer, then start filling up the first layer. Upon completion of the first layer, we start filling up the second layer again. The figure shows: $\mathbf{(a)}$ $\mid\!\!\Gamma_{1}^{(n)}\rangle$, $\mathbf{(b)}$ $\mid\!\!\Gamma_{n+1}^{(n)}\rangle$ and $\mathbf{(c)}$ $\mid\!\!\Gamma_{2n-1}^{(n)}\rangle$ which is the final doped Clifford circuit which we consider in this paper. } 
\label{gammastates} 
\end{figure}

We now proceed to the more difficult task of characterizing a quantum processor capable of preparing entangled states. Starting from the computational basis state $\!\mid\!\!\!0\rangle^{\otimes n}$, i.e. the input state of the quantum processor, we first apply a layer of Hadamard  $H$-gates to obtain $\!\mid\!\!\!+\rangle^{\otimes n}=H^{\otimes n}\!\mid\!\!0\rangle^{\otimes n}$. Then, we apply $T$-gates on $n_1$ qubits, with $n_{1}=0,\dots,n$. The $T$-gates inject magic into the system. For $n_1=n$, the state obtained is the maximal magic product state achievable. If one wants to pump more magic into the system, one needs to create some entanglement between the qubits. To do so, we apply a layer of CX-gates, i.e. Clifford entangling $2$-qubit gates defined as $CX_{i,j}=I_{i}\otimes( I_{j}+X_j)+Z_{i}\otimes (I_{j}-X_{j})$ and nested in the following way: $CX_{n-1,n}CX_{n-2,n-1}\cdots CX_{1,2}$. Then we can inject some more magic in the system by applying another layer of $n_{2}$ $T$-gates with $n_{2}=1,\dots,n-1$ followed by another layer of $CX$: $CX_{1,2}\cdots CX_{n-2,n-1}CX_{n,n-1}$. For the pictorial representation of the previously described architecture see Fig. \ref{gammastates}.
At the end of the state preparation, the magic seeds in the circuit are $t=n_{1}+n_{2}$ and the state prepared is the $\mid\!\!\Gamma^{(n)}_{(n_1,n_2)}\rangle$-state, where $1\le t\le 2n-1$. In the following, we fill in $T$-gates starting from $(0,1)$, then $(n_1, 1)$ with $(n_1, =1,\ldots , n; n_2=1)$, and finally $(n,n_2)$, with $n_2=2,\ldots,n-1$. With this prescription, the label $t$ uniquely describes the circuit. For example, $t=4$ on a system with $n=6$ qubits means three $T$-gates on the first layer and one $T$-gate on the second layer, see Fig.~\ref{gammastates}. The optimal number of $N_U,N_M$ for a system with $n=3,4,5$ qubits can be found in Supplementary Table I in Supplementary Note $4$ and Fig. \ref{fig: numberofresources} in Methods.   

\begin{figure}[H]
    \centering
    \includegraphics[scale=0.3]{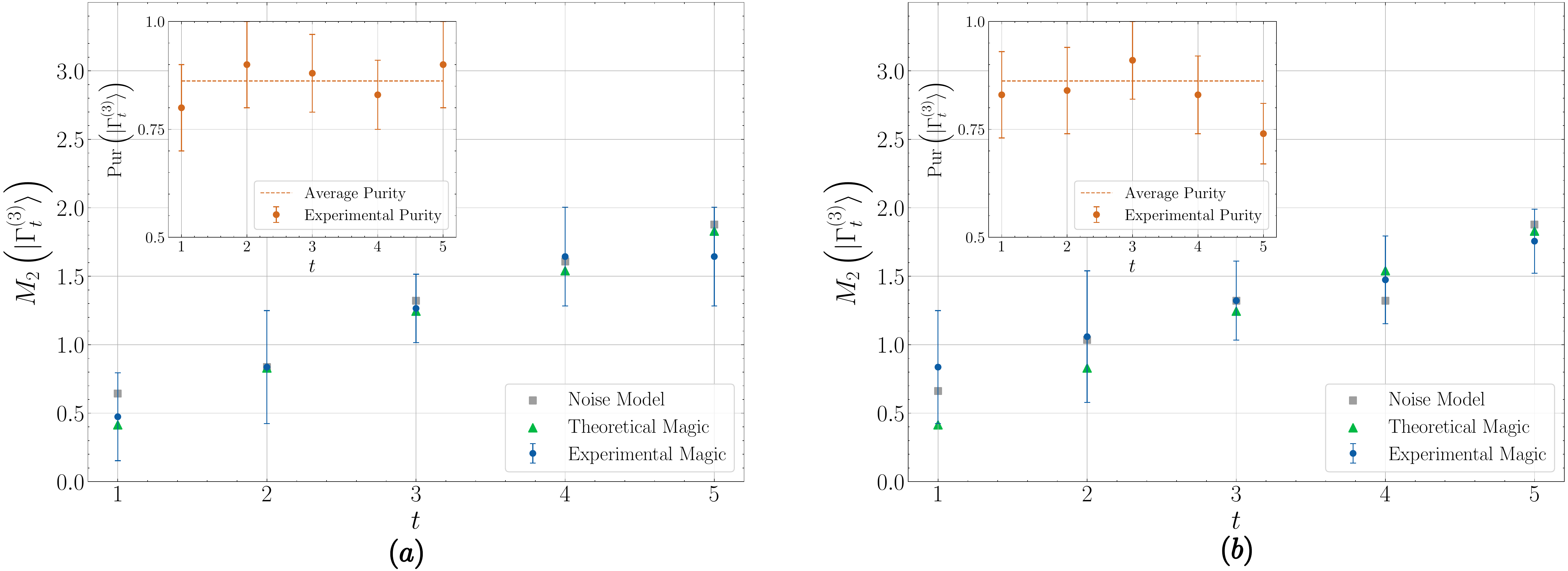}
 \caption{{\bf Stabilizer $2$-R\'enyi entropy for $n=3$.} Plot of the stabilizer $2$-R\'enyi entropy for $\mathbf{(a)}$ ibmq$\_$quito and $\mathbf{(b)}$ ibmq$\_$casablanca. Both figures contain the experimental values (blue dots), the theoretical values (green triangles) of the magic for the desired pure state, and the noise model values (grey squares) of the magic for the mixed state prepared on the quantum processor. The values of the magic of the $\mid\!\!\Gamma^{(3)}_{t}\rangle$-states for $t=1,\dots, 5$ are plotted as functions of the number of $T$-gates $t$ in the doped random Clifford circuits (see Supplementary Tables III and IV in the Supplementary Note $4$ for the data). See Fig. \ref{gammastates} for the preparation of such states. Both figures contain in the upper left corner the purity (orange dots) of the output state prepared on the quantum processor and its average value (dashed line). Here the number of resources $N_{TOT}\equiv N_{U}\times N_{M}$ depends on the number of $T$-gates $t$ thrown in the circuits as $N_{TOT}=2^{A_{3}+B_{3}(5-t)}$, where $A_{3}=10.6\pm0.3$, $B_{3}=0.56\pm 0.08$, see Supplementary Table I in the Supplementary Note $4$ and Fig. \ref{fig: numberofresources} in Methods. Note that the experimentally observed magic can be -- and typically is -- higher than the theoretically predicted magic.   This is because imperfectly performed Clifford gates are no longer exactly Clifford and can inject uncontrolled/unwanted magic into the system. This effect is enhanced for more qubits and deeper circuits. }
 \label{fig:magic3qubit}  
\end{figure}

\begin{figure}[H]
    \centering
    \includegraphics[scale=0.3]{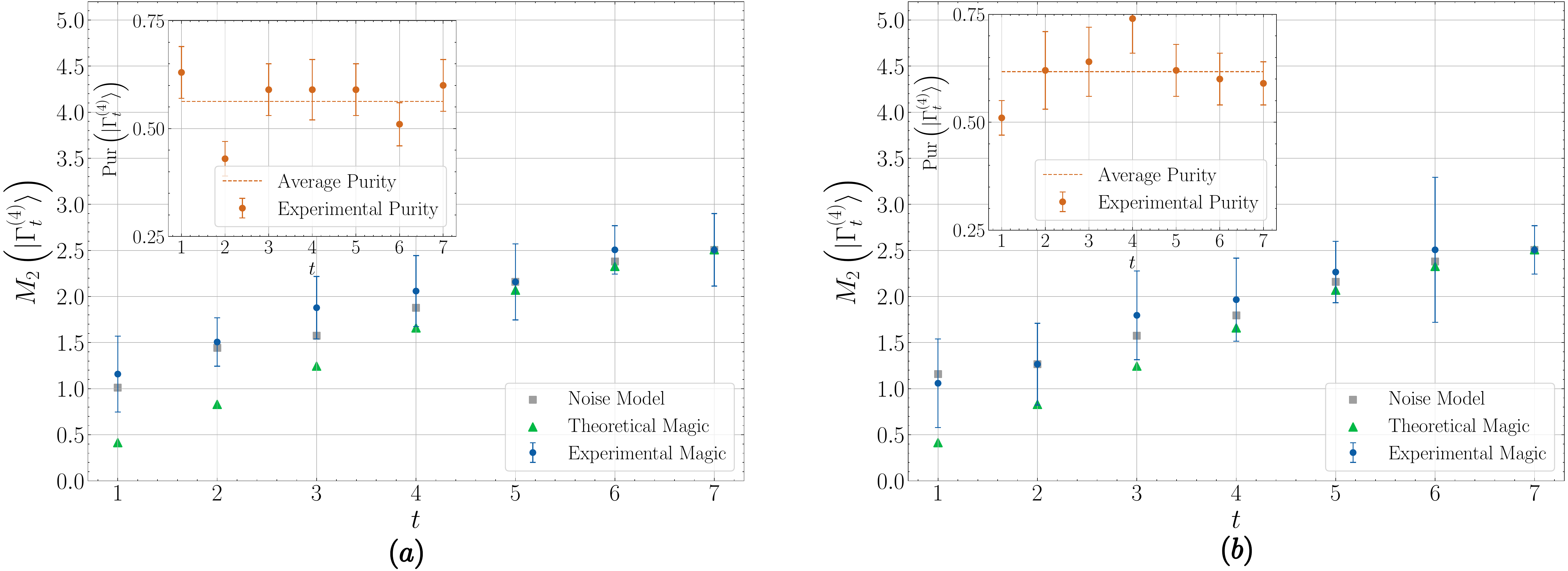}
\caption{{\bf Stabilizer $2$-R\'enyi entropy for $n=4$.} Plot of the stabilizer $2$-R\'enyi entropy for  $\mathbf{(a)}$ ibmq$\_$quito and $\mathbf{(b)}$ ibmq$\_$casablanca. Both figures contain the experimental values (blue dots), the theoretical values (green triangles) of the magic for the desired pure state, and the noise model values (grey squares) of the magic for the mixed state prepared on the quantum processor. The values of the magic of the $\mid\!\!\Gamma^{(4)}_{t}\rangle$-states for $t=1,\dots, 7$ are plotted as functions of the number of $T$-gates $t$ in the doped random Clifford circuits, (see Supplementary Tables III and IV in the Supplementary Note $4$ for the data). See Fig. \ref{gammastates} for the preparation of such states. Both figures contain in the upper left corner the purity (orange dots) of the output state prepared on the quantum processor and its average value (dashed line). Here the number of resources $N_{TOT}\equiv N_{U}\times N_{M}$ depends on the number of $T$-gates $t$ thrown in the circuits as $N_{TOT}=2^{A_{4}+B_{4}(7-t)}$, where $A_{4}=11.3\pm0.3$, $B_{4}=0.49\pm 0.05$, see Supplementary Table I in Supplementary Note $4$ and Fig.~\ref{fig: numberofresources} in Methods. Note that the experimentally observed magic can be -- and typically is -- higher than the theoretically predicted magic. See the caption of Fig. \ref{fig:magic3qubit} for an explanation. }
\label{Exp: 4qubit} 
\end{figure}

\begin{figure}[H]
    \centering
    \includegraphics[scale=0.3]{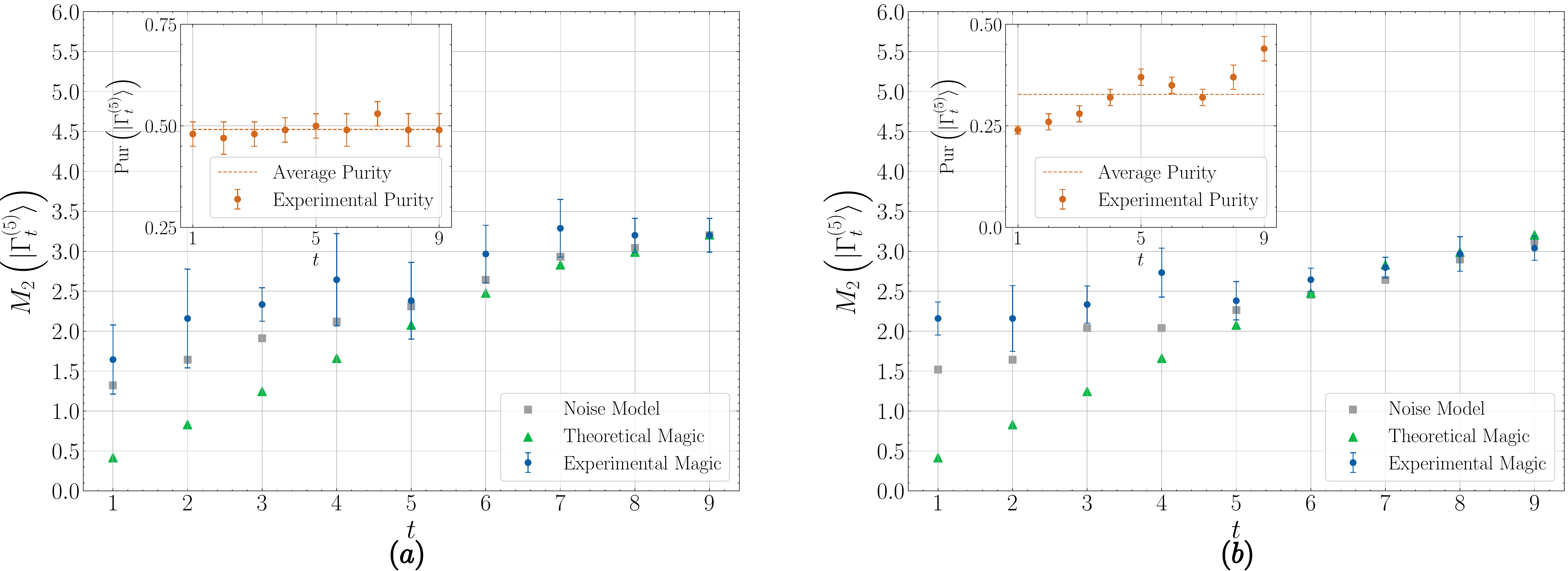}
\caption{{\bf Stabilizer $2$-R\'enyi entropy for $n=5$.} Plot of the stabilizer $2$-R\'enyi for $\mathbf{(a)}$ ibmq$\_$quito and $\mathbf{(b)}$ ibmq$\_$casablanca. Both figures contain the experimental values (blue dots), the theoretical values (green triangles) of the magic for the pure state one would have liked to obtain, and the noise model values (grey squares) of the magic for the mixed state prepared on the quantum processor. The values of the magic of the $\mid\!\!\Gamma^{(5)}_{t}\rangle$-states for $t=1,\dots, 9$ are plotted as functions of the number of $T$-gates $t$ in the doped random Clifford circuits,(see Supplementary Tables III and IV in the Supplementary Note $4$ for the data). See Fig. \ref{gammastates} for the preparation of such states. Both figures contain in the upper left corner the purity (orange dots) of the output state prepared on the quantum processor and its average value (dashed line). Here the number of resources $N_{TOT}\equiv N_{U}\times N_{M}$ depends on the number of $T$-gates $t$ thrown in the circuits as $N_{TOT}=2^{A_{5}+B_{5}(9-t)}$, where $A_{5}=13.7 \pm 0.1$, $B_{5}=0.047\pm 0.002$, see Supplementary Table I in the Supplementary Note and Fig. \ref{fig: numberofresources} in Methods. Note that the experimentally observed magic can be -- and typically is -- higher than the theoretically predicted magic. See the caption of Fig. \ref{fig:magic3qubit} for an explanation.}
\label{Exp: 5qubit} 
\end{figure}

In a system with $n$ qubits we can prepare the states $\mid\!\!\Gamma^{(n)}_t\rangle$ with $t=1,\ldots, 2n-1$. The results of the experiment for $n=3,4,5$ are shown in Figs. \ref{fig:magic3qubit}, \ref{Exp: 4qubit},\ref{Exp: 5qubit}, respectively.
We can see that, for larger values of $n$, the purity of the prepared state is compromised, due to decoherence. The measured experimental values of magic shoot off the theoretical prediction, especially for low magic states.  Somewhat counterintuitively, the experimental value of magic is higher than the theoretical one.  As we mentioned above,  spurious injection (or subtraction) of magic can happen for two reasons. Inaccurate implementation of the Clifford gates - and thus turning them into non-free resources -  or noise affecting them, or decoherence.  That is, our experimental characterization of how magic is created in a quantum circuit tests not only the quantity of magic, but the accuracy with which the desired magic is created.
The fact that the circuit must not only create magic, but must do it so with a certain accuracy, allows us to use the experimental data obtained from our protocol to characterize the noise affecting the system. A first insight comes from the realization, see Figs.~\ref{fig:magic3qubit},~\ref{Exp: 4qubit},~\ref{Exp: 5qubit} that the noise is affecting more the preparation of low-magic states than that of high-magic states, mostly because of imperfection in the implementation of the resource-free Clifford gates like the CX gate. Let us see how we can characterize the noise affecting the system. A very general error model for the target state $\psi$ is through a quantum channel $\mathcal{E}(\psi):=\sum_{i}q_{i}P_{i}\psi P_{i}$. Random states are a good model for high-magic states\cite{leone2021renyi} and thus, to understand why the noise affecting the system does not disturb the magic injected in high-magic states, we compute the average difference in magic between a random state $\psi$ and the noisy state $\mathcal{E}(\psi)$ as:
$\langle\delta M\rangle_{Haar}:=\langle\mid\!M(\mathcal{E}(\psi))-M(\psi)\!\mid\rangle_{Haar}$. Calculation shows (see Supplementary Note 2) that $\langle\delta M\rangle_{Haar}=\mathcal{O}(S_{2}(\mathbf{q}))$. In other words, at high levels of magic, this quantity is robust under the noise model provided that the distribution $\mathbf{q}=\{ q_i\}$ is low in entropy $S_{2}(\mathbf{q})$. 
Guided by this result, we model the noise in two factors (i) noise in state preparation due to decoherence and (ii) imperfection in the realization of the $c_i$ gates in the randomized measurement. This latter error is unitary. We then tune the factors quantifying the noise in our model to match the difference between the experimentally measured and the theoretically predicted amounts of magic.

The ansatz for the (non-unitary) quantum channel $\mathcal N_p$ affecting the state preparation is
\begin{equation}
\psi_{p}\equiv \mathcal N_p(\mid\!\!\psi\rangle\langle\psi\!\!\mid):= p\mid\!\!\psi\rangle\langle\psi\!\!\mid+\frac{(1-p)}{n}\sum_{i=1}^{n}Z_{i}\mid\!\!\psi\rangle\langle\psi\!\!\mid Z_{i}
\label{errormodel0}
\end{equation}
where $Z_{i}$ is a phase flip error on the $i$-th qubit happening with probability $(1-p)/n$.  This channel is not the simple phase-flip channel as the probability $p$ in principle depends on the target state $\mid\!\psi\rangle$.  The imperfection in the gates $c_i$ is modeled by the unitary phase displacement $c_{i}\rightarrow c_{i}^{\varepsilon}\equiv c_{i}P_{\varepsilon}c_{i}^{\prime}$, where use the $P_{\varepsilon}$-gate described above. 
The measured stabilizer purity will be denoted by $W_{exp}(\mid\!\psi\rangle)$. 

Our ansatz on how the noise affects the measurement results is then $W_{exp}(\mid\!\psi\rangle)\stackrel{\mathclap{\normalfont\mbox{!}}}{=} \operatorname{tr}(\psi_{p}^{\otimes 4}Q_{2}^{\varepsilon\otimes n})$
where  $Q_{2}^{\varepsilon}$ represents the correction to the projector onto the single-qubit stabilizer code  due to the gate imperfection error $\varepsilon$.  
The two free parameters $p$ and $\varepsilon$ can be determined experimentally, see Supplementary Note $2$. 

Several points are in order here.  First, notice that the purity $\operatorname{tr}{\psi_{p}^{2}}$ is protected against gate imperfection errors, so it can be measured independently.  Second, one can measure the $\varepsilon$ error directly by measuring the purity of the initial state $\mid\!\! 0\rangle^{\otimes n}$, thus avoiding the decoherence effect altogether.  The values of the stabilizer $2$-R\'enyi entropy given by the noise model are represented by the Grey squares in Figs.\ref{fig:magic3qubit}, \ref{Exp: 4qubit} and \ref{Exp: 5qubit} which show that they provide a better approximation to the experimental data, an approximation which in fact improves as the number of T gates in the circuit increases. By measuring the stabilizer $2$-R\'enyi entropy, thus, we provide a characterization of the noise model and an estimate of its parameters $p,\varepsilon$.

\section*{Discussion}
Magic is a quantity of central importance for quantum computation: no quantum advantage can be obtained without it.   This paper showed how to measure the amount of magic produced by a quantum circuit in terms of stabilizer R\'enyi entropy, and evaluated experimentally how that amount of magic scales as a function of the number of T-gates in the circuit.   A central result of our experimental demonstration is  that it is not enough just to create magic: the circuit must create an ``accurate amount'' of magic. Imperfectly implemented Clifford gates inject or subtract uncontrolled/unwanted magic into the circuit: just as excess entanglement can hinder the ability of a quantum circuit to perform some desired task\cite{Gross2009ent}, uncontrolled excess magic can result in the degradation of the performance of a quantum computation.   Generating the correct amount of stabilizer R\'enyi entropy is thus an important component of the certification process for quantum hardware.  More generally, in a quantum device, e.g. a circuit based  on superconducting qubits, there can be errors beyond decoherence, like gate implementation errors, or other unitary errors. Typically, these errors are investigated through gate fidelity while the loss of purity is a  good figure of merit to quantify decoherence. However, not always gate fidelity is available as a tool. As we can see in Figs.~\ref{fig:magic3qubit},~\ref{Exp: 5qubit}, an inaccurate level of magic compared to the theoretical one signals the presence of unitary errors: a measurement of magic can then be used as a further tool to evaluate the accuracy of an experimental setup.

\subsection*{Methods}

\subsubsection*{Theoretical framework}
In \cite{leone2021renyi} we proved that a global randomized measurements protocol can be employed to measure the stabilizer $2$-R\'enyi entropy for multiqubit states. 

{Here, we make a decisive improvement by establishing a protocol that only requires local measurements. Local measurements are usually the best measurements in terms of quantum control an experimenter has access to. Let us  recall the definition of stabilizer $2$-R\'enyi entropy: for $\psi$  a $n$-qubit quantum state,  the stabilizer $2$-R\'enyi entropy of $\psi$ is defined as $M_{2}(\psi):=-\log_{2} W(\psi)-S_{2}(\psi)-\log d$, where $W(\psi):=\operatorname{tr}(Q{\psi}^{\otimes 4})$, $S_{2}(\psi)=\operatorname{tr}(T\psi^{\otimes 2})$ and the operator $T$ is the swap operator while $Q:={d^{-2}}\sum_{P}P^{\otimes 4}$. 
The local  randomized measurements protocol we introduce here aims at measuring $W(\psi)$ and $P(\psi)$ by only using   single qubit gates and then measuring the qubits in the computational basis. In this way, we reduce access to multi-qubit gates that are typically noisier and whose control is poorer. The logic behind any randomized measurement protocol is to reconstruct operators (e.g. the swap operator for the purity or higher order permutations for higher order purities, see \cite{Enk2012measure,Elben2018Entropies,Elben2019probing,Vermersch2019meas,Elben2020randomized,Elben2020cross,Elben2020topological}) from correlations between randomized measurements. The measurement is randomized by means of Clifford single qubit gates. It is fundamental to use Clifford gates as magic is invariant under these unitary operations. The ideal experimental protocol for measuring simultaneously $W(\psi)$ and $P(\psi)$ is (see Fig.1 for a pictorial schematization):
\vspace{0.4cm}
\begin{enumerate}[label=(\Roman*)]
    \item pick $N_{U}$ random local Clifford operators $C=\bigotimes_{i=1}^{n}c_{i}$ where $c_{i}\in\mathcal{C}_{1}$ are single qubit Clifford gates. For each $C$ do:
    \vspace{0.3cm}
    \begin{enumerate}[label=(\roman*)]
    \item obtain the desired state $\psi$ from the quantum circuit $U$,
    \vspace{0.2cm}
    \item apply $C$  on the state $\psi_{C}\equiv C\psi C^{\dagger}$,
    \vspace{0.2cm}
    \item measure in the computational basis,
    \vspace{0.2cm}
    \item redo steps $(i)$, $(ii)$ and $(iii)$ $N_{M}$ times to estimate the occupation probabilities $Pr(\psi_{C}\!\mid\! s)\equiv \operatorname{tr}(\mid\!\!s\rangle\langle s\!\!\mid \psi_{C})$ for $s=1,\dots, 2^n$,
    \end{enumerate}
    \vspace{0.3cm}    
   \item Estimate the probabilities $\tilde{Pr}(\psi_{C}\!\mid\! s)$ by measuring the frequencies of obtaining the bit-string $s_2$  in the state $\psi_{C}$. The estimator $\tilde{Pr}(\psi_{C}\!\mid\! s)$ for such probability converges to the true probability $Pr(\psi_{C}\!\mid\! s)$ in the limit $N_{M}\rightarrow\infty$. 
 \item  Obtain  $P(\psi)$ and $W(\psi)$ can be computed from the ideal probabilities $Pr(\psi_{C}\!\mid\! s)$ by:
    \begin{eqnarray}
    P(\psi)&=&\frac{1}{24^n}\sum_{C\in\mathcal{C}_{1}^{\otimes n}}\sum_{s_{1},s_{2}=1}^{2^n}O_{2}(s_{1},s_{2})Pr(\psi_{C}\!\mid\! s_1)Pr(\psi_{C}\!\mid\! s_{2})\label{randomizedmeasurementsequations}\\
    W(\psi)\!&=&\!\!\frac{1}{24^n}\!\!\!\sum_{C\in\mathcal{C}_{1}^{\otimes n}}\sum_{s_{1},\dots,s_{4}=1}^{2^n}\!\!\!O_{4}(s_{1},s_{2},s_{3},s_{4})Pr(\psi_{C}\!\!\mid\!\! s_1)Pr(\psi_{C}\!\!\mid\!\! s_{2})Pr(\psi_{C}\!\!\mid\!\! s_{3})Pr(\psi_{C}\!\!\mid\!\! s_{4})\nonumber\\\label{randomizedmeasurementsequations2}
    \end{eqnarray}
\end{enumerate}
 the  weighting coefficients $O_{2}(s_{1},s_{2})$ and $O_{4}(s_{1},s_{2},s_3,s_4)$ are obtained in the following way. First, define  two diagonal operators defined in $\mathcal{H}^{\otimes 2}$ and $\mathcal{H}^{\otimes 4}$ respectively:
 \begin{eqnarray}
\hat{O}_{2}&:=&\sum_{s_1,s_2}O_{2}(s_{1},s_{2})\mid\!\!s_1s_2\rangle\langle s_1s_2\!\!\mid\\
\hat{O}_{4}&:=&\sum_{s_1,s_2,s_3,s_4}O_{4}(s_{1},s_{2},s_3,s_4)\mid\!\!s_1s_2s_3s_4\rangle\langle s_1 s_2 s_3 s_4 \!\!\mid
\end{eqnarray}
Let us now prove  Eqs. \eqref{randomizedmeasurementsequations} and \eqref{randomizedmeasurementsequations2} and show the exact form of $O_{2}$ and $O_{4}$. 
 Let us rewrite Eqs. \eqref{randomizedmeasurementsequations} and \eqref{randomizedmeasurementsequations2} writing the purity as $P(\psi)=\operatorname{tr}(T{\psi}^{\otimes 2})$ and $W(\psi)=\operatorname{tr}(Q{\psi}^{\otimes 4})$
\begin{eqnarray}
    \operatorname{tr}(T{\psi}^{\otimes 2})&=&\frac{1}{24^n}\sum_{C}\operatorname{tr}(C^{\dagger\otimes 2}\hat{O}_2C^{\otimes 2}{\psi}^{\otimes 2})\\
    \operatorname{tr}(Q{\psi}^{\otimes 4})&=&\frac{1}{24^n}\sum_{C}\operatorname{tr}(C^{\dagger\otimes 4}\hat{O}_4C^{\otimes 4}{\psi}^{\otimes 4})\nonumber
\end{eqnarray}
from the above equation is clear that the task is to find two diagonal operators $\hat{O}_{2}$ and $\hat{O}_{4}$ whose local Clifford average gives $T$ and $Q$ respectively. Recalling that $T=\frac{1}{2^{n}}(\mathchoice {\rm 1\mskip-4mu l} {\rm 1\mskip-4mu l} {\rm 1\mskip-4.5mu l} {\rm 1\mskip-5mu l}^{\otimes 2}+X^{\otimes 2}+Y^{\otimes 2}+Z^{\otimes 2})^{\otimes n}$, and $Q=\frac{1}{4^{n}}(\mathchoice {\rm 1\mskip-4mu l} {\rm 1\mskip-4mu l} {\rm 1\mskip-4.5mu l} {\rm 1\mskip-5mu l}^{\otimes 4}+X^{\otimes 4}+Y^{\otimes 4}+Z^{\otimes 4})^{\otimes n}$, it is sufficient to find two single qubit diagonal operators $\hat{o}_{2}$ and $\hat{o}_{4}$ living in $\mathbb{C}^{2\otimes 2}$ and $\mathbb{C}^{2\otimes 4}$ respectively, such that their Clifford average gives $T_{1}\equiv\frac{1}{2}(\mathchoice {\rm 1\mskip-4mu l} {\rm 1\mskip-4mu l} {\rm 1\mskip-4.5mu l} {\rm 1\mskip-5mu l}^{\otimes 2}+X^{\otimes 2}+Y^{\otimes 2}+Z^{\otimes 2})$ and $Q_{1}\equiv\frac{1}{4}(\mathchoice {\rm 1\mskip-4mu l} {\rm 1\mskip-4mu l} {\rm 1\mskip-4.5mu l} {\rm 1\mskip-5mu l}^{\otimes 4}+X^{\otimes 4}+Y^{\otimes 4}+Z^{\otimes 4})$ respectively. At this point, it is straightforward to verify that one should choose $\hat{o}_{2}, \hat{o}_{4}$ to be
\begin{eqnarray}
\hat{o}_{2}&\equiv& \frac{\mathchoice {\rm 1\mskip-4mu l} {\rm 1\mskip-4mu l} {\rm 1\mskip-4.5mu l} {\rm 1\mskip-5mu l}^{\otimes 2}}{2}+\frac{3}{2}Z^{\otimes 2}\\
\hat{o}_{4}&\equiv& \frac{\mathchoice {\rm 1\mskip-4mu l} {\rm 1\mskip-4mu l} {\rm 1\mskip-4.5mu l} {\rm 1\mskip-5mu l}^{\otimes 4}}{4}+\frac{3}{4}Z^{\otimes 4}
\end{eqnarray}
To conclude the proof is sufficient to write $\hat{o}_{2}$ and $\hat{o}_{4}$ in the computational basis to restore the forms of Eqs. \eqref{randomizedmeasurementsequations} and  \eqref{randomizedmeasurementsequations2}. It's easy to verify:
\begin{eqnarray}
O_{2}(s_{1},s_{2})&=&(-2)^{-\sum_{i=1}^{n}s^{i}_{1}\oplus s^{i}_{2}}\nonumber\\
O_{4}(s_{1},s_{2},s_3,s_4)&=&(-2)^{-\sum_{i=1}^{n}s^{i}_{1}\oplus s^{i}_{2}\oplus s^{i}_{3}\oplus s^{i}_{4}}
\end{eqnarray}
where $s_{k}=s_{k}^{1}s_{k}^{2}\dots s_{k}^{n}$ a $n$-length bit string for $k=1,2,3,4$ and $\oplus$ is the logic sum between bits.
\medskip

\subsubsection*{Statistical analysis}\label{Sec: statanalysis}

\smallskip

In this section, we discuss the effect of a finite number of realizations of the experiment. In our scheme,  statistical errors arise from two factors: $(i)$ a partial sampling of the local (single qubit) Clifford group, that is,  $N_U<24^n$, and $(ii)$ the finite number of measurement shots $N_M$ per unitary selected to estimate the occupation probability $\tilde{Pr}(\psi_{C}\!\mid\! s)$, introduced in the previous section, that converge to the true probability only in the limit $N_M\rightarrow\infty$. The total number of resources is thus $N_U\times N_M$. We assume that different rounds of random local unitary and different shots for a given unitary are generated independently and identically distributed. One describes the $i$-th shot for a given sampled unitary $C$ as $\tilde{s}_{C}(i)$ which takes value $\mid\!\!s\rangle\langle s\!\!\mid$ with probability $Pr(\psi_C\!\!\mid\!\!s)\equiv \operatorname{tr}(\mid\!\!s\rangle\langle s\!\!\mid C\psi C^{\dagger})$. An unbiased estimator for the stabilizer purity is given by:
\begin{equation}
\tilde{W}(\psi)=\frac{1}{N_U}\sum_C\tilde{W}_{C}(\psi)
\end{equation}
where $\tilde{W}_C(\psi):=\binom{N_M}{4}^{-1}\sum_{i<j<k<l}\operatorname{tr}(\tilde{s}_{C}(i)\otimes \tilde{s}_{C}(j)\otimes \tilde{s}_{C}(k)\otimes \tilde{s}_{C}(l) \hat{O}_4 )$. Let us prove that it is an unbiased estimator:
\begin{eqnarray}
\mathbb{E}_C\mathbb{E}_{s}\tilde{W}(\psi)&=&\mathbb{E}_C\binom{N_M}{4}^{-1}\sum_{i<j<k<l}\operatorname{tr}(\mathbb{E}_{s}\tilde{s}_{C}(i)\otimes \mathbb{E}_{s}\tilde{s}_{C}(j)\otimes \mathbb{E}_{s}\tilde{s}_{C}(k)\otimes \mathbb{E}_{s}\tilde{s}_{C}(l) \hat{O}_4)\nonumber\\
&=&\mathbb{E}_C\operatorname{tr}(C^{\otimes 4}\psi^{\otimes 4} C^{\dagger\otimes 4}\hat{O}_4)=W(\psi)
\end{eqnarray}
where we used the fact that $\mathbb{E}_s\tilde{s}_U(i)\equiv\sum_{s}\mid\!\!s\rangle\langle s\!\!\mid Pr(\psi_C\!\!\mid \!\!s)=\psi_C$. Our task now is to bound the number of resources needed to estimate $W$ within an error $\epsilon$. We compute the variance given a finite number of shot measurements $N_M$ and a finite sample $N_U$ of the local Clifford group. The variance of the estimator $\tilde{W}(\psi)$ can be written as:
\begin{eqnarray}
\operatorname{Var}[\tilde{W}(\psi)]&=&\frac{1}{N_U}\operatorname{Var}[\tilde{W}_C(\psi)]\nonumber\\
&=&\frac{1}{N_U}\mathbb{E}_C\left[\mathbb{E}_{\mathbf{s}}(\tilde{W}^2_C(\psi)\!\!\mid\!\!C)\right]-\frac{1}{N_U}\left[\mathbb{E}_C\mathbb{E}_\mathbf{s}(\tilde{W}_C(\psi)\!\!\mid\!\!C)\right]^2 
\end{eqnarray}
After some lengthy algebra (see Supplementary Note $3$) it is possible to bound the above variance as:
\begin{eqnarray}
\operatorname{Var}[\tilde{W}(\psi)]&\le&\frac{1}{N_U}\Big[ \frac{8}{\sqrt{d}}+\frac{192}{d^{1/3}N_M^4}+\frac{6792}{d^{1/2}N_M^4}\nonumber\\&+&\frac{5056}{N_M^3}+\frac{8179}{d^{1/2}N_M^2}+\frac{128}{N_M} - \operatorname{tr}[Q\psi^{\otimes 4}]^2\Big]
\label{Eq: variancestabpurity}
\end{eqnarray}
Finally, we make use of Bernstein’s inequality to bound the probability of an estimation within an error $\epsilon$:
\begin{equation}
\operatorname{Pr}\left(\mid\!\! \tilde{W}(\psi)-W(\psi)\!\!\mid\ge \epsilon\right)\le 2^{-\frac{N_U\epsilon^2}{\mathbf{Var}(\tilde{W}_U(\psi))+\frac{2\epsilon}{3}}}
\end{equation}
In the regime of interest, i.e. $d\gg 1$ and $N_M\gg1$ the variance scales like $\operatorname{Var}[\tilde{W}(\psi)]\lesssim \frac{1}{N_U}\left(\frac{c_1}{\sqrt{d}}+\frac{c_2}{N_M}\right)$, where $c_1,c_2$ are two constants. Setting $N_{M}=\mathcal{O}(\sqrt{d})$, in order to have an error $\epsilon$, and an exponentially small probability to fail, the total number of resources $N_U\times N_M$ for the stabilizer purity scales as $\mathcal{O}(\epsilon^{-2})$.}

At this point, a comment is necessary. The stabilizer purity is bounded between $d^{-2}\le \tilde{W}(\psi)\le d^{-1}$, which means that, to have a faithful measurement of $W(\psi)$, the error $\epsilon$ must be (at least) exponentially small in the number of qubits, $\epsilon \sim d^{-1}$. This makes the number of resources exponentially large in $n$. Similarly, for the purity  $P(\psi)$ (see Supplementary Note $3$), setting $N_M=\mathcal{O}(\sqrt{d})$, the variance is $\operatorname{Var}[\tilde{\operatorname{Pur}}(\psi)]= \mathcal{O}(d N_{U}^{-1})$. Thus the number of resources to estimate the purity up to an error $\epsilon$ scales as $\mathcal{O}(\epsilon^{-2}\sqrt{d^3})$.

Therefore, employing a total number of resources 
\begin{equation}
N_{U}\times N_{M}=\mathcal{O}(\epsilon^{-2}d^2)    
\end{equation}
the randomized measurement protocol is able to estimate the purity within an error $\le \epsilon $ and the stabilizer purity within an error $\epsilon d^{-1}$. In the next section, we describe the experimental protocol used to perform the experiments on IBM quantum processors.

\medskip

{\bf Experimental protocol}\label{Sec: experimentalprotocol}

\smallskip

\begin{figure}
    \centering
    \includegraphics[scale=0.3]{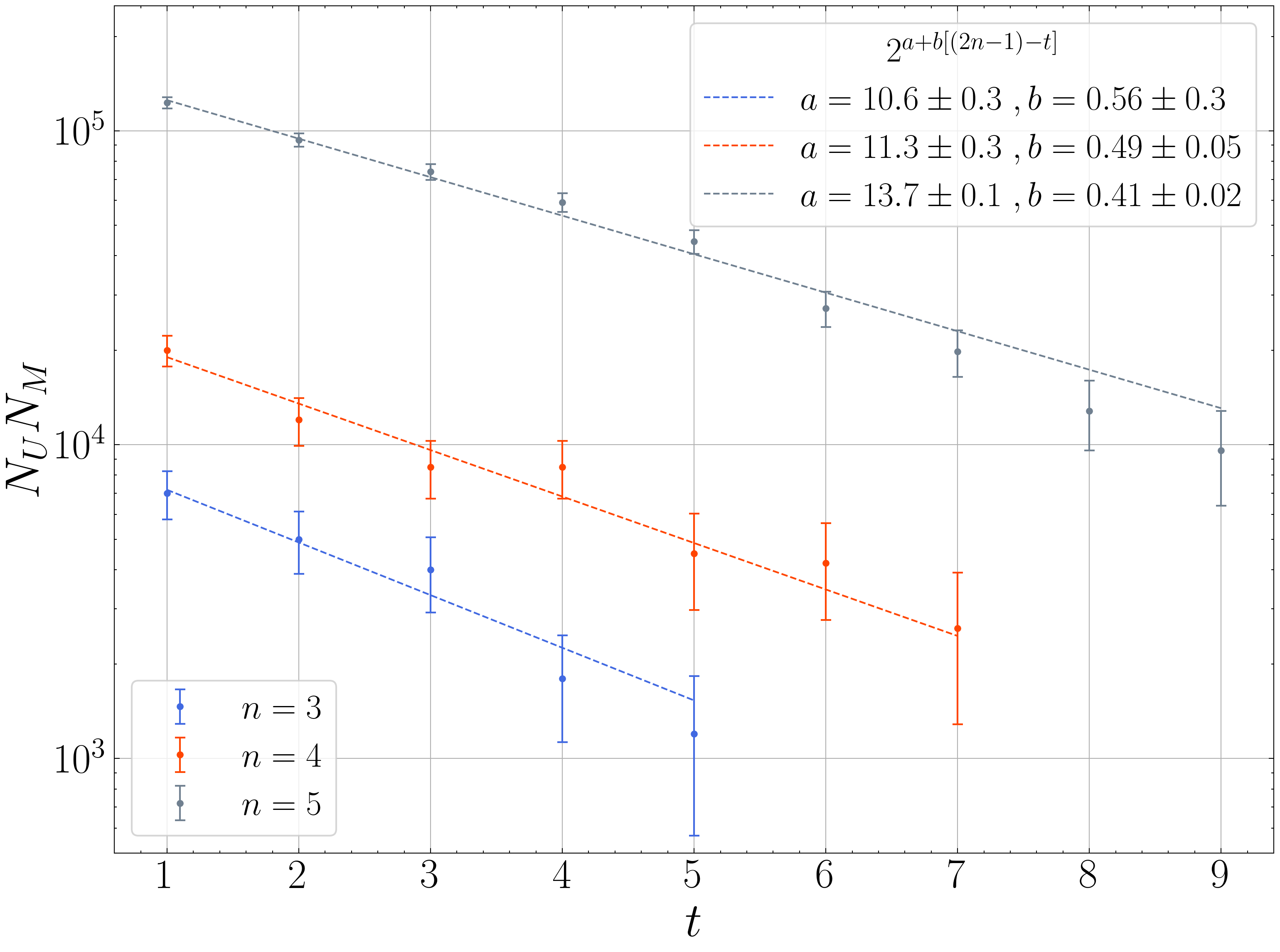}
\caption{{\bf Number of optimal resources $N_U$ and $N_M$.} The figure shows the log-plot of the optimal number of resources $N_{TOT}=N_{U}\times N_{M}$ for $n=3$ (blue dots), $n=4$ (orange dots) and $n=5$ (grey dots) as a function of the number of $T$-gates $t$  injected in the circuit. The dashed lines represent the fit $N_{TOT}=2^{a+b[(2n-1)-t]}$; the values for $a$ and $b$ with the respective errors are reported in the bottom-left corner. The fitted curves are in perfect agreement with the experimental data, whose error bars are due to the finite resolution of the grid: the $R$-squared parameters are $R^{2}_{(3)}=0.985$ for $n=3$, $R^{2}_{(4)}=0.985$ for $n=4$ and $R^{2}_{(5)}=0.995$ for $n=5$. }
\label{fig: numberofresources}  
\end{figure}

To measure the magic of multiqubit states on a quantum processor via statistical correlations between randomized measurements we need three steps: $(i)$ state preparation, $(ii)$ the application of $N_{U}$ random local Clifford unitaries to sample the local $n$-qubit Clifford group, whose dimension is $\mid\!\! C_{loc}(2^{n})\!\!\mid=24^{n}$, and $(iii)$ $N_{M}$ projective measurements to estimate  the probabilities $\tilde{P}_{M}(\psi_{C}\mid s)$. Then, the experimental purity and stabilizer purity are measured as:
\begin{eqnarray}
P({\psi})&=&\frac{1}{N_{U}}\sum_{C}\sum_{s_{1},s_{2}=1}^{2^n}O_{2}(s_{1},s_{2})\tilde{P}_{M}(\psi_{C}\!\!\mid\!\! s_1)\tilde{P}_{M}(\psi_{C}\!\!\mid\!\! s_{2})\\
W({\psi})\!&=&\!\!\frac{1}{N_{U}}\sum_{C}\sum_{s_{1},\dots,s_{4}}^{2^n}\!\!\!O_{4}(s_{1},s_{2},s_{3},s_{4})\tilde{P}_{M}(\psi_{C}\!\!\mid\!\! s_1)\tilde{P}_{M}(\psi_{C}\!\!\mid\!\! s_{2})\tilde{P}_{M}(\psi_{C}\!\!\mid\!\! s_{3})\tilde{P}_{M}(\psi_{C}\!\!\mid\!\! s_{4})\nonumber
\end{eqnarray}
We proved that one needs $N_U\times N_M=\mathcal{O}(\epsilon^{-2}d^2)$ total measurements to estimate the stabilizer purity within a error $\epsilon^{-1}d^{-1}$. Since we obtained such an accuracy guarantee through crude bounds, we expect fewer resources to be spent. We thus follow the protocol employed in \cite{Brydges2019randomized} to find the optimal number of unitaries $N_{U}$ and measurements $N_{M}$. We first build a preliminary $10\times 10$ grid and make $100$ numerical simulation for $10$ different values of $N_{U}=8,\dots,1024$ and $10$ different values $N_{M}=32,\dots,1024$ (the latter taken with logarithmic spacing) for $2$ extreme states, namely the input state $\!\mid\!\!0\rangle^{\otimes n}$ and the final doped Clifford state $\mid\!\!\Gamma_{2n-1}^{(n)}\rangle$, see Fig. \ref{gammastates}. Then, for each value of $N_{U}$ and $N_{M}$ we compute the average $\overline{W_{N_U,N_M}(\mid\!\psi\rangle)}$ over $100$ different realizations, the average purity $\overline{P_{N_U,N_M}(\mid\!\psi\rangle)}$ and the average percent distance $\delta_{N_U,N_M}$ from the average
\begin{equation}
\delta_{N_U,N_M}(\mid\!\psi\rangle):=\frac{\overline{\mid\!\!W_{N_U,N_M}(\mid\!\psi\rangle)-\overline{W_{N_U,N_M}(\mid\!\psi\rangle)}\!\!\mid}}{\overline{W_{N_U,N_M}(\mid\!\psi\rangle)}}
\end{equation}
To obtain the optimal number of $N_{U}$ and $N_{M}$ for the given states $\!\mid\!\!0\rangle^{\otimes n}$ and $\mid\!\!\Gamma_{2n-1}^{(n)}\rangle$, we set a threshold on the average distance $\delta_{N_U,N_M}$ and on the average purity $\overline{P_{N_U,N_M}(\mid\!\psi\rangle)}$: 
\vspace{0.3cm}
\begin{enumerate}[label=(\roman*)]
    \item $\delta_{N_U,N_M}(\mid\!\psi\rangle)<12\%$
    \vspace{0.2cm}
    \item $\mid\!\!\overline{P_{N_U,N_M}(\mid\!\psi\rangle)}-1\!\!\mid< 12\%$
    \vspace{0.4cm}
\end{enumerate}

and pick the pair of $N_{U}, N_{M}$ satisfying conditions $(i)$ and $(ii)$ minimizing their product $N_{U}N_{M}$, i.e. the optimal number of resources. Indeed the product of $N_{U}N_{M}$ is the number of physical times that one redoes the actual experiment and thus the number of necessary resources to perform an experiment. Remarkably, the number of unitaries $N_{U}$ and the number of measurements $N_{M}$ do depend on the state of interest $\mid\!\psi\rangle$. In particular, denoting $N_{U}^{t=2n-1},N_{M}^{t=2n-1}$ and $N_{U}^{t=0},N_{M}^{t=0}$ the number of resources for $\mid\!\!\Gamma_{2n-1}^{(n)}\rangle$ and $\!\mid\!\!0\rangle^{\otimes n}$ respectively, we find $N_{U}^{t=2n-1}<N_{U}^{t=0}$ and $N_{M}^{t=2n-1}<N_{M}^{t=0}$. These findings suggest that the optimal number of resources $N_{U}\times N_{M}$ do depend on the number $t$ of \textit{magic seeds}, i.e. $T$-gates, thrown in the circuit. Thus, in order to find optimal values for $N_{U}$ and $N_{M}$ for all the states of interest $\mid\!\!\Gamma_{t}^{(n)}\rangle$ $t=1,\dots, 2n-1$, we build a linear spaced $10\times 10$ grid for $10$ different value of $N_{U}$ ranging in $[N_{U}^{t=2n-1},\dots, N_{U}^{t=0}]$ and $10$ different values of $N_{M}$ ranging in $[N_{M}^{t=2n-1},\dots, N_{M}^{t=0}]$ for fixed $n$; then we make $100$ numerical simulations and pick the optimal number of resources satisfying conditions (i) and (ii). In this way, we are able to determine the optimal number of resources state by state, see Supplementary Table I in Supplementary Note $4$ for the results. The data are fitted to depend exponentially upon the number $t$ of magic-seeds, as $N_{TOT}=2^{a+b[(2n-1)-t]}$, see Fig. \ref{fig: numberofresources}. The experimental errors on the estimated $P(\psi)$ and $W(\psi)$ are chosen to be the standard error of the average over $N_{U}$, i.e. over the local Clifford operators used to estimate these two quantities from randomized measurements (see Supplementary Note $4$).

\smallskip

{\bf Data Availability}
The authors declare that the main data supporting the findings of this study are available within the article and its Supplementary Information files. Extra data sets are available upon reasonable request.

\smallskip

{\bf Acknowledgments}
The authors thank Benoît Vermersch, Agata Branczyk, Roberto Schiattarella, and Aurora Langella for inspiring discussions and comments. L.L., S.F.E.O. and A.H. acknowledge support from NSF award no. 2014000. The authors also thank the anonymous referees for their constructive comments and suggestions that helped to improve the manuscript. The work of L.L. and S.F.E.O. was supported in part by College of Science and Mathematics Dean’s Doctoral Research Fellowship through fellowship support from Oracle, project ID R20000000025727.  S.L. was supported by DARPA, AFOSR, and ARO under a Blue Sky grant. 


\begin{thebibliography}{55}%
\makeatletter
\providecommand \@ifxundefined [1]{%
 \@ifx{#1\undefined}
}%
\providecommand \@ifnum [1]{%
 \ifnum #1\expandafter \@firstoftwo
 \else \expandafter \@secondoftwo
 \fi
}%
\providecommand \@ifx [1]{%
 \ifx #1\expandafter \@firstoftwo
 \else \expandafter \@secondoftwo
 \fi
}%
\providecommand \natexlab [1]{#1}%
\providecommand \enquote  [1]{``#1''}%
\providecommand \bibnamefont  [1]{#1}%
\providecommand \bibfnamefont [1]{#1}%
\providecommand \citenamefont [1]{#1}%
\providecommand \href@noop [0]{\@secondoftwo}%
\providecommand \href [0]{\begingroup \@sanitize@url \@href}%
\providecommand \@href[1]{\@@startlink{#1}\@@href}%
\providecommand \@@href[1]{\endgroup#1\@@endlink}%
\providecommand \@sanitize@url [0]{\catcode `\\12\catcode `\$12\catcode
  `\&12\catcode `\#12\catcode `\^12\catcode `\_12\catcode `\%12\relax}%
\providecommand \@@startlink[1]{}%
\providecommand \@@endlink[0]{}%
\providecommand \url  [0]{\begingroup\@sanitize@url \@url }%
\providecommand \@url [1]{\endgroup\@href {#1}{\urlprefix }}%
\providecommand \urlprefix  [0]{URL }%
\providecommand \Eprint [0]{\href }%
\providecommand \doibase [0]{https://doi.org/}%
\providecommand \selectlanguage [0]{\@gobble}%
\providecommand \bibinfo  [0]{\@secondoftwo}%
\providecommand \bibfield  [0]{\@secondoftwo}%
\providecommand \translation [1]{[#1]}%
\providecommand \BibitemOpen [0]{}%
\providecommand \bibitemStop [0]{}%
\providecommand \bibitemNoStop [0]{.\EOS\space}%
\providecommand \EOS [0]{\spacefactor3000\relax}%
\providecommand \BibitemShut  [1]{\csname bibitem#1\endcsname}%
\let\auto@bib@innerbib\@empty
\bibitem [{\citenamefont {Preskill}(2018)}]{Preskill2018quantumcomputing}%
  \BibitemOpen
  \bibfield  {author} {\bibinfo {author} {\bibfnamefont {J.}~\bibnamefont
  {Preskill}},\ }\href {https://doi.org/10.22331/q-2018-08-06-79} {\bibfield
  {journal} {\bibinfo  {journal} {{Quantum}}\ }\textbf {\bibinfo {volume}
  {2}},\ \bibinfo {pages} {79} (\bibinfo {year} {2018})}\BibitemShut {NoStop}%
\bibitem [{\citenamefont {Leone}\ \emph {et~al.}(2022)\citenamefont {Leone},
  \citenamefont {Oliviero},\ and\ \citenamefont {Hamma}}]{leone2021renyi}%
  \BibitemOpen
  \bibfield  {author} {\bibinfo {author} {\bibfnamefont {L.}~\bibnamefont
  {Leone}}, \bibinfo {author} {\bibfnamefont {S.~F.~E.}\ \bibnamefont
  {Oliviero}},\ and\ \bibinfo {author} {\bibfnamefont {A.}~\bibnamefont
  {Hamma}},\ }\href {https://doi.org/10.1103/PhysRevLett.128.050402} {\bibfield
   {journal} {\bibinfo  {journal} {Physical Review Letters}\ }\textbf {\bibinfo
  {volume} {128}},\ \bibinfo {pages} {050402} (\bibinfo {year}
  {2022})}\BibitemShut {NoStop}%
\bibitem [{\citenamefont {Campbell}\ and\ \citenamefont
  {Browne}(2010)}]{campbell2010bound}%
  \BibitemOpen
  \bibfield  {author} {\bibinfo {author} {\bibfnamefont {E.~T.}\ \bibnamefont
  {Campbell}}\ and\ \bibinfo {author} {\bibfnamefont {D.~E.}\ \bibnamefont
  {Browne}},\ }\href {https://doi.org/10.1103/PhysRevLett.104.030503}
  {\bibfield  {journal} {\bibinfo  {journal} {Physical Review Letters}\
  }\textbf {\bibinfo {volume} {104}},\ \bibinfo {pages} {030503} (\bibinfo
  {year} {2010})}\BibitemShut {NoStop}%
\bibitem [{\citenamefont {Campbell}(2011)}]{Campbell2011Catalysis}%
  \BibitemOpen
  \bibfield  {author} {\bibinfo {author} {\bibfnamefont {E.~T.}\ \bibnamefont
  {Campbell}},\ }\href {https://doi.org/10.1103/PhysRevA.83.032317} {\bibfield
  {journal} {\bibinfo  {journal} {Physical Review A}\ }\textbf {\bibinfo
  {volume} {83}},\ \bibinfo {pages} {032317} (\bibinfo {year}
  {2011})}\BibitemShut {NoStop}%
\bibitem [{\citenamefont {Campbell}\ \emph
  {et~al.}(2012{\natexlab{a}})\citenamefont {Campbell}, \citenamefont {Anwar},\
  and\ \citenamefont {Browne}}]{campbell2012magic}%
  \BibitemOpen
  \bibfield  {author} {\bibinfo {author} {\bibfnamefont {E.~T.}\ \bibnamefont
  {Campbell}}, \bibinfo {author} {\bibfnamefont {H.}~\bibnamefont {Anwar}},\
  and\ \bibinfo {author} {\bibfnamefont {D.~E.}\ \bibnamefont {Browne}},\
  }\href {https://doi.org/10.1103/PhysRevX.2.041021} {\bibfield  {journal}
  {\bibinfo  {journal} {Physical Review X}\ }\textbf {\bibinfo {volume} {2}},\
  \bibinfo {pages} {041021} (\bibinfo {year} {2012}{\natexlab{a}})}\BibitemShut
  {NoStop}%
\bibitem [{\citenamefont {Campbell}\ \emph {et~al.}(2017)\citenamefont
  {Campbell}, \citenamefont {Terhal},\ and\ \citenamefont
  {Vuillot}}]{Campbell2017fault}%
  \BibitemOpen
  \bibfield  {author} {\bibinfo {author} {\bibfnamefont {E.~T.}\ \bibnamefont
  {Campbell}}, \bibinfo {author} {\bibfnamefont {B.~M.}\ \bibnamefont
  {Terhal}},\ and\ \bibinfo {author} {\bibfnamefont {C.}~\bibnamefont
  {Vuillot}},\ }\href {https://doi.org/10.1038/nature23460} {\bibfield
  {journal} {\bibinfo  {journal} {Nature}\ }\textbf {\bibinfo {volume} {549}},\
  \bibinfo {pages} {172} (\bibinfo {year} {2017})}\BibitemShut {NoStop}%
\bibitem [{\citenamefont {Campbell}(2014)}]{campbell2014enhanced}%
  \BibitemOpen
  \bibfield  {author} {\bibinfo {author} {\bibfnamefont {E.~T.}\ \bibnamefont
  {Campbell}},\ }\href {https://doi.org/10.1103/PhysRevLett.113.230501}
  {\bibfield  {journal} {\bibinfo  {journal} {Physical Review Letters}\
  }\textbf {\bibinfo {volume} {113}},\ \bibinfo {pages} {230501} (\bibinfo
  {year} {2014})}\BibitemShut {NoStop}%
\bibitem [{\citenamefont {Campbell}\ and\ \citenamefont
  {Howard}(2017{\natexlab{a}})}]{campbell2017unified}%
  \BibitemOpen
  \bibfield  {author} {\bibinfo {author} {\bibfnamefont {E.~T.}\ \bibnamefont
  {Campbell}}\ and\ \bibinfo {author} {\bibfnamefont {M.}~\bibnamefont
  {Howard}},\ }\href {https://doi.org/10.1103/PhysRevA.95.022316} {\bibfield
  {journal} {\bibinfo  {journal} {Physical Review A}\ }\textbf {\bibinfo
  {volume} {95}},\ \bibinfo {pages} {022316} (\bibinfo {year}
  {2017}{\natexlab{a}})}\BibitemShut {NoStop}%
\bibitem [{\citenamefont {Campbell}\ and\ \citenamefont
  {Howard}(2017{\natexlab{b}})}]{campbell2017unifying}%
  \BibitemOpen
  \bibfield  {author} {\bibinfo {author} {\bibfnamefont {E.~T.}\ \bibnamefont
  {Campbell}}\ and\ \bibinfo {author} {\bibfnamefont {M.}~\bibnamefont
  {Howard}},\ }\href {https://doi.org/10.1103/PhysRevLett.118.060501}
  {\bibfield  {journal} {\bibinfo  {journal} {Physical Review Letters}\
  }\textbf {\bibinfo {volume} {118}},\ \bibinfo {pages} {060501} (\bibinfo
  {year} {2017}{\natexlab{b}})}\BibitemShut {NoStop}%
\bibitem [{\citenamefont {Knill}\ and\ \citenamefont
  {Laflamme}(1997)}]{Knill1996codes}%
  \BibitemOpen
  \bibfield  {author} {\bibinfo {author} {\bibfnamefont {E.}~\bibnamefont
  {Knill}}\ and\ \bibinfo {author} {\bibfnamefont {R.}~\bibnamefont
  {Laflamme}},\ }\href {https://doi.org/10.1103/PhysRevA.55.900} {\bibfield
  {journal} {\bibinfo  {journal} {Physical Review A}\ }\textbf {\bibinfo
  {volume} {55}},\ \bibinfo {pages} {900} (\bibinfo {year} {1997})}\BibitemShut
  {NoStop}%
\bibitem [{\citenamefont {Gottesman}(1998{\natexlab{a}})}]{Gottesman:1998hu}%
  \BibitemOpen
  \bibfield  {author} {\bibinfo {author} {\bibfnamefont {D.}~\bibnamefont
  {Gottesman}},\ }\href {https://doi.org/10.48550/arXiv.quant-ph/9807006}
  {\bibinfo {title} {The {{Heisenberg Representation}} of {{Quantum
  Computers}}}} (\bibinfo {year} {1998}{\natexlab{a}}),\ \Eprint
  {https://arxiv.org/abs/quant-ph/9807006} {arXiv:quant-ph/9807006}
  \BibitemShut {NoStop}%
\bibitem [{\citenamefont
  {Gottesman}(1998{\natexlab{b}})}]{gottesmann1998fault}%
  \BibitemOpen
  \bibfield  {author} {\bibinfo {author} {\bibfnamefont {D.}~\bibnamefont
  {Gottesman}},\ }\href {https://doi.org/10.1103/PhysRevA.57.127} {\bibfield
  {journal} {\bibinfo  {journal} {Physical Review A}\ }\textbf {\bibinfo
  {volume} {57}},\ \bibinfo {pages} {127} (\bibinfo {year}
  {1998}{\natexlab{b}})}\BibitemShut {NoStop}%
\bibitem [{\citenamefont {Aaronson}\ and\ \citenamefont
  {Gottesman}(2004)}]{aaronson2004improved}%
  \BibitemOpen
  \bibfield  {author} {\bibinfo {author} {\bibfnamefont {S.}~\bibnamefont
  {Aaronson}}\ and\ \bibinfo {author} {\bibfnamefont {D.}~\bibnamefont
  {Gottesman}},\ }\href {https://doi.org/10.1103/PhysRevA.70.052328} {\bibfield
   {journal} {\bibinfo  {journal} {Physical Review A}\ }\textbf {\bibinfo
  {volume} {70}},\ \bibinfo {pages} {052328} (\bibinfo {year}
  {2004})}\BibitemShut {NoStop}%
\bibitem [{\citenamefont {Hebenstreit}\ \emph {et~al.}(2019)\citenamefont
  {Hebenstreit}, \citenamefont {Jozsa}, \citenamefont {Kraus}, \citenamefont
  {Strelchuk},\ and\ \citenamefont {Yoganathan}}]{Hebenstreit2019gaussian}%
  \BibitemOpen
  \bibfield  {author} {\bibinfo {author} {\bibfnamefont {M.}~\bibnamefont
  {Hebenstreit}}, \bibinfo {author} {\bibfnamefont {R.}~\bibnamefont {Jozsa}},
  \bibinfo {author} {\bibfnamefont {B.}~\bibnamefont {Kraus}}, \bibinfo
  {author} {\bibfnamefont {S.}~\bibnamefont {Strelchuk}},\ and\ \bibinfo
  {author} {\bibfnamefont {M.}~\bibnamefont {Yoganathan}},\ }\href
  {https://doi.org/10.1103/PhysRevLett.123.080503} {\bibfield  {journal}
  {\bibinfo  {journal} {Physical Review Letters}\ }\textbf {\bibinfo {volume}
  {123}},\ \bibinfo {pages} {080503} (\bibinfo {year} {2019})}\BibitemShut
  {NoStop}%
\bibitem [{\citenamefont {Hebenstreit}\ \emph {et~al.}(2020)\citenamefont
  {Hebenstreit}, \citenamefont {Jozsa}, \citenamefont {Kraus},\ and\
  \citenamefont {Strelchuk}}]{Hebenstreit2020match}%
  \BibitemOpen
  \bibfield  {author} {\bibinfo {author} {\bibfnamefont {M.}~\bibnamefont
  {Hebenstreit}}, \bibinfo {author} {\bibfnamefont {R.}~\bibnamefont {Jozsa}},
  \bibinfo {author} {\bibfnamefont {B.}~\bibnamefont {Kraus}},\ and\ \bibinfo
  {author} {\bibfnamefont {S.}~\bibnamefont {Strelchuk}},\ }\href
  {https://doi.org/10.1103/PhysRevA.102.052604} {\bibfield  {journal} {\bibinfo
   {journal} {Physical Review A}\ }\textbf {\bibinfo {volume} {102}},\ \bibinfo
  {pages} {052604} (\bibinfo {year} {2020})}\BibitemShut {NoStop}%
\bibitem [{\citenamefont {Veitch}\ \emph {et~al.}(2014)\citenamefont {Veitch},
  \citenamefont {Mousavian}, \citenamefont {Gottesman},\ and\ \citenamefont
  {Emerson}}]{Veitch2014resource}%
  \BibitemOpen
  \bibfield  {author} {\bibinfo {author} {\bibfnamefont {V.}~\bibnamefont
  {Veitch}}, \bibinfo {author} {\bibfnamefont {S.~A.~H.}\ \bibnamefont
  {Mousavian}}, \bibinfo {author} {\bibfnamefont {D.}~\bibnamefont
  {Gottesman}},\ and\ \bibinfo {author} {\bibfnamefont {J.}~\bibnamefont
  {Emerson}},\ }\href {https://doi.org/10.1088/1367-2630/16/1/013009}
  {\bibfield  {journal} {\bibinfo  {journal} {New Journal of Physics}\ }\textbf
  {\bibinfo {volume} {16}},\ \bibinfo {pages} {013009} (\bibinfo {year}
  {2014})}\BibitemShut {NoStop}%
\bibitem [{\citenamefont {Howard}\ and\ \citenamefont
  {Campbell}(2017)}]{howard2017application}%
  \BibitemOpen
  \bibfield  {author} {\bibinfo {author} {\bibfnamefont {M.}~\bibnamefont
  {Howard}}\ and\ \bibinfo {author} {\bibfnamefont {E.}~\bibnamefont
  {Campbell}},\ }\href {https://doi.org/10.1103/PhysRevLett.118.090501}
  {\bibfield  {journal} {\bibinfo  {journal} {Physical Review Letters}\
  }\textbf {\bibinfo {volume} {118}},\ \bibinfo {pages} {090501} (\bibinfo
  {year} {2017})}\BibitemShut {NoStop}%
\bibitem [{\citenamefont {Ahmadi}\ \emph {et~al.}(2018)\citenamefont {Ahmadi},
  \citenamefont {Dang}, \citenamefont {Gour},\ and\ \citenamefont
  {Sanders}}]{Ahmadi2018magic}%
  \BibitemOpen
  \bibfield  {author} {\bibinfo {author} {\bibfnamefont {M.}~\bibnamefont
  {Ahmadi}}, \bibinfo {author} {\bibfnamefont {H.~B.}\ \bibnamefont {Dang}},
  \bibinfo {author} {\bibfnamefont {G.}~\bibnamefont {Gour}},\ and\ \bibinfo
  {author} {\bibfnamefont {B.~C.}\ \bibnamefont {Sanders}},\ }\href
  {https://doi.org/10.1103/PhysRevA.97.062332} {\bibfield  {journal} {\bibinfo
  {journal} {Physical Review A}\ }\textbf {\bibinfo {volume} {97}},\ \bibinfo
  {pages} {062332} (\bibinfo {year} {2018})}\BibitemShut {NoStop}%
\bibitem [{\citenamefont {Wang}\ \emph {et~al.}(2019)\citenamefont {Wang},
  \citenamefont {Wilde},\ and\ \citenamefont {Su}}]{Wang2019magic}%
  \BibitemOpen
  \bibfield  {author} {\bibinfo {author} {\bibfnamefont {X.}~\bibnamefont
  {Wang}}, \bibinfo {author} {\bibfnamefont {M.~M.}\ \bibnamefont {Wilde}},\
  and\ \bibinfo {author} {\bibfnamefont {Y.}~\bibnamefont {Su}},\ }\href
  {https://doi.org/10.1088/1367-2630/ab451d} {\bibfield  {journal} {\bibinfo
  {journal} {New Journal of Physics}\ }\textbf {\bibinfo {volume} {21}},\
  \bibinfo {pages} {103002} (\bibinfo {year} {2019})}\BibitemShut {NoStop}%
\bibitem [{\citenamefont {Seddon}\ and\ \citenamefont
  {Campbell}(2019)}]{seddon2019magic}%
  \BibitemOpen
  \bibfield  {author} {\bibinfo {author} {\bibfnamefont {J.~R.}\ \bibnamefont
  {Seddon}}\ and\ \bibinfo {author} {\bibfnamefont {E.~T.}\ \bibnamefont
  {Campbell}},\ }\href {https://doi.org/10.1098/rspa.2019.0251} {\bibfield
  {journal} {\bibinfo  {journal} {Proceedings of the Royal Society A:
  Mathematical, Physical and Engineering Sciences}\ }\textbf {\bibinfo {volume}
  {475}},\ \bibinfo {pages} {20190251} (\bibinfo {year} {2019})}\BibitemShut
  {NoStop}%
\bibitem [{\citenamefont {Liu}\ and\ \citenamefont
  {Winter}(2022)}]{liu2020many}%
  \BibitemOpen
  \bibfield  {author} {\bibinfo {author} {\bibfnamefont {Z.-W.}\ \bibnamefont
  {Liu}}\ and\ \bibinfo {author} {\bibfnamefont {A.}~\bibnamefont {Winter}},\
  }\href {https://doi.org/10.1103/PRXQuantum.3.020333} {\bibfield  {journal}
  {\bibinfo  {journal} {PRX Quantum}\ }\textbf {\bibinfo {volume} {3}},\
  \bibinfo {pages} {020333} (\bibinfo {year} {2022})}\BibitemShut {NoStop}%
\bibitem [{\citenamefont {Seddon}\ \emph {et~al.}(2021)\citenamefont {Seddon},
  \citenamefont {Regula}, \citenamefont {Pashayan}, \citenamefont {Ouyang},\
  and\ \citenamefont {Campbell}}]{seddon2021quantifying}%
  \BibitemOpen
  \bibfield  {author} {\bibinfo {author} {\bibfnamefont {J.~R.}\ \bibnamefont
  {Seddon}}, \bibinfo {author} {\bibfnamefont {B.}~\bibnamefont {Regula}},
  \bibinfo {author} {\bibfnamefont {H.}~\bibnamefont {Pashayan}}, \bibinfo
  {author} {\bibfnamefont {Y.}~\bibnamefont {Ouyang}},\ and\ \bibinfo {author}
  {\bibfnamefont {E.~T.}\ \bibnamefont {Campbell}},\ }\href
  {https://doi.org/10.1103/PRXQuantum.2.010345} {\bibfield  {journal} {\bibinfo
   {journal} {PRX Quantum}\ }\textbf {\bibinfo {volume} {2}},\ \bibinfo {pages}
  {010345} (\bibinfo {year} {2021})}\BibitemShut {NoStop}%
\bibitem [{\citenamefont {White}\ \emph {et~al.}(2021)\citenamefont {White},
  \citenamefont {Cao},\ and\ \citenamefont {Swingle}}]{White2021cft}%
  \BibitemOpen
  \bibfield  {author} {\bibinfo {author} {\bibfnamefont {C.~D.}\ \bibnamefont
  {White}}, \bibinfo {author} {\bibfnamefont {C.}~\bibnamefont {Cao}},\ and\
  \bibinfo {author} {\bibfnamefont {B.}~\bibnamefont {Swingle}},\ }\href
  {https://doi.org/10.1103/PhysRevB.103.075145} {\bibfield  {journal} {\bibinfo
   {journal} {Physical Review B}\ }\textbf {\bibinfo {volume} {103}},\ \bibinfo
  {pages} {075145} (\bibinfo {year} {2021})}\BibitemShut {NoStop}%
\bibitem [{\citenamefont {Qassim}\ \emph {et~al.}(2021)\citenamefont {Qassim},
  \citenamefont {Pashayan},\ and\ \citenamefont {Gosset}}]{qassim2021improved}%
  \BibitemOpen
  \bibfield  {author} {\bibinfo {author} {\bibfnamefont {H.}~\bibnamefont
  {Qassim}}, \bibinfo {author} {\bibfnamefont {H.}~\bibnamefont {Pashayan}},\
  and\ \bibinfo {author} {\bibfnamefont {D.}~\bibnamefont {Gosset}},\ }\href
  {https://doi.org/10.22331/q-2021-12-20-606} {\bibfield  {journal} {\bibinfo
  {journal} {{Quantum}}\ }\textbf {\bibinfo {volume} {5}},\ \bibinfo {pages}
  {606} (\bibinfo {year} {2021})}\BibitemShut {NoStop}%
\bibitem [{\citenamefont {Koukoulekidis}\ and\ \citenamefont
  {Jennings}(2022)}]{Kouko2021magic}%
  \BibitemOpen
  \bibfield  {author} {\bibinfo {author} {\bibfnamefont {N.}~\bibnamefont
  {Koukoulekidis}}\ and\ \bibinfo {author} {\bibfnamefont {D.}~\bibnamefont
  {Jennings}},\ }\href {https://doi.org/10.1038/s41534-022-00551-1} {\bibfield
  {journal} {\bibinfo  {journal} {npj Quantum Information}\ }\textbf {\bibinfo
  {volume} {8}},\ \bibinfo {pages} {1} (\bibinfo {year} {2022})}\BibitemShut
  {NoStop}%
\bibitem [{\citenamefont {Hahn}\ \emph {et~al.}(2022)\citenamefont {Hahn},
  \citenamefont {Ferraro}, \citenamefont {Hultquist}, \citenamefont {Ferrini}
  \emph {et~al.}}]{Hahn2021magic}%
  \BibitemOpen
  \bibfield  {author} {\bibinfo {author} {\bibfnamefont {O.}~\bibnamefont
  {Hahn}}, \bibinfo {author} {\bibfnamefont {A.}~\bibnamefont {Ferraro}},
  \bibinfo {author} {\bibfnamefont {L.}~\bibnamefont {Hultquist}}, \bibinfo
  {author} {\bibfnamefont {G.}~\bibnamefont {Ferrini}}, \emph {et~al.},\ }\href
  {https://doi.org/10.1103/PhysRevLett.128.210502} {\bibfield  {journal}
  {\bibinfo  {journal} {Physical Review Letters}\ }\textbf {\bibinfo {volume}
  {128}},\ \bibinfo {pages} {210502} (\bibinfo {year} {2022})}\BibitemShut
  {NoStop}%
\bibitem [{\citenamefont {Saxena}\ and\ \citenamefont
  {Gour}(2022)}]{Saxena2022stab}%
  \BibitemOpen
  \bibfield  {author} {\bibinfo {author} {\bibfnamefont {G.}~\bibnamefont
  {Saxena}}\ and\ \bibinfo {author} {\bibfnamefont {G.}~\bibnamefont {Gour}},\
  }\href {https://doi.org/10.1103/PhysRevA.106.042422} {\bibfield  {journal}
  {\bibinfo  {journal} {Physical Review A}\ }\textbf {\bibinfo {volume}
  {106}},\ \bibinfo {pages} {042422} (\bibinfo {year} {2022})}\BibitemShut
  {NoStop}%
\bibitem [{\citenamefont {Anwar}\ \emph {et~al.}(2012)\citenamefont {Anwar},
  \citenamefont {Campbell},\ and\ \citenamefont {Browne}}]{anwar2012qutrit}%
  \BibitemOpen
  \bibfield  {author} {\bibinfo {author} {\bibfnamefont {H.}~\bibnamefont
  {Anwar}}, \bibinfo {author} {\bibfnamefont {E.~T.}\ \bibnamefont
  {Campbell}},\ and\ \bibinfo {author} {\bibfnamefont {D.~E.}\ \bibnamefont
  {Browne}},\ }\href {https://doi.org/10.1088/1367-2630/14/6/063006} {\bibfield
   {journal} {\bibinfo  {journal} {New Journal of Physics}\ }\textbf {\bibinfo
  {volume} {14}},\ \bibinfo {pages} {063006} (\bibinfo {year}
  {2012})}\BibitemShut {NoStop}%
\bibitem [{\citenamefont {Campbell}\ \emph
  {et~al.}(2012{\natexlab{b}})\citenamefont {Campbell}, \citenamefont {Anwar},\
  and\ \citenamefont {Browne}}]{Campbell2012dist}%
  \BibitemOpen
  \bibfield  {author} {\bibinfo {author} {\bibfnamefont {E.~T.}\ \bibnamefont
  {Campbell}}, \bibinfo {author} {\bibfnamefont {H.}~\bibnamefont {Anwar}},\
  and\ \bibinfo {author} {\bibfnamefont {D.~E.}\ \bibnamefont {Browne}},\
  }\href {https://doi.org/10.1103/PhysRevX.2.041021} {\bibfield  {journal}
  {\bibinfo  {journal} {Physical Review X}\ }\textbf {\bibinfo {volume} {2}},\
  \bibinfo {pages} {041021} (\bibinfo {year} {2012}{\natexlab{b}})}\BibitemShut
  {NoStop}%
\bibitem [{\citenamefont {Bravyi}\ and\ \citenamefont
  {Haah}(2012)}]{Bravyi2012magic}%
  \BibitemOpen
  \bibfield  {author} {\bibinfo {author} {\bibfnamefont {S.}~\bibnamefont
  {Bravyi}}\ and\ \bibinfo {author} {\bibfnamefont {J.}~\bibnamefont {Haah}},\
  }\href {https://doi.org/10.1103/PhysRevA.86.052329} {\bibfield  {journal}
  {\bibinfo  {journal} {Physical Review A}\ }\textbf {\bibinfo {volume} {86}},\
  \bibinfo {pages} {052329} (\bibinfo {year} {2012})}\BibitemShut {NoStop}%
\bibitem [{\citenamefont {Dawkins}\ and\ \citenamefont
  {Howard}(2015)}]{Dawkins2015dist}%
  \BibitemOpen
  \bibfield  {author} {\bibinfo {author} {\bibfnamefont {H.}~\bibnamefont
  {Dawkins}}\ and\ \bibinfo {author} {\bibfnamefont {M.}~\bibnamefont
  {Howard}},\ }\href {https://doi.org/10.1103/PhysRevLett.115.030501}
  {\bibfield  {journal} {\bibinfo  {journal} {Physical Review Letters}\
  }\textbf {\bibinfo {volume} {115}},\ \bibinfo {pages} {030501} (\bibinfo
  {year} {2015})}\BibitemShut {NoStop}%
\bibitem [{\citenamefont {Bravyi}\ \emph {et~al.}(2016)\citenamefont {Bravyi},
  \citenamefont {Smith},\ and\ \citenamefont {Smolin}}]{bravyi2016resources}%
  \BibitemOpen
  \bibfield  {author} {\bibinfo {author} {\bibfnamefont {S.}~\bibnamefont
  {Bravyi}}, \bibinfo {author} {\bibfnamefont {G.}~\bibnamefont {Smith}},\ and\
  \bibinfo {author} {\bibfnamefont {J.~A.}\ \bibnamefont {Smolin}},\ }\href
  {https://doi.org/10.1103/PhysRevX.6.021043} {\bibfield  {journal} {\bibinfo
  {journal} {Physical Review X}\ }\textbf {\bibinfo {volume} {6}},\ \bibinfo
  {pages} {021043} (\bibinfo {year} {2016})}\BibitemShut {NoStop}%
\bibitem [{\citenamefont {Hastings}\ and\ \citenamefont
  {Haah}(2018)}]{Hastings2018dist}%
  \BibitemOpen
  \bibfield  {author} {\bibinfo {author} {\bibfnamefont {M.~B.}\ \bibnamefont
  {Hastings}}\ and\ \bibinfo {author} {\bibfnamefont {J.}~\bibnamefont
  {Haah}},\ }\href {https://doi.org/10.1103/PhysRevLett.120.050504} {\bibfield
  {journal} {\bibinfo  {journal} {Physical Review Letters}\ }\textbf {\bibinfo
  {volume} {120}},\ \bibinfo {pages} {050504} (\bibinfo {year}
  {2018})}\BibitemShut {NoStop}%
\bibitem [{\citenamefont {Gross}\ \emph {et~al.}(2009)\citenamefont {Gross},
  \citenamefont {Flammia},\ and\ \citenamefont {Eisert}}]{Gross2009ent}%
  \BibitemOpen
  \bibfield  {author} {\bibinfo {author} {\bibfnamefont {D.}~\bibnamefont
  {Gross}}, \bibinfo {author} {\bibfnamefont {S.~T.}\ \bibnamefont {Flammia}},\
  and\ \bibinfo {author} {\bibfnamefont {J.}~\bibnamefont {Eisert}},\ }\href
  {https://doi.org/10.1103/PhysRevLett.102.190501} {\bibfield  {journal}
  {\bibinfo  {journal} {Physical Review Letters}\ }\textbf {\bibinfo {volume}
  {102}},\ \bibinfo {pages} {190501} (\bibinfo {year} {2009})}\BibitemShut
  {NoStop}%
\bibitem [{\citenamefont {Beverland}\ \emph {et~al.}(2020)\citenamefont
  {Beverland}, \citenamefont {Campbell}, \citenamefont {Howard},\ and\
  \citenamefont {Kliuchnikov}}]{beverland2019lower}%
  \BibitemOpen
  \bibfield  {author} {\bibinfo {author} {\bibfnamefont {M.}~\bibnamefont
  {Beverland}}, \bibinfo {author} {\bibfnamefont {E.}~\bibnamefont {Campbell}},
  \bibinfo {author} {\bibfnamefont {M.}~\bibnamefont {Howard}},\ and\ \bibinfo
  {author} {\bibfnamefont {V.}~\bibnamefont {Kliuchnikov}},\ }\href
  {https://doi.org/10.1088/2058-9565/ab8963} {\bibfield  {journal} {\bibinfo
  {journal} {Quantum Science and Technology}\ }\textbf {\bibinfo {volume}
  {5}},\ \bibinfo {pages} {035009} (\bibinfo {year} {2020})}\BibitemShut
  {NoStop}%
\bibitem [{\citenamefont {van Enk}\ and\ \citenamefont
  {Beenakker}(2012)}]{Enk2012measure}%
  \BibitemOpen
  \bibfield  {author} {\bibinfo {author} {\bibfnamefont {S.~J.}\ \bibnamefont
  {van Enk}}\ and\ \bibinfo {author} {\bibfnamefont {C.~W.~J.}\ \bibnamefont
  {Beenakker}},\ }\href {https://doi.org/10.1103/PhysRevLett.108.110503}
  {\bibfield  {journal} {\bibinfo  {journal} {Physical Review Letters}\
  }\textbf {\bibinfo {volume} {108}},\ \bibinfo {pages} {110503} (\bibinfo
  {year} {2012})}\BibitemShut {NoStop}%
\bibitem [{\citenamefont {Tran}\ \emph {et~al.}(2016)\citenamefont {Tran},
  \citenamefont {Daki\ifmmode~\acute{c}\else \'{c}\fi{}}, \citenamefont
  {Laskowski},\ and\ \citenamefont {Paterek}}]{Tran2016Correlations}%
  \BibitemOpen
  \bibfield  {author} {\bibinfo {author} {\bibfnamefont {M.~C.}\ \bibnamefont
  {Tran}}, \bibinfo {author} {\bibfnamefont {B.}~\bibnamefont
  {Daki\ifmmode~\acute{c}\else \'{c}\fi{}}}, \bibinfo {author} {\bibfnamefont
  {W.}~\bibnamefont {Laskowski}},\ and\ \bibinfo {author} {\bibfnamefont
  {T.}~\bibnamefont {Paterek}},\ }\href
  {https://doi.org/10.1103/PhysRevA.94.042302} {\bibfield  {journal} {\bibinfo
  {journal} {Physical Review A}\ }\textbf {\bibinfo {volume} {94}},\ \bibinfo
  {pages} {042302} (\bibinfo {year} {2016})}\BibitemShut {NoStop}%
\bibitem [{\citenamefont {Elben}\ \emph {et~al.}(2018)\citenamefont {Elben},
  \citenamefont {Vermersch}, \citenamefont {Dalmonte}, \citenamefont {Cirac},\
  and\ \citenamefont {Zoller}}]{Elben2018Entropies}%
  \BibitemOpen
  \bibfield  {author} {\bibinfo {author} {\bibfnamefont {A.}~\bibnamefont
  {Elben}}, \bibinfo {author} {\bibfnamefont {B.}~\bibnamefont {Vermersch}},
  \bibinfo {author} {\bibfnamefont {M.}~\bibnamefont {Dalmonte}}, \bibinfo
  {author} {\bibfnamefont {J.~I.}\ \bibnamefont {Cirac}},\ and\ \bibinfo
  {author} {\bibfnamefont {P.}~\bibnamefont {Zoller}},\ }\href
  {https://doi.org/10.1103/PhysRevLett.120.050406} {\bibfield  {journal}
  {\bibinfo  {journal} {Physical Review Letters}\ }\textbf {\bibinfo {volume}
  {120}},\ \bibinfo {pages} {050406} (\bibinfo {year} {2018})}\BibitemShut
  {NoStop}%
\bibitem [{\citenamefont {Elben}\ \emph {et~al.}(2019)\citenamefont {Elben},
  \citenamefont {Vermersch}, \citenamefont {Roos},\ and\ \citenamefont
  {Zoller}}]{Elben2019probing}%
  \BibitemOpen
  \bibfield  {author} {\bibinfo {author} {\bibfnamefont {A.}~\bibnamefont
  {Elben}}, \bibinfo {author} {\bibfnamefont {B.}~\bibnamefont {Vermersch}},
  \bibinfo {author} {\bibfnamefont {C.~F.}\ \bibnamefont {Roos}},\ and\
  \bibinfo {author} {\bibfnamefont {P.}~\bibnamefont {Zoller}},\ }\href
  {https://doi.org/10.1103/PhysRevA.99.052323} {\bibfield  {journal} {\bibinfo
  {journal} {Physical Review A}\ }\textbf {\bibinfo {volume} {99}},\ \bibinfo
  {pages} {052323} (\bibinfo {year} {2019})}\BibitemShut {NoStop}%
\bibitem [{\citenamefont {Vermersch}\ \emph {et~al.}(2019)\citenamefont
  {Vermersch}, \citenamefont {Elben}, \citenamefont {Sieberer}, \citenamefont
  {Yao},\ and\ \citenamefont {Zoller}}]{Vermersch2019meas}%
  \BibitemOpen
  \bibfield  {author} {\bibinfo {author} {\bibfnamefont {B.}~\bibnamefont
  {Vermersch}}, \bibinfo {author} {\bibfnamefont {A.}~\bibnamefont {Elben}},
  \bibinfo {author} {\bibfnamefont {L.~M.}\ \bibnamefont {Sieberer}}, \bibinfo
  {author} {\bibfnamefont {N.~Y.}\ \bibnamefont {Yao}},\ and\ \bibinfo {author}
  {\bibfnamefont {P.}~\bibnamefont {Zoller}},\ }\href
  {https://doi.org/10.1103/PhysRevX.9.021061} {\bibfield  {journal} {\bibinfo
  {journal} {Physical Review X}\ }\textbf {\bibinfo {volume} {9}},\ \bibinfo
  {pages} {021061} (\bibinfo {year} {2019})}\BibitemShut {NoStop}%
\bibitem [{\citenamefont {Brydges}\ \emph {et~al.}(2019)\citenamefont
  {Brydges}, \citenamefont {Elben}, \citenamefont {Jurcevic}, \citenamefont
  {Vermersch}, \citenamefont {Maier}, \citenamefont {Lanyon}, \citenamefont
  {Zoller}, \citenamefont {Blatt},\ and\ \citenamefont
  {Roos}}]{Brydges2019randomized}%
  \BibitemOpen
  \bibfield  {author} {\bibinfo {author} {\bibfnamefont {T.}~\bibnamefont
  {Brydges}}, \bibinfo {author} {\bibfnamefont {A.}~\bibnamefont {Elben}},
  \bibinfo {author} {\bibfnamefont {P.}~\bibnamefont {Jurcevic}}, \bibinfo
  {author} {\bibfnamefont {B.}~\bibnamefont {Vermersch}}, \bibinfo {author}
  {\bibfnamefont {C.}~\bibnamefont {Maier}}, \bibinfo {author} {\bibfnamefont
  {B.~P.}\ \bibnamefont {Lanyon}}, \bibinfo {author} {\bibfnamefont
  {P.}~\bibnamefont {Zoller}}, \bibinfo {author} {\bibfnamefont
  {R.}~\bibnamefont {Blatt}},\ and\ \bibinfo {author} {\bibfnamefont {C.~F.}\
  \bibnamefont {Roos}},\ }\href {https://doi.org/10.1126/science.aau4963}
  {\bibfield  {journal} {\bibinfo  {journal} {Science}\ }\textbf {\bibinfo
  {volume} {364}},\ \bibinfo {pages} {260} (\bibinfo {year}
  {2019})}\BibitemShut {NoStop}%
\bibitem [{\citenamefont {Ketterer}\ \emph {et~al.}(2019)\citenamefont
  {Ketterer}, \citenamefont {Wyderka},\ and\ \citenamefont
  {G\"uhne}}]{Ketterer2019ranmeas}%
  \BibitemOpen
  \bibfield  {author} {\bibinfo {author} {\bibfnamefont {A.}~\bibnamefont
  {Ketterer}}, \bibinfo {author} {\bibfnamefont {N.}~\bibnamefont {Wyderka}},\
  and\ \bibinfo {author} {\bibfnamefont {O.}~\bibnamefont {G\"uhne}},\ }\href
  {https://doi.org/10.1103/PhysRevLett.122.120505} {\bibfield  {journal}
  {\bibinfo  {journal} {Physical Review Letters}\ }\textbf {\bibinfo {volume}
  {122}},\ \bibinfo {pages} {120505} (\bibinfo {year} {2019})}\BibitemShut
  {NoStop}%
\bibitem [{\citenamefont {Elben}\ \emph
  {et~al.}(2020{\natexlab{a}})\citenamefont {Elben}, \citenamefont {Kueng},
  \citenamefont {Huang}, \citenamefont {van Bijnen}, \citenamefont {Kokail},
  \citenamefont {Dalmonte}, \citenamefont {Calabrese}, \citenamefont {Kraus},
  \citenamefont {Preskill}, \citenamefont {Zoller},\ and\ \citenamefont
  {Vermersch}}]{Elben2020randomized}%
  \BibitemOpen
  \bibfield  {author} {\bibinfo {author} {\bibfnamefont {A.}~\bibnamefont
  {Elben}}, \bibinfo {author} {\bibfnamefont {R.}~\bibnamefont {Kueng}},
  \bibinfo {author} {\bibfnamefont {H.-Y.~R.}\ \bibnamefont {Huang}}, \bibinfo
  {author} {\bibfnamefont {R.}~\bibnamefont {van Bijnen}}, \bibinfo {author}
  {\bibfnamefont {C.}~\bibnamefont {Kokail}}, \bibinfo {author} {\bibfnamefont
  {M.}~\bibnamefont {Dalmonte}}, \bibinfo {author} {\bibfnamefont
  {P.}~\bibnamefont {Calabrese}}, \bibinfo {author} {\bibfnamefont
  {B.}~\bibnamefont {Kraus}}, \bibinfo {author} {\bibfnamefont
  {J.}~\bibnamefont {Preskill}}, \bibinfo {author} {\bibfnamefont
  {P.}~\bibnamefont {Zoller}},\ and\ \bibinfo {author} {\bibfnamefont
  {B.}~\bibnamefont {Vermersch}},\ }\href
  {https://doi.org/10.1103/PhysRevLett.125.200501} {\bibfield  {journal}
  {\bibinfo  {journal} {Physical Review Letters}\ }\textbf {\bibinfo {volume}
  {125}},\ \bibinfo {pages} {200501} (\bibinfo {year}
  {2020}{\natexlab{a}})}\BibitemShut {NoStop}%
\bibitem [{\citenamefont {Knips}\ \emph {et~al.}(2020)\citenamefont {Knips},
  \citenamefont {Dziewior}, \citenamefont {K{\l}obus}, \citenamefont
  {Laskowski}, \citenamefont {Paterek}, \citenamefont {Shadbolt}, \citenamefont
  {Weinfurter},\ and\ \citenamefont {Meinecke}}]{Knips2020entanglement}%
  \BibitemOpen
  \bibfield  {author} {\bibinfo {author} {\bibfnamefont {L.}~\bibnamefont
  {Knips}}, \bibinfo {author} {\bibfnamefont {J.}~\bibnamefont {Dziewior}},
  \bibinfo {author} {\bibfnamefont {W.}~\bibnamefont {K{\l}obus}}, \bibinfo
  {author} {\bibfnamefont {W.}~\bibnamefont {Laskowski}}, \bibinfo {author}
  {\bibfnamefont {T.}~\bibnamefont {Paterek}}, \bibinfo {author} {\bibfnamefont
  {P.~J.}\ \bibnamefont {Shadbolt}}, \bibinfo {author} {\bibfnamefont
  {H.}~\bibnamefont {Weinfurter}},\ and\ \bibinfo {author} {\bibfnamefont
  {J.~D.~A.}\ \bibnamefont {Meinecke}},\ }\href
  {https://doi.org/10.1038/s41534-020-0281-5} {\bibfield  {journal} {\bibinfo
  {journal} {npj Quantum Information}\ }\textbf {\bibinfo {volume} {6}},\
  \bibinfo {pages} {51} (\bibinfo {year} {2020})}\BibitemShut {NoStop}%
\bibitem [{\citenamefont {Ketterer}\ \emph {et~al.}(2022)\citenamefont
  {Ketterer}, \citenamefont {Imai}, \citenamefont {Wyderka},\ and\
  \citenamefont {G{\ifmmode\ddot{u}\else\"{u}\fi}hne}}]{Ketterer2020randmeas}%
  \BibitemOpen
  \bibfield  {author} {\bibinfo {author} {\bibfnamefont {A.}~\bibnamefont
  {Ketterer}}, \bibinfo {author} {\bibfnamefont {S.}~\bibnamefont {Imai}},
  \bibinfo {author} {\bibfnamefont {N.}~\bibnamefont {Wyderka}},\ and\ \bibinfo
  {author} {\bibfnamefont {O.}~\bibnamefont
  {G{\ifmmode\ddot{u}\else\"{u}\fi}hne}},\ }\href
  {https://doi.org/10.1103/PhysRevA.106.L010402} {\bibfield  {journal}
  {\bibinfo  {journal} {Physical Review A}\ }\textbf {\bibinfo {volume}
  {106}},\ \bibinfo {pages} {L010402} (\bibinfo {year} {2022})}\BibitemShut
  {NoStop}%
\bibitem [{\citenamefont {Zhou}\ \emph
  {et~al.}(2020{\natexlab{a}})\citenamefont {Zhou}, \citenamefont {Zeng},\ and\
  \citenamefont {Liu}}]{Zhou2020Estimation}%
  \BibitemOpen
  \bibfield  {author} {\bibinfo {author} {\bibfnamefont {Y.}~\bibnamefont
  {Zhou}}, \bibinfo {author} {\bibfnamefont {P.}~\bibnamefont {Zeng}},\ and\
  \bibinfo {author} {\bibfnamefont {Z.}~\bibnamefont {Liu}},\ }\href
  {https://doi.org/10.1103/PhysRevLett.125.200502} {\bibfield  {journal}
  {\bibinfo  {journal} {Physical Review Letters}\ }\textbf {\bibinfo {volume}
  {125}},\ \bibinfo {pages} {200502} (\bibinfo {year}
  {2020}{\natexlab{a}})}\BibitemShut {NoStop}%
\bibitem [{\citenamefont {Cian}\ \emph {et~al.}(2021)\citenamefont {Cian},
  \citenamefont {Dehghani}, \citenamefont {Elben}, \citenamefont {Vermersch},
  \citenamefont {Zhu}, \citenamefont {Barkeshli}, \citenamefont {Zoller},\ and\
  \citenamefont {Hafezi}}]{Cian2021chern}%
  \BibitemOpen
  \bibfield  {author} {\bibinfo {author} {\bibfnamefont {Z.-P.}\ \bibnamefont
  {Cian}}, \bibinfo {author} {\bibfnamefont {H.}~\bibnamefont {Dehghani}},
  \bibinfo {author} {\bibfnamefont {A.}~\bibnamefont {Elben}}, \bibinfo
  {author} {\bibfnamefont {B.}~\bibnamefont {Vermersch}}, \bibinfo {author}
  {\bibfnamefont {G.}~\bibnamefont {Zhu}}, \bibinfo {author} {\bibfnamefont
  {M.}~\bibnamefont {Barkeshli}}, \bibinfo {author} {\bibfnamefont
  {P.}~\bibnamefont {Zoller}},\ and\ \bibinfo {author} {\bibfnamefont
  {M.}~\bibnamefont {Hafezi}},\ }\href
  {https://doi.org/10.1103/PhysRevLett.126.050501} {\bibfield  {journal}
  {\bibinfo  {journal} {Physical Review Letters}\ }\textbf {\bibinfo {volume}
  {126}},\ \bibinfo {pages} {050501} (\bibinfo {year} {2021})}\BibitemShut
  {NoStop}%
\bibitem [{\citenamefont {Imai}\ \emph {et~al.}(2021)\citenamefont {Imai},
  \citenamefont {Wyderka}, \citenamefont {Ketterer},\ and\ \citenamefont
  {G\"uhne}}]{Satoya2021ranmeas}%
  \BibitemOpen
  \bibfield  {author} {\bibinfo {author} {\bibfnamefont {S.}~\bibnamefont
  {Imai}}, \bibinfo {author} {\bibfnamefont {N.}~\bibnamefont {Wyderka}},
  \bibinfo {author} {\bibfnamefont {A.}~\bibnamefont {Ketterer}},\ and\
  \bibinfo {author} {\bibfnamefont {O.}~\bibnamefont {G\"uhne}},\ }\href
  {https://doi.org/10.1103/PhysRevLett.126.150501} {\bibfield  {journal}
  {\bibinfo  {journal} {Physical Review Letters}\ }\textbf {\bibinfo {volume}
  {126}},\ \bibinfo {pages} {150501} (\bibinfo {year} {2021})}\BibitemShut
  {NoStop}%
\bibitem [{\citenamefont {Rath}\ \emph {et~al.}(2021)\citenamefont {Rath},
  \citenamefont {van Bijnen}, \citenamefont {Elben}, \citenamefont {Zoller},\
  and\ \citenamefont {Vermersch}}]{Rath2021sampling}%
  \BibitemOpen
  \bibfield  {author} {\bibinfo {author} {\bibfnamefont {A.}~\bibnamefont
  {Rath}}, \bibinfo {author} {\bibfnamefont {R.}~\bibnamefont {van Bijnen}},
  \bibinfo {author} {\bibfnamefont {A.}~\bibnamefont {Elben}}, \bibinfo
  {author} {\bibfnamefont {P.}~\bibnamefont {Zoller}},\ and\ \bibinfo {author}
  {\bibfnamefont {B.}~\bibnamefont {Vermersch}},\ }\href
  {https://doi.org/10.1103/PhysRevLett.127.200503} {\bibfield  {journal}
  {\bibinfo  {journal} {Physical Review Letters}\ }\textbf {\bibinfo {volume}
  {127}},\ \bibinfo {pages} {200503} (\bibinfo {year} {2021})}\BibitemShut
  {NoStop}%
\bibitem [{\citenamefont {{IBM Quantum}}(2021)}]{IBMquantum}%
  \BibitemOpen
  \bibfield  {author} {\bibinfo {author} {\bibnamefont {{IBM Quantum}}},\
  }\href@noop {} {\bibinfo {title}
  {https://quantum-computing.ibm.com/\href{https://quantum-computing.ibm.com/}}}
  (\bibinfo {year} {2021})\BibitemShut {NoStop}%
\bibitem [{\citenamefont {Zhou}\ \emph
  {et~al.}(2020{\natexlab{b}})\citenamefont {Zhou}, \citenamefont {Yang},
  \citenamefont {Hamma},\ and\ \citenamefont {Chamon}}]{zhou2020single}%
  \BibitemOpen
  \bibfield  {author} {\bibinfo {author} {\bibfnamefont {S.}~\bibnamefont
  {Zhou}}, \bibinfo {author} {\bibfnamefont {Z.-C.}\ \bibnamefont {Yang}},
  \bibinfo {author} {\bibfnamefont {A.}~\bibnamefont {Hamma}},\ and\ \bibinfo
  {author} {\bibfnamefont {C.}~\bibnamefont {Chamon}},\ }\href
  {https://doi.org/10.21468/SciPostPhys.9.6.087} {\bibfield  {journal}
  {\bibinfo  {journal} {SciPost Physics}\ }\textbf {\bibinfo {volume} {9}},\
  \bibinfo {pages} {87} (\bibinfo {year} {2020}{\natexlab{b}})}\BibitemShut
  {NoStop}%
\bibitem [{\citenamefont {Leone}\ \emph {et~al.}(2021)\citenamefont {Leone},
  \citenamefont {Oliviero}, \citenamefont {Zhou},\ and\ \citenamefont
  {Hamma}}]{leone2021quantum}%
  \BibitemOpen
  \bibfield  {author} {\bibinfo {author} {\bibfnamefont {L.}~\bibnamefont
  {Leone}}, \bibinfo {author} {\bibfnamefont {S.~F.~E.}\ \bibnamefont
  {Oliviero}}, \bibinfo {author} {\bibfnamefont {Y.}~\bibnamefont {Zhou}},\
  and\ \bibinfo {author} {\bibfnamefont {A.}~\bibnamefont {Hamma}},\ }\href
  {https://doi.org/10.22331/q-2021-05-04-453} {\bibfield  {journal} {\bibinfo
  {journal} {{Quantum}}\ }\textbf {\bibinfo {volume} {5}},\ \bibinfo {pages}
  {453} (\bibinfo {year} {2021})}\BibitemShut {NoStop}%
\bibitem [{\citenamefont {Oliviero}\ \emph {et~al.}(2021)\citenamefont
  {Oliviero}, \citenamefont {Leone},\ and\ \citenamefont
  {Hamma}}]{oliviero2021transitions}%
  \BibitemOpen
  \bibfield  {author} {\bibinfo {author} {\bibfnamefont {S.~F.}\ \bibnamefont
  {Oliviero}}, \bibinfo {author} {\bibfnamefont {L.}~\bibnamefont {Leone}},\
  and\ \bibinfo {author} {\bibfnamefont {A.}~\bibnamefont {Hamma}},\ }\href
  {https://doi.org/https://doi.org/10.1016/j.physleta.2021.127721} {\bibfield
  {journal} {\bibinfo  {journal} {Physics Letters A}\ }\textbf {\bibinfo
  {volume} {418}},\ \bibinfo {pages} {127721} (\bibinfo {year}
  {2021})}\BibitemShut {NoStop}%
\bibitem [{\citenamefont {Elben}\ \emph
  {et~al.}(2020{\natexlab{b}})\citenamefont {Elben}, \citenamefont {Vermersch},
  \citenamefont {van Bijnen}, \citenamefont {Kokail}, \citenamefont {Brydges},
  \citenamefont {Maier}, \citenamefont {Joshi}, \citenamefont {Blatt},
  \citenamefont {Roos},\ and\ \citenamefont {Zoller}}]{Elben2020cross}%
  \BibitemOpen
  \bibfield  {author} {\bibinfo {author} {\bibfnamefont {A.}~\bibnamefont
  {Elben}}, \bibinfo {author} {\bibfnamefont {B.}~\bibnamefont {Vermersch}},
  \bibinfo {author} {\bibfnamefont {R.}~\bibnamefont {van Bijnen}}, \bibinfo
  {author} {\bibfnamefont {C.}~\bibnamefont {Kokail}}, \bibinfo {author}
  {\bibfnamefont {T.}~\bibnamefont {Brydges}}, \bibinfo {author} {\bibfnamefont
  {C.}~\bibnamefont {Maier}}, \bibinfo {author} {\bibfnamefont {M.~K.}\
  \bibnamefont {Joshi}}, \bibinfo {author} {\bibfnamefont {R.}~\bibnamefont
  {Blatt}}, \bibinfo {author} {\bibfnamefont {C.~F.}\ \bibnamefont {Roos}},\
  and\ \bibinfo {author} {\bibfnamefont {P.}~\bibnamefont {Zoller}},\ }\href
  {https://doi.org/10.1103/PhysRevLett.124.010504} {\bibfield  {journal}
  {\bibinfo  {journal} {Physical Review Letters}\ }\textbf {\bibinfo {volume}
  {124}},\ \bibinfo {pages} {010504} (\bibinfo {year}
  {2020}{\natexlab{b}})}\BibitemShut {NoStop}%
\bibitem [{\citenamefont {Elben}\ \emph
  {et~al.}(2020{\natexlab{c}})\citenamefont {Elben}, \citenamefont {Yu},
  \citenamefont {Zhu}, \citenamefont {Hafezi}, \citenamefont {Pollmann},
  \citenamefont {Zoller},\ and\ \citenamefont
  {Vermersch}}]{Elben2020topological}%
  \BibitemOpen
  \bibfield  {author} {\bibinfo {author} {\bibfnamefont {A.}~\bibnamefont
  {Elben}}, \bibinfo {author} {\bibfnamefont {J.}~\bibnamefont {Yu}}, \bibinfo
  {author} {\bibfnamefont {G.}~\bibnamefont {Zhu}}, \bibinfo {author}
  {\bibfnamefont {M.}~\bibnamefont {Hafezi}}, \bibinfo {author} {\bibfnamefont
  {F.}~\bibnamefont {Pollmann}}, \bibinfo {author} {\bibfnamefont
  {P.}~\bibnamefont {Zoller}},\ and\ \bibinfo {author} {\bibfnamefont
  {B.}~\bibnamefont {Vermersch}},\ }\href
  {https://doi.org/10.1126/sciadv.aaz3666} {\bibfield  {journal} {\bibinfo
  {journal} {Science Advances}\ }\textbf {\bibinfo {volume} {6}},\ \bibinfo
  {pages} {eaaz3666} (\bibinfo {year} {2020}{\natexlab{c}})}\BibitemShut
  {NoStop}%
\end{thebibliography}
%


\end{document}


\setcounter{secnumdepth}{3}

\title{Supplementary Information for ``Measuring magic on a quantum processor''}

\author{Salvatore F.E. Oliviero}\email{s.oliviero001@umb.edu}
\affiliation{Physics Department,  University of Massachusetts Boston,  02125, USA}
\author{Lorenzo Leone}
\affiliation{Physics Department,  University of Massachusetts Boston,  02125, USA}
\author{Alioscia Hamma}
\affiliation{Physics Department,  University of Massachusetts Boston,  02125, USA}
\affiliation{Dipartimento di Fisica Ettore Pancini, Universit\`a degli Studi di Napoli Federico II, Via Cinthia, 80126 Fuorigrotta, Napoli NA, Italy}
\affiliation{INFN, Sezione di Napoli, Italy}
\author{Seth Lloyd}
\affiliation{Department of Mechanical Engineering, Massachusetts Institute of Technology, Cambridge, MA, USA}
\affiliation{Turing Inc., Brooklyn, NY, USA}
\maketitle


\section{Supplementary Note 1: The stabilizer R\'enyi entropy: a wrap-up of the results}\label{Sec: stabrenyientropy}
The stabilizer R\'enyi entropy has been introduced and defined in Ref.~\cite{leone2021renyi}. This section is devoted to a brief review of the quantity, and its relation to quantum chaos and quantum certification. As an original result, here we prove that the stabilizer R\'enyi entropy quantifies the resources necessary to extract useful information from a given quantum state.

Let $\psi$ be a $n$-qubit state and $\mathbb{P}(n)$ the Pauli group on $n$ qubit. Define the following probability distribution
\begin{equation}
    \Xi_{\psi}:=\{d^{-1}P(\psi)^{-1}\operatorname{tr}^{2}(P\psi)\,\mid P\in\mathbb{P}(n)\}
\end{equation}
where $P(\psi):=\operatorname{tr}\psi^2$ is the purity of the state $\psi$. The above is a probability distribution because $\Xi_{\psi}(P)\ge 0$ and $\sum_{P}\Xi_{\psi}(P)=1$. A family of magic monotones is given by the $\alpha$-R\'enyi entropy of $\Xi_{\psi}$ plus the log of the purity of the state $\psi$:
\begin{equation}
M_{\alpha}(\psi):=S_{\alpha}(\Xi_{\psi})-\log P(\psi)-\log d    
\end{equation}
where $S_{\alpha}(\Xi_{\psi}):=\frac{1}{1-\alpha}\log \sum_{P}\Xi_{\psi}(P)$. $M_{\alpha}(\psi)$ is called $\alpha$-Stabilizer R\'enyi entropy. The above family of measures obeys the following properties:
\vspace{0.3cm}
\begin{enumerate}[label=(\roman*)]
    \item It follows a hierarchy, i.e. $M_{\alpha}(\psi)\ge M_{\alpha^{\prime}}(\psi)$ for $\alpha<\alpha^{\prime}$.
    \vspace{0.3cm}
    \item It is faithful, i.e. $M_{\alpha}(\psi)=0$ iff $\psi=d^{-1}\sum_{P\in G}\psi_PP$, where $G$ is a commuting subset of $\mathbb{P}$ and $\psi_P=\pm 1$.
    \vspace{0.3cm}
    \item It is invariant under Clifford rotations $M_{\alpha}(C\psi C^{\dag})=M_{\alpha}(\psi)$.
    \vspace{0.3cm}
    \item It is additive $M_{\alpha}(\psi\otimes \phi)=M_{\alpha}(\psi)+M_{\alpha}(\phi)$.
    \vspace{0.3cm}
    \item It is bounded $0\le M_{\alpha}(\psi)\le \log d$
    \vspace{0.3cm}
    \item It bounds other measures of magic. For pure states only, it bounds the Robustness of Magic\cite{howard2017application} $M_{\alpha}(\psi)<2\log R(\psi)$ for $\alpha\ge 1/2$, and the stabilizer nullity $M_{\alpha}(\psi)\le\nu(\psi)$\cite{beverland2019lower}.
    \vspace{0.3cm}
\end{enumerate}
Across this family of magic measures, the $2$-Stabilizer R\'enyi entropy plays an important role. Indeed it distinguishes itself from the others because it can be experimentally measured via statistical correlations between randomized measurements, see Methods. Explicit calculations show that the $2$-Stabilizer R\'enyi entropy can be expressed in terms of the stabilizer purity -- $W(\psi):=\operatorname{tr}(Q\psi^{\otimes 4})$ where $Q=d^{-2}\sum_{P}P^{\otimes 4}$ is the projector onto the stabilizer code -- and can be written as in Eq.~$1$. 

The introduction of the stabilizer R\'enyi entropy revealed the intriguing connection between the resource theory of magic states and chaos. In these settings, a unitary operator $U$ is said to be chaotic iff attains the Haar value of general multipoint $(2k)$ Out-Of-Time-Order correlators (OTOCs) defined as:
\begin{equation}
 \operatorname{OTOC}_{2k}=\frac{1}{d}\operatorname{tr}(A_{1}(U)BA_{2}(U)B\cdots A_{k}(U)B) .
\end{equation}
where $A(U):=U^{\dag}AU$. In order to see such a connection, consider a unitary operator $U$, and its Choi isomorphism~\cite{choi1975completely}, i.e. consider two copies of the Hilbert space and apply $\mathchoice {\rm 1\mskip-4mu l} {\rm 1\mskip-4mu l} {\rm 1\mskip-4.5mu l} {\rm 1\mskip-5mu l}\otimes U$ on the Bell state $\mid\!\!I\rangle:=\frac{1}{d}\sum_{i}\mid\!\!i\rangle\otimes \mid\!\!i\rangle\in\mathcal{H}^{\otimes 2}$. The Choi state is defined as:
\begin{equation}
    \mid\!\!U\rangle:=\mathchoice {\rm 1\mskip-4mu l} {\rm 1\mskip-4mu l} {\rm 1\mskip-4.5mu l} {\rm 1\mskip-5mu l}\otimes U\mid\!\!I\rangle=\frac{1}{d}\sum_{i}\mid\!\!i\rangle\otimes U\mid\!\!i\rangle
\end{equation}
A lemma proved in Ref.~\cite{leone2022magic} shows that the $\alpha$-Stabilizer R\'enyi entropy of $\mid\!\!U\rangle$ is proportional to the log of the general $4\alpha$-point OTOC:
\begin{equation}
    M_{\alpha}( \mid\!\!U\rangle)=\frac{1}{1-\alpha}\log \overline{\operatorname{OTOC}}_{4\alpha}(U)
    \label{otocsandmagic}
\end{equation}
where the above OTOCs are average and nasty OTOCs defined as $OTOC_{2\alpha}:={d^{-2}}\sum_{P,P^{\prime}}otoc_{2\alpha}(P,P^\prime)$, and $d \times otoc_{4\alpha}(P,P^\prime):=\operatorname{tr}[\langle P_{4\alpha}\prod_{i=1}^{4\alpha}P(U)P^{\prime}P_{i-1}P_{i}\rangle]$ with $P_{0}\equiv \mathchoice {\rm 1\mskip-4mu l} {\rm 1\mskip-4mu l} {\rm 1\mskip-4.5mu l} {\rm 1\mskip-5mu l}$,$\langle{\cdot}\rangle$ the average over $P_{1},\ldots, P_{4\alpha}$ and $P(U)\equiv UPU^{\dagger}$. Let us comment Eq.~\eqref{otocsandmagic}: $(i)$ it establishes a direct connection to the theory of magic states and chaos. The more the magic in the Choi state associated with $U$, the more chaotic the evolution generated by $U$. And $(ii)$, for $\alpha=2$ allows the direct measurement of the $8$ point OTOC via statistical correlations between randomized measurements, the protocol analyzed and proven in the present paper. 

The stabilizer R\'enyi entropy is nothing but an entropy in the operator basis of Pauli operators, let us say the \textit{operator-computational basis}. The more a state $\psi$ is spread in the Pauli basis, the more the Stabilizer R\'enyi entropy (and, consequently, the more the magic). Here is the thing: a state too spread in the Pauli basis cannot be used for a fruitful quantum computation when measured in the computational basis. Intuitively, in the case of almost maximal spreading, an exponential number of measurement shots is needed to distinguish the state $\psi$ from a random state. This means that a state possessing an excessive amount of magic could be worse for quantum computation, as already pointed out in Ref.~\cite{liu2020many}. Here we want to show that a similar conclusion can also be made by looking at stabilizer R\'enyi entropies. The above intuition is formalized as follows: consider a state $\psi$ and sample a Pauli operator $P$, to be measured at the end of a quantum computation, according to the state-dependent probability distribution $\Xi_{\psi}$. This choice of probability distribution has the following operational meaning: across all the Pauli operators, $\Xi_{\psi}$ promotes the collection of Pauli operators having a large component on $\psi$. The probability that $\mid\!\!\operatorname{tr}(P\psi)\!\!\mid\ge \epsilon$ is upper bounded by $\epsilon$ depends on the $2$-stabilizer R\'enyi entropy $M_{2}(\psi)$
\begin{equation}
\operatorname{Pr}[\mid\!\!\operatorname{tr}(P\psi)\!\!\mid\ge \epsilon]\le \epsilon^{-1} 2^{-M_{2}(\psi)/2}
\end{equation}
where we used Markov's inequality, and the fact that $\langle\mid\!\!\operatorname{tr}(P\psi)\!\!\mid\rangle_{\Xi_{\psi}}\le \sqrt{\langle\operatorname{tr}^{2}(P\psi)\rangle_{\Xi_{\psi}}}=2^{-M_{2}(\psi)/2}$. Thus, given an extensive amount of magic $M_{2}(\psi)=\alpha n$, for some $0<\alpha<1$, the probability to pick a Pauli operator $P$ such that $\mid\!\!\operatorname{tr}(P\psi)\!\!\mid\le 2^{-\frac{\alpha}{4}n}$ is overwhelming $\sim 1-2^{-\frac{\alpha}{4}n}$. This simple fact puts the above considerations in a rigorous fashion, showing that an excess amount of unwanted magic makes the task of distinguishing the state $\psi$ from a random state an exponentially difficult task. We can prove the converse statement also: the less distinguishable a state $\psi$ is from a random state, the more magic the state $\psi$ contains. Consider a pure state $\psi$. We define the \textit{amount of distinguishability} as the maximum expectation value over the set of Pauli operators, i.e. $\max_{P\neq \mathchoice {\rm 1\mskip-4mu l} {\rm 1\mskip-4mu l} {\rm 1\mskip-4.5mu l} {\rm 1\mskip-5mu l}}\mid\!\!\operatorname{tr}(P\psi)\!\!\mid$, which quantifies the minimum number of resources necessary to distinguish $\psi$ from a random state. Indeed, if $\max_{P\neq \mathchoice {\rm 1\mskip-4mu l} {\rm 1\mskip-4mu l} {\rm 1\mskip-4.5mu l} {\rm 1\mskip-5mu l}}\mid\!\operatorname{tr}(P\psi)\!\mid=\epsilon$ then a number $\mathcal{O}(\epsilon^{-2})$ of measurement shots is necessary for the evaluation of any expectation value of Pauli operators $P$. We have the following bound:
\begin{equation}
\max_{P\neq \mathchoice {\rm 1\mskip-4mu l} {\rm 1\mskip-4mu l} {\rm 1\mskip-4.5mu l} {\rm 1\mskip-5mu l}}\mid\!\operatorname{tr}(P\psi)\!\mid\ge 2^{-(M_{2}(\psi)+1)/2}
\label{regularizedeq}
\end{equation}
see Supplementary Note $3$.~\ref{Sec: proofs} for a proof. The above inequality tells us that, if $\max_{P\neq \mathchoice {\rm 1\mskip-4mu l} {\rm 1\mskip-4mu l} {\rm 1\mskip-4.5mu l} {\rm 1\mskip-5mu l}}\mid\!\operatorname{tr}(P\psi)\!\mid=2^{-\alpha n}$, i.e. the state is not distinguishable from a random state with a polynomial number of shot measurements, then the $2$-Stabilizer R\'enyi entropy $M_{2}(\psi)=\mathcal{O}(n)$ is maximal. 

The task of distinguishing a state from a random one is central in the theory of quantum certification. A quantum certificate is a guarantee of the correct preparation of a given quantum state. Among the plethora of proposed certification protocols, one of the more efficient protocols to directly measure the fidelity between the desired state $\psi$ and the prepared one $\tilde{\psi}$, i.e. $\operatorname{tr}(\psi\tilde{\psi})$, is the Monte Carlo fidelity estimation introduced by Flammia and Liu\cite{flammia2011fidelity}. The core of their protocol is to sample Pauli operators $P$ according to the probability distribution $\Xi_{\psi}$ and measure them. Let $N_{\psi}$ the total number of resources -- including both the number of Pauli operators extracted and the finite number of shot measurements -- necessary to compute the fidelity up to an error $\epsilon$, then:
\begin{equation}
    \Omega(\exp[M_{2}(\psi)])\le N_{\psi}\le \Omega(\exp[M_{0}(\psi)])
\end{equation}

i.e. the resources are directly quantified by stabilizer R\'enyi entropies (see~\cite{leone2022magic} for more details).
\section{Supplementary Note 2: Noise model and corrections}\label{Sec: noisemodel}
\subsection{Magic is robust under noisy preparations}
In this section, we explain why the amount of magic in high magical states is protected against a noisy state preparation. This result is completely inspired by experimental results. Indeed, looking at Figs.~$3,4,5$ one can note that the theoretical magic, the experimental value, and the noise model get closer and closer the more magic seeds are injected into the circuit. Here we prove that, for high magical and entangled states, this is indeed the case, provided that the noise model is enough well-behaving. Let $\psi=\mid\!\!\psi\rangle\langle\psi\!\!\mid$ the state one aims to prepare on the quantum processor; we can model (almost) any noisy state preparation with the following quantum channel\cite{Knill2005noisy}:
\begin{equation}
\mathcal{E}(\psi):=\sum_{i}q_{i}P_{i}\psi P_{i}
\end{equation}
where $q_{i}$s form a probability distribution and $P_{i}$ are Pauli strings. Note that the noise model employed will fit the above definition. We model a high magical and entangled state as a Haar random state $\psi$, and evaluate the (average) difference in magic due to a noisy state preparation:
\begin{equation}
\langle\delta M\rangle_{Haar}:=\left\langle\mid\!\!M(\mathcal{E}(\psi))-M(\psi)\!\!\mid\right\rangle_{Haar}=\left\langle -\log \frac{\operatorname{tr}(Q\psi^{\otimes 4})\operatorname{Pur}(\mathcal{E}(\psi))}{\operatorname{tr}(Q\mathcal{E}(\psi)^{\otimes 4})}\right\rangle_{Haar}
\end{equation}
We can exploit the typicality of $\operatorname{Pur}(\mathcal{E}(\psi))$, $\operatorname{tr}(Q\psi^{\otimes 4})$ and $\operatorname{tr}(Q\mathcal{E}(\psi)^{\otimes 4})$ (see \cite{leone2021renyi}),  and take the average of every single term in the log, committing an error exponentially small in $n$:
\begin{equation}
\langle\delta M\rangle_{Haar}\simeq -\log \frac{\langle\operatorname{tr}(Q\psi^{\otimes 4})\rangle_{Haar}\langle\operatorname{Pur}(\mathcal{E}(\psi))\rangle_{Haar}}{\langle\operatorname{tr}(Q\mathcal{E}(\psi)^{\otimes 4})\rangle_{Haar}}
\end{equation}
Let us evaluate term by term, starting from $\langle\operatorname{Pur}(\mathcal{E}(\psi))\rangle_{Haar}$:
\begin{equation}
\langle\operatorname{Pur}(\mathcal{E}(\psi))\rangle_{haar}=\frac{1}{d(d+1)}\sum_{i,j}q_{i}q_{j}\operatorname{tr}(P_{i}P_{j}\otimes P_{j}P_{i}\Pi_{\text{sym}}^{(2)})=\frac{d\sum_{i}q_{i}^{2}+1}{d+1}
\end{equation}
where $\Pi_{\text{sym}}^{(2)}:=\mathchoice {\rm 1\mskip-4mu l} {\rm 1\mskip-4mu l} {\rm 1\mskip-4.5mu l} {\rm 1\mskip-5mu l}+T$. Then:
\begin{eqnarray}
\langle\operatorname{tr}(Q\mathcal{E}(\psi)^{\otimes 4})\rangle_{Haar}=&&\frac{1}{d^3(d+1)(d+2)(d+3)}\\\times&&\!\!\sum_{i,j,k,l,P}q_{i}q_{j}q_{k}q_{l}\operatorname{tr}(P_{i}PP_{i}\otimes\ldots\otimes P_{l}PP_{l}\Pi_{\text{sym}}^{(4)})\nonumber
\end{eqnarray}
note that the term $P_{i}PP_{i}\otimes\ldots\otimes P_{l}PP_{l}$ is invariant under the conjugacy classes of $\mathcal{S}_{4}$ and therefore:
\begin{equation}
\langle\operatorname{tr}(Q\mathcal{E}(\psi)^{\otimes 4})\rangle_{Haar}=\frac{1}{d(d+1)(d+2)(d+3)}\left(d^2+6d+8+3X+\frac{6}{d}X\right)
\end{equation}
where we defined $X:=\sum_{P}\left(d^{-1}\sum_{i}q_{i}\operatorname{tr}(P_{i}PP_{i}P)\right)^{4}$. Note that $1\le X\le d^2$ and $X=d^2$ iff $q_{i}=1$ for some $i$ (i.e. in presence of unitary stabilizer noise) and $X=1$ iff $\mathcal{E}$ is the completely dephasing channel $\mathcal{E}(\psi)\propto\mathchoice {\rm 1\mskip-4mu l} {\rm 1\mskip-4mu l} {\rm 1\mskip-4.5mu l} {\rm 1\mskip-5mu l}$ (i.e. for bad noise). Thus, in the large $d$ limit we have $\langle\operatorname{tr}(Q\mathcal{E}(\psi)^{\otimes 4})\rangle_{Haar}\simeq \alpha d^{-2}$ where $1\le \alpha\le 4$. Putting it all together we have:
\begin{equation}
\langle\delta M\rangle_{Haar}\simeq -\log \frac{4}{\alpha}-\log \frac{d\sum_{i}q_{i}^{2}+1}{d+1}
\end{equation}
Neglecting the $\mathcal{O}(1)$ factor and expanding the log, we can write the following relation:
\begin{equation}
\langle\delta M\rangle_{Haar}=\mathcal{O}(S_{2}(\mathbf{q}))
\end{equation}
where $\mathbf{q}$ is the probability distribution of the $q_i$s and $S_{2}(\mathbf{q})$ its $2$-R\'enyi entropy. Thus, the experimental data are telling us that the noise affecting the hardware features $S_{2}(\mathbf{q})\le \mathcal{O}(poly(\log n))$. In the following we set up a noise model having $S_{2}(\mathbf{q})=\mathcal{O}(\log n)$.

\subsection{Noise model}
In what follows we introduce a noise model aimed to correct experimental values of magic, see Fig.~$1$ in the main text for a pictorial representation. A single run of our experiment consists of three steps: (i) state preparation, (ii) application of a random Clifford gate on each qubit, and (iii) local projective measurements in the computational basis. We aim to measure the magic in the quantum state at the end of the state preparation. We keep track of decoherence in the system, by measuring the purity of the output state $P(\psi)$ along with $W(\psi)$. We observe that the purity is more than $30\%$ less than one, revealing the presence of errors with non-negligible probability. Let us first discuss errors during the state preparation, i.e. step (i). We model the effect of decoherence in the state preparation by a \textit{state-aware}, self-correcting phase flip error occurring on every qubit with probability $(1-p)/n$. Suppose one aims to prepare the state $\mid\!\psi\rangle$. Because of noise, the state actually prepared on the quantum computer is mixed and we postulate it to be:
\begin{equation}
\psi_{p}\equiv\mathcal{N}_{p}(\mid\!\!\psi\rangle\langle\psi\!\!\mid):=p\mid\!\!\psi\rangle\langle\psi\!\!\mid+\frac{(1-p)}{n}\sum_{i=1}^{n}Z_{i}\mid\!\!\psi\rangle\langle\psi\!\!\mid Z_{i}
\end{equation}
where $Z_{i}:=\mathchoice {\rm 1\mskip-4mu l} {\rm 1\mskip-4mu l} {\rm 1\mskip-4.5mu l} {\rm 1\mskip-5mu l}\otimes\dots \otimes Z\otimes \dots\otimes \mathchoice {\rm 1\mskip-4mu l} {\rm 1\mskip-4mu l} {\rm 1\mskip-4.5mu l} {\rm 1\mskip-5mu l}$. Id est, our ansatz is that during the state preparation phase flip occurs on every qubit with the same probability $(1-p)/n$. Here $0\le p\le 1$ is a state-dependent (and run-dependent) constant that will be experimental measured for each state $\mid\!\psi\rangle$ from the outcome probabilities, as explained in what follows. 

In step $(ii)$, we apply $n$ local Clifford gates, one on each qubit. Contrary to the case of universal gates (cfr. \cite{Brydges2019randomized}), Clifford gates are fine-tuned and this can be a problem during an experiment aimed to measure magic. To understand this, consider a simple Clifford gate, e.g. phase-gate $S:=\!\mid\!\!0\rangle\langle 0\!\!\mid\!+e^{i\pi/2}\!\mid\!\!1\rangle\langle 1\!\mid\!$. It does belong to the Clifford group, but a small displacement of the $\pi/2$ angle makes $S_{\pm\varepsilon}:=\!\mid\!\!0\rangle\langle 0\!\!\mid\!+e^{i\pi/2\pm \varepsilon}\!\mid\!\!1\rangle\langle 1\!\mid\!$ not belonging to the Clifford group anymore. Although $S_{\pm\varepsilon}$ is $\varepsilon$-away from being a Clifford gate, a small error $\varepsilon$ in the gate implementation can result in affecting the results substantially.
Indeed, since only Clifford operators are magic-preserving transformations, the application of a non-Clifford gate (despite being $\varepsilon$-away from being Clifford) would result in a biased measurement of the magic of the state $\mid\!\psi\rangle$. By applying $n$ Clifford gates before collecting the outcome probabilities, also a small gate-imperfection error can pump magic into the system reflecting in an erroneous measurement of magic. This gate-imperfection error is more visible in low-magic states, compared to high-magic states, and the reason why is clear: while pumping some magic in low-magic states comes easy, it becomes harder and harder the more the state becomes a high-magic state. To collect unbiased measurements of the magic of quantum states prepared by the quantum processor, we need to get rid of these spurious contributions only due to the experimental apparatus. In what follows we build up a model that helps us to correct the experimental value of the magic. Let $C=\bigotimes_{i=1}^{n}c_{i}$, where $c_{i}\in\mathcal{C}(2)$ are random local Clifford operators applied after the state $\psi_{p}$ is prepared on the quantum processor and before the collection of the outcomes.
To take into account gate-imperfection errors, let us suppose that each Clifford gate $c_{i}$ is affected by the same small phase displacement $\varepsilon$:
\begin{equation}
c_{i}\rightarrow c_{i}^{\varepsilon}\equiv c_{i}P_{\varepsilon}c_{i}^{\prime}
\label{gateimperfection}
\end{equation}
where $P_{\varepsilon}=\!\mid\!\!0\rangle\langle 0\!\!\mid\!+e^{i\varepsilon}\!\mid\!\!1\rangle\langle 1\!\mid\!$ is a $\varepsilon$-phase gate, that aims to model the phase imperfections when applying $S$-gates. The outcome probabilities are therefore: 
\begin{equation}
P(\psi_{p}^{C^{\varepsilon}}\mid s)=\operatorname{tr}(C^{\varepsilon\dagger}\psi_{p}C^{\varepsilon}\mid\!\!s\rangle\langle s\!\!\mid)
\end{equation}
where $C^{\varepsilon}:=\bigotimes_{i=1}^{n}c_{i}^{\varepsilon}$. Recall that the magic is computed by statistical correlations between measurements, averaging over the local Clifford operators $C$ applied at each run. Because of gate-imperfection errors modeled by $C^{\varepsilon}$ and the decoherence in the state preparation modeled by $\psi_{p}$, the stabilizer purity computed is:
\begin{equation}
W_{\varepsilon}(\psi_{p})=\langle\sum_{\mathbf{s}}O_{4}(\mathbf{s}) \operatorname{tr}(C^{\varepsilon\dagger\otimes 4}\psi_{p}^{\otimes 4}C^{\varepsilon\otimes 4}\mid\!\!\mathbf{s}\rangle\langle\mathbf{s}\!\!\mid)\rangle_{C^{\varepsilon}}\equiv \langle\operatorname{tr}(C^{\varepsilon\dagger\otimes 4}\psi_{p}^{\otimes 4}C^{\varepsilon\otimes 4}Q)\rangle_{C^{\varepsilon}}
\label{errormodel1}
\end{equation}
where $\mathbf{s}\equiv (s_{1},s_{2},s_{3},s_{4})$. Note that the average is no longer taken on the single qubit Clifford group, but rather in the $P_{\varepsilon}$-doped Clifford gates, defined in \cite{leone2021quantum,oliviero2021transitions}. Now, Eq.~\eqref{errormodel1} can be written as:
\begin{equation}
W_{\varepsilon}(\psi_{p})=\operatorname{tr}(\psi_{p}^{\otimes 4}Q_{1}^{\varepsilon\otimes n})
\label{errormodel2}
\end{equation}
where we exploited the cyclic property of the trace, the locality of the doped Clifford operator $C^{\varepsilon}$ and the fact that $Q=Q_{1}^{\otimes n}$, where $Q_{1}=\frac{1}{4}(\mathchoice {\rm 1\mskip-4mu l} {\rm 1\mskip-4mu l} {\rm 1\mskip-4.5mu l} {\rm 1\mskip-5mu l}^{\otimes 4}+X^{\otimes 4}+Y^{\otimes 4}+Z^{\otimes 4})$; then we defined:
\begin{equation}
Q_{1}^{\varepsilon}:=\langle c^{\dagger\otimes 4}P_{\varepsilon}^{\dagger\otimes 4}Q_{1}P_{\varepsilon}^{\otimes 4}c^{\otimes 4}\rangle_{c}
\end{equation}
By Theorem $1$ in \cite{leone2021quantum} one can compute $Q^{\varepsilon}_{1}$ as
\begin{equation}
Q_{1}^{\varepsilon}=\frac{1}{6}(5+\cos(4\varepsilon))Q_{1}-\frac{1}{24}\sin^{2}(2\varepsilon)Q_{1}(T^{(1)}_{(ij)}+T_{(ijkl)}^{(1)})+\frac{1}{12}\sin^{2}(2\varepsilon)(\mathchoice {\rm 1\mskip-4mu l} {\rm 1\mskip-4mu l} {\rm 1\mskip-4.5mu l} {\rm 1\mskip-5mu l}^{\otimes 4}+T^{(1)}_{(ij)(kl)})
\label{errormodel6}
\end{equation}
where $T^{(1)}_{(ij)},T^{(1)}_{(ij)(kl)}, T^{(1)}_{(ijkl)}$ are permutation operators defined on $4$ copies of $\mathbb{C}^{2}$ and $T_{(ij)}:=T_{(12)}+T_{(23)}+T_{(34)}+T_{(13)}+T_{(14)}+T_{(24)}$ is a fast notation for the summation over the full conjugacy class, similarly for $T^{(1)}_{(ij)(kl)}$ and $ T^{(1)}_{(ijkl)}$. Then, by making the $n$-th tensor power to reconstruct $Q_{1}^{\varepsilon\otimes n}$, the only term containing $tr(Q\psi_{p}^{\otimes 4})$ is the one (coming from the $n$-th tensor power) with coefficient $g(\varepsilon):=\frac{1}{6^{n}}(5+\cos(4\varepsilon))^{n}$; the other contributions constitute a correction to $W(\psi_{p})$ and depend on the state $\mid\!\psi\rangle$, the shift-angle $\epsilon$ and the decoherence parameter $p$. We define this contribution as:
\begin{equation}
\Omega(\varepsilon,\mid\!\psi\rangle,p):=W_{\varepsilon}(\psi_{p})-g(\varepsilon)W(\psi_{p})
\label{errormodel3}
\end{equation}
Thus, according to our noise model, $W(\psi)$ measured in the experiment $W_{\varepsilon}(\psi_{p})$ is a combination of the stabilizer purity of the noisy-prepared mixed state $W(\psi_{p})$ and an error term depending on the shift angle $\varepsilon$, which constitutes a spurious contribution to the magic due to the measurement apparatus. Finally, making the ansatz:
\begin{equation}
W_{exp}(\mid\!\psi\rangle) \stackrel{\mathclap{\normalfont\mbox{!}}}{=} W_{\varepsilon}(\psi_{p})
\end{equation}
we can estimate the corrected experimental stabilizer purity $W_{exp}^{corr}(\mid\!\psi\rangle)$ of the state $\psi$ prepared on the quantum computer as:
\begin{equation}
W_{exp}^{corr}(\mid\!\psi\rangle)=\frac{1}{g(\varepsilon)}(W_{exp}(\mid\!\psi\rangle)-\Omega(p,\varepsilon,\mid\!\psi\rangle))
\end{equation}
where $\Omega(p,\varepsilon,\mid\!\psi\rangle)$ is defined in Eq.~\eqref{errormodel3}. 

So far, so good, but what about $p$ and $\varepsilon$? Alongside with the magic, having collected the outcome probabilities $P(\psi_{p}^{C}\mid s)$ allows us to compute the purity as:
\begin{equation}
P_{\varepsilon}(\psi_{p})=\langle\sum_{s_1,s_2}O_{2}(s_{1},s_{2})P(\psi_{p}^{C^{\varepsilon}}\!\!\mid\!\! s_{1})P(\psi_{p}^{C^{\varepsilon}}\!\!\mid\!\! s_{2})\rangle_{C^{\varepsilon}}
\end{equation}
again, note that the average is taken on the $P_{\varepsilon}$-doped Clifford group. The purity though involves just the second tensor power of the doped Clifford average, and since the Clifford group forms a unitary $3$-design\cite{scott2008optimizing,webb2016clifford,zhu2017multiqubit} the $P_{\varepsilon}$-doping gets absorbed thanks to the left/right invariance of the Haar measure over groups\cite{collins2003moments,collins2006integration}. Thus, we can simply write:
\begin{equation}
P_{\varepsilon}(\psi_{p})=\operatorname{tr}(T\psi_{p}^{\otimes 2})=\operatorname{tr}(\psi_{p}^{2})
\label{errormodel4}
\end{equation}
in other words, the estimation of the purity via statistical correlations between randomized measurements is protected against gate-imperfection errors due to the non-exact implementation of Clifford gates. This feature makes the purity a perfect candidate to estimate the state-aware parameter $p$ governing the noise model during the state preparation. Simple algebra leads to
\begin{equation}
P(\psi_{p})=p^{2}+\frac{(1-p)^{2}}{n}+2p\frac{(1-p)}{n}\sum_{i}\operatorname{tr}^{2}(\psi Z_{i})
\end{equation}
Then, making the ansatz that:
\begin{equation}
P(\psi_{p})\stackrel{\mathclap{\normalfont\mbox{!}}}{=} P_{exp}(\mid\!\psi\rangle)
\end{equation}
one can determine $p$ as the positive solution of the above second-order equation:
\begin{equation}
p=\frac{1-Z+\sqrt{n}\sqrt{P_{exp}(\mid\!\psi\rangle)(1-2Z+n)+\frac{Z^2}{n}-1}}{1-2Z+n}
\label{psolution}
\end{equation}
where $Z:=\sum_{i}\operatorname{tr}^{2}(Z_{i}\psi)$. As it is clear from the above equation, the constant $p$ does depend on the state $\mid\!\psi\rangle$ and can be computed once having measured the purity of the outcome state $P_{exp}(\mid\!\psi\rangle)$. Further note that if the experimental purity is one, i.e. the state preparation has not been affected by decoherence, Eq.~\eqref{psolution} gives $p=1$. 

Now, what about the error $\varepsilon$? It does not depend on the state preparation, but rather on the experimental apparatus. We can therefore determine it once and for all from the experimental data coming from the input state $\!\mid\!\!0\rangle^{\otimes n}$. Since this state, unlike any other state, does not need to be prepared, according to our noise model, the measurements in such a state are not affected by decoherence. As usual, we apply $n$ Clifford gates $c_{i}$, one for each qubit $i=1,\dots, n$, then we estimate the occupation probabilities $P(\bigotimes_{i=1}^{n}(c_{i}\!\mid\!\!0\rangle\langle 0\!\!\mid c_{i}^{\dagger})\!\!\mid \mathbf{s})$ and compute $P(\psi)$ and $W(\psi)$ via statistical correlations between randomized measurements. Modeling the error in Clifford gates implementation through the model introduced before (cfr. Eq.~\eqref{gateimperfection}):
\begin{equation}
W_{\varepsilon}(\mid\!\!0\rangle^{\otimes n})=\operatorname{tr}(\mid\!\!0\rangle\langle 0\!\!\mid^{\otimes4 n}Q_{2}^{\varepsilon\otimes n})
\end{equation}
since $W(\psi)$ is multiplicative, note that $W_{\varepsilon}(\mid\!\!0\rangle^{\otimes n})=(W_{\varepsilon}(\mid\!\!0\rangle))^{n}$ and thus we can just work on $W_{\varepsilon}(\mid\!\!0\rangle)$. Thanks to the absence of the state preparation, we expect the purity computed via statistical correlations to be one; unfortunately, we also observe a small discrepancy of the experimental results with respect to one which reveals some error occurring during the projective measurements, i.e. \textit{read-out} error during measurements. We model the readout error with a non-null probability $(1-q)$ that just after the application of the Clifford gate $c_{i}^{\varepsilon}$ and before the measurement the bit is flipped, see Fig.~$1$ in the main text. Since we are dealing with a product state, we can just work on the single qubit state $\!\mid\!\!0\rangle$. The parameter $q$ can be estimated, just as the parameter $p$, from a purity measurement. The single qubit state just before the measurement is:
\begin{equation}
\chi_{q}^{\varepsilon}:=qc^{\varepsilon}\!\mid\!\!0\rangle\langle 0\!\!\mid c^{\varepsilon\dagger}+(1-q)Xc^{\varepsilon}\!\mid\!\!0\rangle\langle 0\!\!\mid^{\otimes n}c^{\varepsilon\dagger}X
\end{equation}
note that the spin flip $X$ occur after the $P_{\varepsilon}$-doped Clifford gate has been applied to the input state $\!\mid\!\!0\rangle$. The probability to find the outcome $\mid\!\!s\rangle$ is:
\begin{equation}
Pr(\chi_{q}^{\varepsilon}\!\!\mid\!\! s)=q Pr(c^{\varepsilon}\!\mid\!\!0\rangle\langle 0\!\!\mid c^{\varepsilon\dagger}\!\!\mid\!\! s)+(1-q)Pr(c^{\varepsilon}\!\mid\!\!0\rangle\langle 0\!\!\mid c^{\varepsilon\dagger}\!\!\mid\!\!\bar{s})
\end{equation}
where $\bar{s}$ is the not of the classical bit $s$ due to the spin-flip $X$. When combining the outcome probabilities to compute the purity via statistical correlations:
\begin{eqnarray}
P(\chi_{q}^{\varepsilon})&=&\langle\sum_{s_1s_2}o_{2}(s_1,s_2)Pr(\chi_{q}^{\varepsilon}\mid s_{1})Pr(\chi_{q}^{\varepsilon}\!\!\mid\!\! s_{2})\rangle_{C^{\varepsilon}}\\
&=&q^2\langle\sum_{s_1s_2}o_{2}(s_1,s_2)Pr(c^{\varepsilon}\!\!\mid\!\!0\rangle\langle 0\!\!\mid c^{\varepsilon\dagger}\!\!\mid\!\! s_{1})Pr(c^{\varepsilon}\!\mid\!\!0\rangle\langle 0\!\!\mid c^{\varepsilon\dagger}\!\!\mid\!\! s_{2})\rangle_{C^{\varepsilon}}\nonumber\\&+& (1-q)^2\langle\sum_{s_1s_2}o_{2}(s_1,s_2)Pr(c^{\varepsilon}\!\mid\!\!0\rangle\langle 0\!\!\mid c^{\varepsilon\dagger}\!\!\mid\!\!\bar{s}_{1})Pr(c^{\varepsilon}\!\mid\!\!0\rangle\langle 0\!\!\mid c^{\varepsilon\dagger}\!\!\mid\!\!\bar{s}_{2})\rangle_{C^{\varepsilon}}\nonumber\\&+&2q(1-q)\langle\sum_{s_1s_2}o_{2}(s_1,s_2)Pr(c^{\varepsilon}\!\mid\!\!0\rangle\langle 0\!\!\mid c^{\varepsilon\dagger}\!\!\mid\!\! s_{1})Pr(c^{\varepsilon}\!\mid\!\!0\rangle\langle 0\!\!\mid c^{\varepsilon\dagger}\!\!\mid\!\!{s}_{2})\rangle_{C^{\varepsilon}}\nonumber\\
&=&(q^{2}+(1-q)^{2})\operatorname{tr}\left(\mid\!\!0\rangle\langle 0\!\!\mid^{\otimes 2} \langle\hat{o}_{2}\rangle_{{\varepsilon}}\right)+2q(1-q)\operatorname{tr}\left(\mid\!\!0\rangle\langle 0\!\!\mid^{\otimes 2} \langle\hat{o}_{2}^{x}\rangle_{{\varepsilon}}\right)\nonumber
\label{errormodel5}
\end{eqnarray}
where we denoted $\langle\cdot\rangle_{\varepsilon}\equiv \langle c^{\varepsilon\otimes 2}\cdot c^{\varepsilon\dagger\otimes 2}\ rangle_{C^{\varepsilon}}$, then recall that:
\begin{equation}
\hat{o}_{2}=\sum_{s_1,s_2}O(s_{1},s_{2})\mid\!\! s_1s_2\rangle\langle s_1s_2\!\!\mid=\frac{\mathchoice {\rm 1\mskip-4mu l} {\rm 1\mskip-4mu l} {\rm 1\mskip-4.5mu l} {\rm 1\mskip-5mu l}^{\otimes 2}}{2}+\frac{3}{2}Z^{\otimes 2}
\end{equation}
$\hat{O}_{2}^{x}:=(\mathchoice {\rm 1\mskip-4mu l} {\rm 1\mskip-4mu l} {\rm 1\mskip-4.5mu l} {\rm 1\mskip-5mu l}\otimes X)\hat{O}_{2}(\mathchoice {\rm 1\mskip-4mu l} {\rm 1\mskip-4mu l} {\rm 1\mskip-4.5mu l} {\rm 1\mskip-5mu l}\otimes X)$ and used the fact that $(X\otimes X)\hat{O}_{2}(X\otimes X)=\hat{O}_{2}$. The doped Clifford averages read:
\begin{eqnarray}
\langle\hat{o}_2\rangle_{\varepsilon}&=&\frac{1}{2}(\mathchoice {\rm 1\mskip-4mu l} {\rm 1\mskip-4mu l} {\rm 1\mskip-4.5mu l} {\rm 1\mskip-5mu l}^{\otimes 2}+X^{\otimes 2}+Y^{\otimes 2}+Z^{\otimes 2})\equiv T_1\nonumber\\
\langle\hat{o}^{x}_{2}\rangle_{\varepsilon}&=&\mathchoice {\rm 1\mskip-4mu l} {\rm 1\mskip-4mu l} {\rm 1\mskip-4.5mu l} {\rm 1\mskip-5mu l}-T_1
\end{eqnarray}
where $T_1$ is the single qubit swap operator. One thus can rewrite Eq.~\eqref{errormodel5} as:
\begin{eqnarray}
P(\chi_{q}^{\varepsilon})\!&=&\!q^2\operatorname{tr}(\mid\!\!0\rangle\langle 0\!\!\mid^{\otimes 2}T_1)+(1-q)^2\operatorname{tr}(\mid\!\!0\rangle\langle 0\!\!\mid^{\otimes 2}T_1)+2q(1-q)\operatorname{tr}(\mid\!\!0\rangle\langle 0\!\!\mid^{\otimes 2}\!\!(\mathchoice {\rm 1\mskip-4mu l} {\rm 1\mskip-4mu l} {\rm 1\mskip-4.5mu l} {\rm 1\mskip-5mu l}-T_1))\nonumber\\
&=&q^2+(1-q)^2
\end{eqnarray}
Finally, making the ansatz 
\begin{equation}
P(\chi_{q}^{\varepsilon})^{n}\stackrel{\mathclap{\normalfont\mbox{!}}}{=} P_{exp}(\mid\!\!0\rangle^{\otimes n})
\end{equation}
one can determine $q$ by the positive solution of the second-order equation:
\begin{equation}
q=\frac{1}{2}\left(1+\sqrt{2P_{exp}(\mid\!\!0\rangle^{\otimes n})^{1/n}-1}\right)
\end{equation}
note that if $P_{exp}(\mid\!\!0\rangle^{\otimes n})=1$, then $q=1$. Now that we know the \textit{read-out} error parameter $q$, we turn to compute $W(\chi_{q}^{\varepsilon})$ to estimate the shift angle $\varepsilon$:
\begin{equation}
W(\chi_{q}^{\varepsilon})=\langle\sum_{s1,s_2,s_3,s_4}o_{4}(s_1,s_2,s_3,s_4)Pr(\chi_{q}^{\varepsilon}\!\!\mid\!\! s_{1})Pr(\chi_{q}^{\varepsilon}\!\!\mid\!\! s_{2})Pr(\chi_{q}^{\varepsilon}\!\!\mid\!\! s_{3})Pr(\chi_{q}^{\varepsilon}\!\!\mid\!\! s_{4})\rangle_{C^{\varepsilon}}
\label{singlequbitstabilizer purity}
\end{equation}
Denoting $\hat{o}_{4}\equiv\sum_{s_1,s_2,s_3,s_4}o_4(s_1,s_2,s_3,s_4)\mid\!\!s_1s_2s_3s_4\rangle\langle s_1s_2s_3s_4\!\!\mid=\frac{1}{4}\mathchoice {\rm 1\mskip-4mu l} {\rm 1\mskip-4mu l} {\rm 1\mskip-4.5mu l} {\rm 1\mskip-5mu l}^{\otimes 4}+\frac{3}{4}Z^{\otimes 4}$, we have the following rules:
\begin{eqnarray}
X^{\otimes 4}\hat{o}_{4}X^{\otimes 4}&=&(X^{\otimes 2}\otimes \mathchoice {\rm 1\mskip-4mu l} {\rm 1\mskip-4mu l} {\rm 1\mskip-4.5mu l} {\rm 1\mskip-5mu l}^{\otimes 2})\hat{o}_{4}(X^{\otimes 2}\otimes \mathchoice {\rm 1\mskip-4mu l} {\rm 1\mskip-4mu l} {\rm 1\mskip-4.5mu l} {\rm 1\mskip-5mu l}^{\otimes 2})=\hat{o}_{4}\nonumber\\ 
(X^{\otimes 3}\otimes \mathchoice {\rm 1\mskip-4mu l} {\rm 1\mskip-4mu l} {\rm 1\mskip-4.5mu l} {\rm 1\mskip-5mu l})\hat{o}_{4}(X^{\otimes 3}\otimes \mathchoice {\rm 1\mskip-4mu l} {\rm 1\mskip-4mu l} {\rm 1\mskip-4.5mu l} {\rm 1\mskip-5mu l})&=&(X\otimes \mathchoice {\rm 1\mskip-4mu l} {\rm 1\mskip-4mu l} {\rm 1\mskip-4.5mu l} {\rm 1\mskip-5mu l}^{\otimes 3})\hat{o}_{4}(X\otimes \mathchoice {\rm 1\mskip-4mu l} {\rm 1\mskip-4mu l} {\rm 1\mskip-4.5mu l} {\rm 1\mskip-5mu l}^{\otimes 3})=\frac{\mathchoice {\rm 1\mskip-4mu l} {\rm 1\mskip-4mu l} {\rm 1\mskip-4.5mu l} {\rm 1\mskip-5mu l}^{\otimes 4}}{2}-\hat{o}_{4}
\end{eqnarray}
from which we can update Eq.~\eqref{singlequbitstabilizer purity} as:
\begin{eqnarray}
W(\chi_{q}^{\varepsilon})=(1-2q)^{4}W_{\varepsilon}(\mid\!\!0\rangle)+2(q^3(1-q)+(1-q)^3q)
\label{singlequbitstabilizer purity2}
\end{eqnarray}
where we defined $W_{\varepsilon}(\mid\!\!0\rangle):=\operatorname{tr}(\mid\!\!0\rangle\langle 0\!\!\mid^{\otimes 4}Q^{\varepsilon}_{1})$. Now, from Eq.~\eqref{errormodel6} we can compute $W_{\varepsilon}(\mid\!\!0\rangle)$ as:
\begin{eqnarray}
W_{\varepsilon}(\mid\!\!0\rangle)=\frac{1}{12}(5+\cos(4\varepsilon)+\sin^2(2\varepsilon))
\label{singlequbitstabilizer purity3}
\end{eqnarray}
and finally, making the ansatz 
\begin{equation}
(W_{exp}(\mid\!\!0\rangle^{\otimes n}))^{1/n}\stackrel{\mathclap{\normalfont\mbox{!}}}{=} W(\chi_{q}^{\varepsilon})
\end{equation}
from Eqs. \eqref{singlequbitstabilizer purity2} and \eqref{singlequbitstabilizer purity3} we can estimate $\varepsilon$:
\begin{equation}
\varepsilon=\pm\frac{1}{4}\cos^{-1}     \left(\frac{-80q^4+160q^3-120q^2+40q+24(W_{exp}(\mid\!\!0\rangle^{\otimes n}))^{1/n}-11}{(2q-1)^4}\right)
\end{equation}
this concludes the section.

\section{Supplementary Note 3: Some proofs}\label{Sec: proofs}
This section is devoted to some lengthy proofs.
\subsection{Proof of Eq.~\eqref{regularizedeq}}
In this section, we prove Eq.~\eqref{regularizedeq}. The first step of the proof is to introduce the following probability distribution (valid for pure states only):
\begin{equation}
\Xi^{reg}_{\psi}:=\{(d-1)^{-1}\operatorname{tr}^{2}(P\psi)\,\mid\, P\neq \mathchoice {\rm 1\mskip-4mu l} {\rm 1\mskip-4mu l} {\rm 1\mskip-4.5mu l} {\rm 1\mskip-5mu l}\}  
\end{equation}
which is obtained by the probability distribution $\Xi_{\psi}$ without the identity component. Define the $\alpha$-R\'enyi entropies of $\Xi^{reg}_{\psi}$, $S_{\alpha}(\Xi^{reg}_{\psi})$, and note that 
\begin{equation}
S_{\infty}(\Xi^{reg}_{\psi})-\log(d-1)=-2\log \max_{P\neq\mathchoice {\rm 1\mskip-4mu l} {\rm 1\mskip-4mu l} {\rm 1\mskip-4.5mu l} {\rm 1\mskip-5mu l}} \mid\!\operatorname{tr}(P\psi)\!\mid 
\end{equation}
By hierarchy of R\'enyi entropies, we can write the following bound:
\begin{equation}
-2\log \max_{P\neq\mathchoice {\rm 1\mskip-4mu l} {\rm 1\mskip-4mu l} {\rm 1\mskip-4.5mu l} {\rm 1\mskip-5mu l}} \mid\!\operatorname{tr}(P\psi)\!\mid \le -\log \sum_{P\neq \mathchoice {\rm 1\mskip-4mu l} {\rm 1\mskip-4mu l} {\rm 1\mskip-4.5mu l} {\rm 1\mskip-5mu l}}\frac{1}{(d-1)}\operatorname{tr}^{4}(P\psi)
\end{equation}
Let us now bound the r.h.s. of the above equation with the stabilizer R\'enyi entropy. 
\begin{eqnarray}
-\log \sum_{P\neq \mathchoice {\rm 1\mskip-4mu l} {\rm 1\mskip-4mu l} {\rm 1\mskip-4.5mu l} {\rm 1\mskip-5mu l}}\frac{1}{(d-1)}\operatorname{tr}^{4}(P\psi)&=&-\log \left(\frac{1}{d}\sum_{P}\operatorname{tr}^{4}(P\psi)-\frac{1}{d}\right)+\log(1-d^{-1})\\
&=& -\log dW(\psi) -\log[1-(d^{2}W(\psi))^{-1}]+\log(1-d^{-1})\nonumber\\
&\le & M_{2}(\psi) -\log[1-(d^{2}W(\psi))^{-1}]
\end{eqnarray}
We are just left to bound the last term. Note that $W(\psi)\ge \frac{2}{d(d-1)}$ (see~\cite{leone2021renyi}), and thus 
\begin{equation}
-\log[1-(d^{2}W(\psi))^{-1}]\le -\log(\frac{1}{2}+\frac{1}{2d})\le \log 2
\end{equation}
which proves the inequality, i.e.
\begin{equation}
-2\log \max_{P\neq\mathchoice {\rm 1\mskip-4mu l} {\rm 1\mskip-4mu l} {\rm 1\mskip-4.5mu l} {\rm 1\mskip-5mu l}} \mid\!\operatorname{tr}(P\psi)\!\mid \le M_{2}(\psi)+1
\end{equation}
\subsection{Bound on the variance of the stabilizer purity}
In this section, we compute the variance $\mathbf{Var}(\tilde{W}(\psi))$. To this aim, let us look at the variance of $\tilde{W}_C(\psi)$:
\begin{equation}
\mathbf{Var}(\tilde{W}_C(\psi))=\mathbb{E}_C\left[\mathbb{E}_{\mathbf{s}}(\tilde{W}^2_C(\psi)\!\!\mid\!\!C)\right]-\left[\mathbb{E}_C\mathbb{E}_\mathbf{s}(\tilde{W}_C(\psi)\!\!\mid\!\!C)\right]^2
\end{equation}
The first term of the variance is equal to
\begin{eqnarray}
\mathbb{E}_C\left[\mathbb{E}_{\mathbf{s}}(\tilde{W}^2_C(\psi)\!\!\mid\!\!C)\right]=\binom{N_m}{4}^{-2}\sum_{\substack{j<k<l<m \\ n<o<p<q}}\hspace{-0.4cm}\mathbb{E}_C\mathbb{E}_\mathbf{s}&&\{\operatorname{tr}\left(\tilde{s}_C(j)\otimes\tilde{s}_C(k)\otimes\tilde{s}_C(l)\otimes\tilde{s}_C(m) \hat{O}_4\right)\nonumber\\\nonumber\times&&\operatorname{tr}\left(\tilde{s}_C(n)\otimes\tilde{s}_C(o)\otimes\tilde{s}_C(p)\otimes\tilde{s}_C(q) \hat{O}_4\right)\}\\
\end{eqnarray}
In the summation, there may be repeated indexes between the sets of indexes $\{j,k,l,m\}$ and $\{n,o,p,q\}$. Let us label with $\alpha$ the number of repeated indexes between the two sets and with $K_{\alpha}$ the term of the sum corresponding to the $\alpha$ repeated indexes. $\mathbb{E}_C\left[\mathbb{E}_{\mathbf{s}}(\tilde{W}^2_C(\psi)\!\!\mid\!\!C)\right]$ can be written in term of $K_\alpha$ as:
\begin{eqnarray}
\mathbb{E}_C\left[\mathbb{E}_{\mathbf{s}}(\tilde{W}^2_C(\psi)\!\!\mid\!\!C)\right]&=&\binom{N_m}{4}^{-2}\left\{\binom{N_M}{8}\binom{8}{0}\binom{8}{4}K_0\right.\\\nonumber&+&\binom{N_M}{7}\binom{7}{1}\binom{6}{3}K_1+\binom{N_M}{6}\binom{6}{2}\binom{4}{2}K_2
\\\nonumber&+&\binom{N_M}{5}\binom{5}{3}\binom{2}{1}K_3+\left.\binom{N_M}{4}\binom{4}{4}\binom{0}{0}K_4\right\}
\end{eqnarray}
Let us compute each term separately, $K_0$ reads:
\begin{eqnarray}
K_0&=&\mathbb{E}_C\operatorname{tr}[\psi_C^{\otimes 4} \hat{O}_4]^2\\
&=&\mathbb{E}_C\operatorname{tr}[\psi^{\otimes 8}C^{\dag\otimes 8}\hat{O}^{\otimes 2}_4C^{\otimes 8}]\nonumber\\
&=&\operatorname{tr}[\psi^{\otimes 8}\frac{1}{4^{n}}(3 P_1 + (Q_1\otimes \mathchoice {\rm 1\mskip-4mu l} {\rm 1\mskip-4mu l} {\rm 1\mskip-4.5mu l} {\rm 1\mskip-5mu l}_4 + \mathchoice {\rm 1\mskip-4mu l} {\rm 1\mskip-4mu l} {\rm 1\mskip-4.5mu l} {\rm 1\mskip-5mu l}_4\otimes Q_1 - \mathchoice {\rm 1\mskip-4mu l} {\rm 1\mskip-4mu l} {\rm 1\mskip-4.5mu l} {\rm 1\mskip-5mu l}_8) )^{\otimes n}]\nonumber
\end{eqnarray}
where $P_n:=\frac{1}{d^2}\sum_{P\in\mathbb \mathbb{P}(n)} P^{\otimes 8}$, $Q_n:=\frac{1}{d^2}\sum_{P\in\mathbb{P}(n)}P^{\otimes 4}$ and $\mathchoice {\rm 1\mskip-4mu l} {\rm 1\mskip-4mu l} {\rm 1\mskip-4.5mu l} {\rm 1\mskip-5mu l}_n:=\mathchoice {\rm 1\mskip-4mu l} {\rm 1\mskip-4mu l} {\rm 1\mskip-4.5mu l} {\rm 1\mskip-5mu l}^{\otimes n}$. It is not difficult to observe that $P_1$ and $(Q_1\otimes\mathchoice {\rm 1\mskip-4mu l} {\rm 1\mskip-4mu l} {\rm 1\mskip-4.5mu l} {\rm 1\mskip-5mu l}_4 + \mathchoice {\rm 1\mskip-4mu l} {\rm 1\mskip-4mu l} {\rm 1\mskip-4.5mu l} {\rm 1\mskip-5mu l}_4\otimes Q_1)$ commutes, meaning that there exists a basis in which both are diagonal. Then if 
\begin{eqnarray}
3 P_1&=&\sum_{i}a_i \mid\!\! i \rangle\langle i \!\!\mid \quad 0\le a_i\le 3 \\
(Q_1\otimes\mathchoice {\rm 1\mskip-4mu l} {\rm 1\mskip-4mu l} {\rm 1\mskip-4.5mu l} {\rm 1\mskip-5mu l}_4 + \mathchoice {\rm 1\mskip-4mu l} {\rm 1\mskip-4mu l} {\rm 1\mskip-4.5mu l} {\rm 1\mskip-5mu l}_4\otimes Q_1))&=&\sum_{i}b_i \mid\!\! i \rangle\langle i \!\!\mid \quad 0\le b_i \le 2\nonumber
\end{eqnarray}
then $K_0$ can be written as 
\begin{eqnarray}
K_0&=& \frac{1}{4^{n}}\sum_{i_1\ldots i_n}(a_{i_1}+b_{i_1}-1)(a_{i_2}+b_{i_2}-1)\cdots(a_{i_n}+b_{i_n}-1)\langle i_1 \ldots i_n\!\!\mid\!\! \rho^{\otimes 8}\!\!\mid\!\! i_1 \ldots i_n \rangle\nonumber\\
&\le& \sum_{i_1\ldots i_n}\!\!\mid\!\!(a_{i_1}+b_{i_1}-1)\!\!\mid\!\!(a_{i_2}+b_{i_2}-1)\cdots\!\!\mid\!\!(a_{i_n}+b_{i_n}-1)\!\!\mid\!\!\langle i_1 \ldots i_n\!\!\mid\!\! \rho^{\otimes 8}\!\!\mid\!\! i_1 \ldots i_n \rangle\nonumber\\
&\le&\sum_{i_1\ldots i_n}(\mid\!\!a_{i_1}\!\!\mid\!\!+\!\!\mid\!\!b_{i_1}-1\!\!\mid)(\mid\!\!a_{i_2}\!\!\mid\!\!+\!\!\mid\!\!b_{i_2}-1\!\!\mid)\cdots(\mid\!\!a_{i_n}\!\!\mid\!\!+\!\!\mid\!\!b_{i_n}-1\!\!\mid)\langle i_1 \ldots i_n\!\!\mid\!\! \rho^{\otimes 8}\!\!\mid\!\! i_1 \ldots i_n \rangle\nonumber\\
&\le&\sum_{i_1\ldots i_n}(\mid\!\!a_{i_1}\!\!\mid\!\!+\!\!\mid\!\!1\!\!\mid)(\mid\!\!a_{i_2}\!\!\mid\!\!+\!\!\mid\!\!1\!\!\mid)\cdots(\mid\!\!a_{i_n}\!\!\mid\!\!+\!\!\mid\!\!1\!\!\mid)\langle i_1 \ldots i_n\!\!\mid\!\! \rho^{\otimes 8}\!\!\mid\!\! i_1 \ldots i_n \rangle\nonumber\\
&=&\operatorname{tr}[\psi^{\otimes 8}\frac{1}{4^n}(3P_1 +\mathchoice {\rm 1\mskip-4mu l} {\rm 1\mskip-4mu l} {\rm 1\mskip-4.5mu l} {\rm 1\mskip-5mu l}_8 )^{\otimes n}]\\&=&\frac{1}{4^n}\sum_{k}\left\{\operatorname{tr}(\psi^{\otimes 8}\mathchoice {\rm 1\mskip-4mu l} {\rm 1\mskip-4mu l} {\rm 1\mskip-4.5mu l} {\rm 1\mskip-5mu l}^{\otimes n- k}_8 3^{k}P_1^{\otimes k}\right\}_k\le \frac{1}{4^n}(1+\frac{3}{2})^{n}\le\frac{1}{2^{n/2}}\nonumber
\end{eqnarray}
where $\{\cdot\}_k$ labels all the possible tensor product of $\mathchoice {\rm 1\mskip-4mu l} {\rm 1\mskip-4mu l} {\rm 1\mskip-4.5mu l} {\rm 1\mskip-5mu l}_8^{\otimes n-k}P_1^{k}$. To prove the bound we used the following properties: positivity of $P_1, (Q_1\otimes \mathchoice {\rm 1\mskip-4mu l} {\rm 1\mskip-4mu l} {\rm 1\mskip-4.5mu l} {\rm 1\mskip-5mu l}_4+\mathchoice {\rm 1\mskip-4mu l} {\rm 1\mskip-4mu l} {\rm 1\mskip-4.5mu l} {\rm 1\mskip-5mu l}_4\otimes Q_1),\mathchoice {\rm 1\mskip-4mu l} {\rm 1\mskip-4mu l} {\rm 1\mskip-4.5mu l} {\rm 1\mskip-5mu l}_8$, eigenvalues of $ (Q_1\otimes \mathchoice {\rm 1\mskip-4mu l} {\rm 1\mskip-4mu l} {\rm 1\mskip-4.5mu l} {\rm 1\mskip-5mu l}_4+\mathchoice {\rm 1\mskip-4mu l} {\rm 1\mskip-4mu l} {\rm 1\mskip-4.5mu l} {\rm 1\mskip-5mu l}_4\otimes Q_1)$ bounded between $0$ and $2$, and that $\operatorname{tr}{P_n\rho^{\otimes 8}}\le \operatorname{tr}{Q_n\rho^{\otimes 4}}\le \frac{1}{2^n}$(the proof of this inequality is a direct consequence of $-1 \le\operatorname{tr}{P\psi}\le 1$).

The second term $K_1$ can be written as:
\begin{eqnarray}
K_1&=&\mathbb{E}_C\sum_{s_1}\operatorname{tr}[\tilde{s}_1 \psi_C]\operatorname{tr}[\tilde{s}_1\otimes\psi_C^{\otimes 3} \hat{O}_4]^2\nonumber\\
&=& \mathbb{E}_C\sum_{\substack{s_1,s_2,s_3,s_4\\
s_5,s_6,s_7}}\langle s_1s_2s_3s_4s_5s_6s_7\!\!\mid\!\!\psi_C^{\otimes 7}\!\!\mid\!\! s_1s_2s_3s_4s_5s_6s_7\rangle O(s_1s_2s_3s_4)O(s_1s_5s_6s_7)\nonumber\\&=&\mathbb{E}_C\operatorname{tr}(\psi^{\otimes 7} C^{\dag\otimes 7}R_{1}C^{\otimes 7})\le 1
\end{eqnarray}
with 
\begin{equation}
R_{1}:=\sum_{\substack{s_1s_2s_3s_4\\s_5s_6s_7}}O(s_1s_2s_3s_4)O(s_1s_5s_6s_7)\!\!\mid\!\! s_1s_2s_3s_4s_5s_6s_7\rangle \langle s_1s_2s_3s_4s_5s_6s_7\!\!\mid 
\end{equation}
that since $\hat{O}_4$ is a diagonal operator, $R_{1}$ is also a diagonal operator and then we can rewrite the term $K_1$ as the average of expectation values $C^{\dag\otimes 7}R_{1}C^{\otimes 7}$ over the state $\psi^{\otimes 7}$. As before, we can upper bound $K_1$ with the highest eigenvalue of $R_{1}$ that, since it is defined as a diagonal operator whose components are given by the product of components of $\hat{O}_4$, is equal to $1$. 
The third term is equal to:
\begin{eqnarray}
K_2&=&\mathbb{E}_C\sum_{s_1s_2}\operatorname{tr}[\tilde{s}_1 \psi_C]\operatorname{tr}[\tilde{s}_2 \psi_C]\operatorname{tr}[\tilde{s}_1\otimes\tilde{s}_2\psi_C^{\otimes 2} \hat{O}_4]^2\nonumber\\
&=&\mathbb{E}_{C}\sum_{\substack{s_1,s_2,s_3\\
s_4,s_5,s_6}}\langle s_1s_2s_3s_4s_5s_6\!\!\mid\!\! \psi_C^{\otimes 7}\!\!\mid\!\! s_1s_2s_3s_4s_5s_6\rangle O(s_1s_2s_3s_4)O(s_1s_2s_5s_6)\nonumber\\
&=&\operatorname{tr}\left(\psi^{\otimes 6}C^{\dag\otimes 6}R_{2}C^{\otimes 6} \right)\le 1
\end{eqnarray}
with 
\begin{equation}
R_{2}:=\sum_{s_1s_2s_3s_4s_5s_6}O(s_1s_2s_3s_4)O(s_1s_2s_5s_6)\!\!\mid\!\! s_1s_2s_3s_4s_5s_6\rangle\langle s_1s_2s_3s_4s_5s_6\!\!\mid 
\end{equation}
The proof is a direct consequence of the arguments given for the second term. The fourth term can be rewritten as:
\begin{eqnarray}
K_3&=&\mathbb{E}_C\sum_{s_1s_2s_3}\operatorname{tr}[\tilde{s}_1 \psi_C]\operatorname{tr}[\tilde{s}_2 \psi_C]\operatorname{tr}[\tilde{s}_3\otimes \psi_C]\operatorname{tr}[\tilde{s}_1\otimes\tilde{s}_2\otimes\tilde{s}_3\otimes\psi_C \hat{O}_4]^2\nonumber\\
&=&\mathbb{E}_{C}\sum_{\substack{s_1,s_2,s_3\\
s_4,s_5}}\langle s_1s_2s_3s_4s_5\mid \psi_C^{\otimes 7}\!\!\mid\!\! s_1s_2s_3s_4s_5\rangle O(s_1s_2s_3s_4)O(s_1s_2s_3s_5)\nonumber\\
&=&\operatorname{tr}\left(\psi^{\otimes 5}C^{\dag\otimes 5}R_{3}C^{\otimes 5} \right)\le 1
\end{eqnarray}
with
\begin{equation}
R_{3}:=\sum_{s_1s_2s_3s_4s_5}O(s_1s_2s_3s_4)O(s_1s_2s_3s_5)\!\!\mid\!\! s_1s_2s_3s_4s_5\rangle\langle s_1s_2s_3s_4s_5\!\!\mid 
\end{equation}
The last term is equal to
\begin{eqnarray} 
K_4&=&\mathbb{E}_C\sum_{s_1s_2s_3s_4}\operatorname{tr}[\tilde{s}_1 \psi_C]\operatorname{tr}[\tilde{s}_2 \psi_C]\operatorname{tr}[\tilde{s}_3 \psi_C]\operatorname{tr}[\tilde{s}_4 \psi_C]\operatorname{tr}[\tilde{s}_1\otimes\tilde{s}_2\otimes\tilde{s}_3\otimes \tilde{s}_4 O]^2\nonumber\\
&=& \operatorname{tr}(\psi^{\otimes 4}C^{\dag\otimes4} \hat{O}^2C^{\otimes 4})
\end{eqnarray}
The last equivalence is due to the fact that $\hat{O}$ is diagonal. Let us take a step back and compute the average of $\hat{o}_4^2$,
\begin{eqnarray} 
\mathbb{E}_{C_2}\left[\hat{o}_4^2\right]&=&\mathbb{E}_C\left[(\frac{1}{4}\mathchoice {\rm 1\mskip-4mu l} {\rm 1\mskip-4mu l} {\rm 1\mskip-4.5mu l} {\rm 1\mskip-5mu l}^{\otimes4}+\frac{3}{4}Z^{\otimes 4})^2\right]\\
&=&\frac{5}{8}\mathchoice {\rm 1\mskip-4mu l} {\rm 1\mskip-4mu l} {\rm 1\mskip-4.5mu l} {\rm 1\mskip-5mu l}^{\otimes4}+\frac{1}{8}(X^{\otimes4}+Y^{\otimes4}+Z^{\otimes4})\nonumber\\
&=&\frac{1}{2}\mathchoice {\rm 1\mskip-4mu l} {\rm 1\mskip-4mu l} {\rm 1\mskip-4.5mu l} {\rm 1\mskip-5mu l}^{\otimes4}+\frac{1}{2}Q_1\nonumber
\end{eqnarray}
then $K_4$ reads:
\begin{eqnarray}
K_4&=&\operatorname{tr}\left(\left(\frac{1}{2}\mathchoice {\rm 1\mskip-4mu l} {\rm 1\mskip-4mu l} {\rm 1\mskip-4.5mu l} {\rm 1\mskip-5mu l}^{\otimes 4}+\frac{1}{2}Q_1\right)^{\otimes n}\psi^{4}\right)\\
&=&\frac{1}{2^n}\sum_{k}\left\{\operatorname{tr}(\psi^{\otimes 4}\mathchoice {\rm 1\mskip-4mu l} {\rm 1\mskip-4mu l} {\rm 1\mskip-4.5mu l} {\rm 1\mskip-5mu l}^{\otimes n- k}Q^{\otimes k}\right\}_k\nonumber\\
&\le&\frac{1}{2^n}\sum_{k}\binom{n}{k}\frac{1}{2^k}=\left(\frac{3}{4}\right)^n\le d^{-1/3}\nonumber
\end{eqnarray}
where with $\{\cdot\}_k$ we labeled all the tensor product permutations of $\mathchoice {\rm 1\mskip-4mu l} {\rm 1\mskip-4mu l} {\rm 1\mskip-4.5mu l} {\rm 1\mskip-5mu l}^{\otimes n-k}\otimes Q_2^{\otimes k}$ and used the following inequality $\operatorname{tr}(Q_d\psi)\le \frac{1}{d}$. 
The term $\mathbb{E}_C\left[\mathbb{E}_{\mathbf{s}}(\tilde{W}^2_C(\psi)\!\!\mid\!\!C)\right]$ reads:
\begin{eqnarray}
\mathbb{E}_C\left[\mathbb{E}_{\mathbf{s}}(\tilde{W}^2_C(\psi)\!\!\mid\!\!C)\right]&=&\binom{N_M}{4}^{-2}\left\{2\binom{N_M}{8}\binom{8}{0}\binom{8}{4}\sqrt{d}^{-1}\right.+\binom{N_M}{7}\binom{7}{1}\binom{6}{3}\\&+&\binom{N_M}{6}\binom{6}{2}\binom{4}{2}+\binom{N_M}{5}\binom{5}{3}\binom{2}{1}+\left.\binom{N_M}{4}\binom{4}{4}\binom{0}{0}d^{-1/3}\right\}\nonumber\\
&\le&\frac{8}{\sqrt{d}}+\frac{192}{d^{1/3}N_M^4}+\frac{6792}{d^{1/2}N_M^4}+\frac{5056}{N_M^3}+\frac{8179}{d^{1/2}N_M^2}+\frac{128}{N_M}\nonumber
\end{eqnarray}
and consequently
\begin{equation}
\mathbf{Var}{ \tilde{W}_C(\psi)}\le \frac{8}{\sqrt{d}}+\frac{192}{d^{1/3}N_M^4}+\frac{6792}{d^{1/2}N_M^4}+\frac{5056}{N_M^3}+\frac{8179}{d^{1/2}N_M^2}+\frac{128}{N_M} - \operatorname{tr}[Q\psi^{\otimes 4}]^2
\end{equation}
When $N_M\ll d$ we have that $\mathbf{Var}{ \tilde{W}_C(\psi)}\le \frac{8}{\sqrt{d}}-\operatorname{tr}[Q\psi^{\otimes 4}]^2$. Then the variance of $\mathbf{Var}(\tilde{W}(\psi))$ reads:
\begin{equation}
\mathbf{Var}(\tilde{W}(\psi))\le\frac{1}{N_U}\left[ \frac{8}{\sqrt{d}}+\frac{192}{d^{1/3}N_M^4}+\frac{6792}{d^{1/2}N_M^4}+\frac{5056}{N_M^3}+\frac{8179}{d^{1/2}N_M^2}+\frac{128}{N_M} - \operatorname{tr}[Q\psi^{\otimes 4}]^2\right]
\end{equation}
\subsection{Bound on the variance of the purity}

In this section, we compute the variance $\mathbf{Var}(\tilde{\operatorname{Pur}}(\psi))$. 
\begin{equation}
\mathbf{Var}(\tilde{\operatorname{Pur}}_C(\psi))=\mathbb{E}_C\left[\mathbb{E}_{\mathbf{s}}(\tilde{\operatorname{Pur}}^2_C(\psi)\!\!\mid\!\!C)\right]-\left[\mathbb{E}_C\mathbb{E}_\mathbf{s}(\tilde{\operatorname{Pur}}_C(\psi)\!\!\mid\!\!C)\right]^2
\end{equation}
As for the stabilizer purity, the first term reads:
\begin{eqnarray}
\mathbb{E}_C\left[\mathbb{E}_{\mathbf{s}}(\tilde{\operatorname{Pur}}^2_C(\psi)\!\!\mid\!\!C)\right]&=&\binom{N_m}{2}^{-2}\hspace{-0.2cm}\sum_{\substack{j<k \\ l<m}}\mathbb{E}_C\mathbb{E}_\mathbf{s}\{\operatorname{tr}\left(\tilde{s}_C(j)\otimes\tilde{s}_C(k) \hat{O}_2\right)\nonumber\\ &\times&\operatorname{tr}\left(\tilde{s}_C(l)\otimes\tilde{s}_C(m) \hat{O}_2\right)\}
\end{eqnarray}
Similarly to what done for the stabilizer purity, we label with $\alpha$ the number of repeated indexes between the two sets and with $J_{\alpha}$ the term of the sum corresponding to the $\alpha$ repeated indexes. $\mathbb{E}_C\left[\mathbb{E}_{\mathbf{s}}(\tilde{\operatorname{Pur}}^2_C(\psi)\!\!\mid\!\!C)\right]$ can be written in term of $J_\alpha$ as:

\begin{eqnarray}
\mathbb{E}_C\left[\mathbb{E}_{\mathbf{s}}(\tilde{\operatorname{Pur}}^2_C(\psi)\!\!\mid\!\!C)\right]&=&\binom{N_m}{2}^{-2}\left\{\binom{N_M}{4}\binom{4}{0}\binom{4}{2}J_0\right.\\\nonumber&+&\left.\binom{N_M}{3}\binom{3}{1}\binom{2}{1}J_1+\binom{N_M}{2}\binom{2}{2}\binom{0}{0}J_2\right\}
\end{eqnarray}
The term $J_0$ reads:
\begin{eqnarray}
J_0&=&\mathbb{E}_C\operatorname{tr}[\psi_C^{\otimes 2} \hat{O}_2]^2\\
&=&\mathbb{E}_C\operatorname{tr}[\psi^{\otimes 4}C^{\dag\otimes 4}\hat{O}^{\otimes 2}_2C^{\otimes 4}]\nonumber\\
&=&\operatorname{tr}[\psi^{\otimes 4}(3 Q_1 + (\frac{T_1\otimes \mathchoice {\rm 1\mskip-4mu l} {\rm 1\mskip-4mu l} {\rm 1\mskip-4.5mu l} {\rm 1\mskip-5mu l}_2 + \mathchoice {\rm 1\mskip-4mu l} {\rm 1\mskip-4mu l} {\rm 1\mskip-4.5mu l} {\rm 1\mskip-5mu l}_2\otimes T_1}{2} - \mathchoice {\rm 1\mskip-4mu l} {\rm 1\mskip-4mu l} {\rm 1\mskip-4.5mu l} {\rm 1\mskip-5mu l}_4) )^{\otimes n}]\nonumber\\
\end{eqnarray}
where $T_1$ is the swap operator acting on two copies of $\mathbb{C}_2$. In order to find an upper-bound to $J_0$, let us introduce the following operator $2Q_1+\mathchoice {\rm 1\mskip-4mu l} {\rm 1\mskip-4mu l} {\rm 1\mskip-4.5mu l} {\rm 1\mskip-5mu l}_4$. The above operator commutes with $3 Q_1 + ((T_1\otimes \mathchoice {\rm 1\mskip-4mu l} {\rm 1\mskip-4mu l} {\rm 1\mskip-4.5mu l} {\rm 1\mskip-5mu l}_2 + \mathchoice {\rm 1\mskip-4mu l} {\rm 1\mskip-4mu l} {\rm 1\mskip-4.5mu l} {\rm 1\mskip-5mu l}_2\otimes T_1)/2 - \mathchoice {\rm 1\mskip-4mu l} {\rm 1\mskip-4mu l} {\rm 1\mskip-4.5mu l} {\rm 1\mskip-5mu l}_4) $, and so they can be written in their spectral decomposition as
\begin{eqnarray}
    (3 Q_1 + (\frac{T_1\otimes \mathchoice {\rm 1\mskip-4mu l} {\rm 1\mskip-4mu l} {\rm 1\mskip-4.5mu l} {\rm 1\mskip-5mu l}_2 + \mathchoice {\rm 1\mskip-4mu l} {\rm 1\mskip-4mu l} {\rm 1\mskip-4.5mu l} {\rm 1\mskip-5mu l}_2\otimes T_1}{2} - \mathchoice {\rm 1\mskip-4mu l} {\rm 1\mskip-4mu l} {\rm 1\mskip-4.5mu l} {\rm 1\mskip-5mu l}_4) )&=&\sum_{i}a_i \mid\!\! i \rangle\langle i \!\!\mid \\
    2Q_1+\mathchoice {\rm 1\mskip-4mu l} {\rm 1\mskip-4mu l} {\rm 1\mskip-4.5mu l} {\rm 1\mskip-5mu l}_4&=&\sum_{i}b_i\mid\!\! i \rangle\langle i \!\!\mid
\end{eqnarray}
By inspection, one can show that $\forall i\quad \mid\!\! a_i\!\!\mid \le \mid \!\! b_i\!\!\mid=b_i$. As a consequence:
\begin{eqnarray}
   J_0&=&\operatorname{tr}[\psi^{\otimes 4}(3 Q_1 + (\frac{T_1\otimes \mathchoice {\rm 1\mskip-4mu l} {\rm 1\mskip-4mu l} {\rm 1\mskip-4.5mu l} {\rm 1\mskip-5mu l}_2 + \mathchoice {\rm 1\mskip-4mu l} {\rm 1\mskip-4mu l} {\rm 1\mskip-4.5mu l} {\rm 1\mskip-5mu l}_2\otimes T_1}{2} - \mathchoice {\rm 1\mskip-4mu l} {\rm 1\mskip-4mu l} {\rm 1\mskip-4.5mu l} {\rm 1\mskip-5mu l}_4) )^{\otimes n}]\\
   &=&\sum_{i_1\ldots i_n}a_{i_1}\ldots a_{i_n}\langle i_1\ldots i_n \!\!\mid\!\! \rho^{\otimes 4} \!\!\mid\!\! i_1 \ldots i_n \rangle\\
   &\le& \sum_{i_1\ldots i_n}\!\!\mid\!\!a_{i_1}\!\!\mid\!\!\ldots \!\!\mid\!\!a_{i_n}\!\!\mid\!\!\langle i_1\ldots i_n \!\!\mid\!\! \rho^{\otimes 4} \!\!\mid\!\! i_1 \ldots i_n \rangle\\
   &\le&\sum_{i_1\ldots i_n}b_{i_1}\ldots b_{i_n}\langle i_1\ldots i_n \!\!\mid\!\! \rho^{\otimes 4} \!\!\mid\!\! i_1 \ldots i_n \rangle\\
   &=&\operatorname{tr}[\psi^{\otimes 4}(2Q_1+\mathchoice {\rm 1\mskip-4mu l} {\rm 1\mskip-4mu l} {\rm 1\mskip-4.5mu l} {\rm 1\mskip-5mu l}_4)^{\otimes n}]\\
   &=&\sum_{k}\{\operatorname{tr}[2^k Q_k\otimes \mathchoice {\rm 1\mskip-4mu l} {\rm 1\mskip-4mu l} {\rm 1\mskip-4.5mu l} {\rm 1\mskip-5mu l}_4^{\otimes n-k}]\}_k\\
   &\le&\sum_{k}\binom{n}{k}=d
\end{eqnarray}
where $\{\}_k$ as already introduced labels all the possible permutation of $Q^{\otimes k}$ and $\otimes\mathchoice {\rm 1\mskip-4mu l} {\rm 1\mskip-4mu l} {\rm 1\mskip-4.5mu l} {\rm 1\mskip-5mu l}_4^{\otimes n-k}$. The proof of the bound follows from the two bounds: the first introduces above $\forall i\quad \mid\!\! a_i\!\!\mid \le \mid \!\! b_i\!\!\mid=b_i$ and that $\operatorname{tr}[Q_k\psi^{\otimes 4}]\le 2^{-k}$
The second term $J_1$ can be written as:
\begin{eqnarray}
J_1&=&\mathbb{E}_C\sum_{s_1}\operatorname{tr}[\tilde{s}_1 \psi_C]\operatorname{tr}[\tilde{s}_1\otimes\psi_C  \hat{O}_2]^2\nonumber\\
&=& \mathbb{E}_C\sum_{\substack{s_1,s_2,s_3}}\langle s_1s_2s_3\!\!\mid\!\!\psi_C^{\otimes 3}\!\!\mid\!\! s_1s_2s_3\rangle O(s_1s_2)O(s_1s_3)\nonumber\\&=&\mathbb{E}_C\operatorname{tr}(\psi^{\otimes 3} C^{\dag\otimes 3}S_{1}C^{\otimes 3})\le 2^n
\end{eqnarray}
with 
\begin{equation}
S_{1}:=\sum_{\substack{s_1s_2s_3}}O(s_1s_2)O(s_1s_3)\!\!\mid\!\! s_1s_2s_3\rangle \langle s_1s_2s_3\!\!\mid 
\end{equation}
 As before, we can upper bound $J_1$ with the highest eigenvalue of $S_{1}$ that, since it is defined as a diagonal operator whose components are given by the product of components of $\hat{O}_2$, is bounded to be less than $2^n$. 
The third term is equal to:
\begin{eqnarray}
J_2&=&\mathbb{E}_C\sum_{s_1s_2}\operatorname{tr}[\tilde{s}_1 \psi_C]\operatorname{tr}[\tilde{s}_2 \psi_C]\operatorname{tr}[\tilde{s}_1\otimes\tilde{s}_2 \hat{O}_2]^2\nonumber\\
&=&\mathbb{E}_C\operatorname{tr}[\psi_{C}^{\otimes 2} \hat{O}_2^{2}]\\
&=&\operatorname{tr}[\psi_{C}^{\otimes 2}\left(2 \mathchoice {\rm 1\mskip-4mu l} {\rm 1\mskip-4mu l} {\rm 1\mskip-4.5mu l} {\rm 1\mskip-5mu l}_2 + T_2\right)^{\otimes n}]
\end{eqnarray}
The calculations for $J_2$ can be developed in a similar fashion to what was done for $K_4$. We obtain
\begin{equation}
J_2\le 3^n
\end{equation}

The term $\mathbb{E}_C\left[\mathbb{E}_{\mathbf{s}}(\tilde{W}^2_C(\psi)\!\!\mid\!\!C)\right]$ reads:
\begin{eqnarray}
\mathbb{E}_C\left[\mathbb{E}_{\mathbf{s}}(\tilde{\operatorname{Pur}}^2_C(\psi)\!\!\mid\!\!C)\right]&=&\binom{N_M}{4}^{-2}\left\{2\binom{N_M}{4}\binom{4}{0}\binom{4}{2}(2)^n\right.\\&+&\binom{N_M}{3}\binom{3}{1}\binom{2}{1}2^n+\left.\binom{N_M}{2}\binom{2}{2}\binom{0}{0}3^n\right\}\nonumber\\
&\le&\left(2^{n+1} + \frac{4}{N_M^2}3^{n}\right)\nonumber
\end{eqnarray}
and consequently
\begin{equation}
\mathbf{Var}( \tilde{\operatorname{Pur}}_C(\psi))\le \left(2^{n+1} + \frac{4}{N_M^2}3^{n}\right) - \operatorname{tr}[T\psi^{\otimes 2}]^2
\end{equation}
Then when $N_M\ll d$ we have that $\mathbf{Var}(\tilde{\operatorname{Pur}}_C(\psi))\le d-\operatorname{tr}[T\psi^{\otimes 2}]^2$. Then the variance of $\mathbf{Var}(\tilde{\operatorname{Pur}}(\psi))$ reads:
\begin{equation}
\mathbf{Var}(\tilde{\operatorname{Pur}}(\psi))\le\left(2^{n+1} + \frac{4}{N_M^2}3^{n}\right) - \operatorname{tr}[T\psi^{\otimes 2}]^2
\end{equation}
Consequently, for $d\gg N_M \gg 1$ we have  
\begin{equation}
    \mathbf{Var}(\tilde{\operatorname{Pur}}(\psi))\lesssim\left(c_1 \frac{d}{N_{U}}+\frac{c_2 d^{\log_2 3}}{N_M^2 N_U}\right).
\end{equation}
where $c_1$ and $c_2$ are constants.

\section{Supplementary Note 4: Tables and data}\label{Sec: tablesanddata}
In this section, we present the numerical and experimental data organized in Tables. In Table \ref{table: numberofresources}, we show the numerical data for the optimal number of total measurement $N_{U}\times N_M$, where $N_U$ is the number of sampling unitaries and $N_M$ the number of measurement shots per unitary selected. 

\begin{table}[H]
    \centering
    \caption{ Number of optimal resources $N_{U}$ and $N_{M}$ for n=1,3,4,5.} 
    \begin{tabular}{cccc}
        
\toprule
    \multicolumn{1}{c}{\# qubits} & $\mid\!\psi\rangle$ & $N_{U}$ & $N_M$ \\
    \midrule
  \multicolumn{1}{c}{\multirow{6}{*}{ $n=1$}} & $\!\mid\!\!+\rangle$ &$24$& $32$\\
  \multicolumn{1}{c}{} & $\mid\!\!P_{\pi/16}\rangle$& $23$& $32$\\
 \multicolumn{1}{c}{}  & $\mid\!\!P_{\pi/8}\rangle$& $20$& $32$\\
  \multicolumn{1}{c}{} & $\mid\!\!P_{\pi/6}\rangle$& $17$&$32$\\
     \multicolumn{1}{c}{} & $\mid\!\!P_{\pi/5}\rangle$& $11$&$32$\\
     \multicolumn{1}{c}{} & $\mid\!\!T\rangle$& $8$&$32$\\
   \bottomrule
   \end{tabular}
         \begin{tabular}{cccc}
      \toprule
          \multicolumn{1}{c}{\# qubits} & $\mid\!\psi\rangle$ & $N_{U}$ & $N_M$ \\
          \midrule
 \multicolumn{1}{c}{ \multirow{5}{*} {$n=3$}} & $\mid\!\!\Gamma^{(3)}_{1}\rangle$& $70$& $100$\\
 \multicolumn{1}{c}{} & $\mid\!\!\Gamma^{(3)}_{2}\rangle$& $50$& $100$\\
 \multicolumn{1}{c}{}  & $\mid\!\!\Gamma^{(3)}_{3}\rangle$& $40$& $100$\\
 \multicolumn{1}{c}{}  & $\mid\!\!\Gamma^{(3)}_{4}\rangle$& $30$& $60$\\
   \multicolumn{1}{c}{}  & $\mid\!\!\Gamma^{(3)}_{5}\rangle$& $20$& $60$\\
   \bottomrule
   \end{tabular}

      \begin{tabular}{cccc}
      \toprule
          \multicolumn{1}{c}{\# qubits} & $\mid\!\psi\rangle$ & $N_{U}$ & $N_M$ \\
    \midrule
     \multicolumn{1}{c}{ \multirow{5}{*} {$n=4$}}
 & $\mid\!\!\Gamma^{(4)}_{1}$& $100$& $200$\\
   \cline{2-4}
 \multicolumn{1}{c}{} & $\mid\!\!\Gamma^{(4)}_{2}\rangle$& $60$& $200$\\
   \cline{2-4}
 \multicolumn{1}{c}{}  & $\mid\!\!\Gamma^{(4)}_{3}\rangle$& $50$& $170$\\
   \cline{2-4}
 \multicolumn{1}{c}{}  & $\mid\!\!\Gamma^{(4)}_{4}\rangle$& $50$& $170$\\
  \cline{2-4}
   \multicolumn{1}{c}{}  & $\mid\!\!\Gamma^{(4)}_{5}\rangle$& $30$& $150$\\
     \cline{2-4}
    \multicolumn{1}{c}{}  & $\mid\!\!\Gamma^{(4)}_{6}\rangle$& $30$& $140$\\
  \cline{2-4}
   \multicolumn{1}{c}{}  & $\mid\!\!\Gamma^{(4)}_{7}\rangle$& $20$& $130$\\
   \bottomrule
   \end{tabular}
         \begin{tabular}{cccc}
      \toprule
          \multicolumn{1}{c}{\# qubits} & $\mid\!\psi\rangle$ & $N_{U}$ & $N_M$ \\
    \midrule
      \multicolumn{1}{c}{ \multirow{9}{*} {$n=5$}} & $\mid\!\!\Gamma^{(5)}_{1}\rangle$ &$300$&$410$  \\
 \multicolumn{1}{c}{} & $\mid\!\!\Gamma^{(5)}_{2}\rangle$& $240$ & $390$  \\
 \multicolumn{1}{c}{}  & $\mid\!\!\Gamma^{(5)}_{3}\rangle$&$190$ &$390$\\
 \multicolumn{1}{c}{}  & $\mid\!\!\Gamma^{(5)}_{4}\rangle$&$160$ & $370$\\
   \multicolumn{1}{c}{}  & $\mid\!\!\Gamma^{(5)}_{5}\rangle$&$120$ &$370$\\
    \multicolumn{1}{c}{}  & $\mid\!\!\Gamma^{(5)}_{6}\rangle$&$80$ &$340$\\
   \multicolumn{1}{c}{}  & $\mid\!\!\Gamma^{(5)}_{7}\rangle$&$60$ &$330$\\
      \multicolumn{1}{c}{}  & $\mid\!\!\Gamma^{(5)}_{8}\rangle$&$40$ &$320$\\
      \multicolumn{1}{c}{}  & $\mid\!\!\Gamma^{(5)}_{9}\rangle$&$30$ &$320$\\
   \bottomrule
    \end{tabular}
    \label{table: numberofresources}
\end{table}

In table~\ref{tableexp: 1qubit} we show the data for the measurement of the magic of single qubit $\mid \!\!P_{\theta}\rangle$. In tables~\ref{tableexp: quito} and~\ref{tableexp: casablanca} we show the experimental data for the measurement of the magic of $\mid \!\!\Gamma_{t}^{(n)}\rangle$-states, obtained from the IBM quantum falcon processors, quito and casablanca respectively. 

The errors on the stabilizer purity $W(\psi)$ and on the purity $P(\psi)$ are chosen to be the standard error of the average over the finite sampling of the local Clifford group. Namely, let $\tilde{W}(\psi)=\frac{1}{N_U}\sum_{C}\tilde{W}_{C}(\psi)$ (introduced in Methods) the estimation of the stabilizer purity given a finite number of Clifford sampling $N_U$. Then the error $\Delta \tilde{W}(\psi)$ on $\tilde{W}(\psi)$ is chosen to be 
\begin{equation}
   \Delta \tilde{W}(\psi)\equiv \sqrt{\frac{1}{N_U(N_U-1)}\sum_{C}(\tilde{W}_{C}(\psi)-\tilde{W}(\psi))^2 } 
\end{equation}
and similarly for the purity $P(\psi)$:
\begin{equation}
       \Delta \tilde{P}(\psi)\equiv \sqrt{\frac{1}{N_U(N_U-1)}\sum_{C}(\tilde{P}_{C}(\psi)-\tilde{P}(\psi))^2 } 
\end{equation}
where the specific value of $N_U$ is state-dependent and it has been estimated by numerical simulations, as described in Methods. All the other errors are obtained by error propagation techniques.

\begin{table}[H]
\caption{ Purity and Magic $\mid\!\! P_{\vartheta}\rangle$-states}
    \centering
    \begin{tabular}{ccc}
    \toprule
     $\mid\!\psi\rangle$   & $P_{exp}(\mid\!\psi\rangle)$ & $W_{exp}(\mid\!\psi\rangle)$ \\
     \midrule
     $\mid\!\!P_{0}\rangle$&$1.0\pm 0.1$ &$0.48\pm 0.06$  \\
     $\mid\!\!P_{\pi/16}\rangle$&$1.0\pm 0.1$ &$0.45\pm 0.06$ \\
     $\mid\!\!P_{\pi/8}\rangle$&$1.0\pm 0.1$ &$0.42\pm 0.05$\\
     $\mid\!\!P_{\pi/6}\rangle$&$1.0\pm 0.1$ &$0.40\pm 0.05$ \\
     $\mid\!\!P_{\pi/5}\rangle$&$0.9\pm 0.1$ &$0.36\pm 0.04$\\
     $\mid\!\!P_{\pi/4}\rangle$&$0.9\pm 0.1$ &$0.34\pm 0.04$ \\
     \bottomrule
    \end{tabular}
    \footnotetext{Table containing experimental values of $P(\mid\!\! P_{\vartheta}\rangle)$ and $W(\mid\!\! P_{\vartheta}\rangle)$ for $\vartheta=0,\pi/16,\dots, \pi/4$, obtained from the quantum processor ibmq$\_$quito.}
    \label{tableexp: 1qubit}
\end{table}
\begin{table}[H]
    \centering
    \caption{ Experimental results of ibmq$\_$quito for n=3,4,5.}
    \begin{tabular}{ccccccc}
    \toprule
    \multicolumn{1}{c}{\# qubits} &$\mid\!\psi\rangle$   & $P_{exp}(\mid\!\psi\rangle)$ & $W_{exp} (\mid\!\psi\rangle)$ & $W_{exp}^{corr}/P_{exp}(\mid\!\psi\rangle)$ & $W(\psi_{p})/P(\psi_{p})$&$W_{th}(\mid\!\psi\rangle) $\\
     \midrule
     {\multirow{6}{*}{ $n=3$}}& $\mid\!\!\Gamma_{1}^{(3)}\rangle$ &$(8\pm 1)$& $(5.3\pm 0.9)$& $(9\pm2)$ &$(8\pm2)$&$\sim 9.4$\\
&$\mid\!\!\Gamma_{2}^{(3)}\rangle$ &$(9\pm 1)$& $(5.1\pm 0.7)$& $(7\pm2)$ &$(7\pm1)$&$\sim 7.0$\\
&$\mid\!\!\Gamma_{3}^{(3)}\rangle$ &$(8.8\pm 0.9)$& $(4.3\pm 0.4)$& $(5.2\pm0.9)$ &$(5.0\pm0.5)$&$\sim 5.2$\\
&$\mid\!\!\Gamma_{4}^{(3)}\rangle$ &$(8.3\pm 0.8)$& $(3.5\pm 0.3)$& $(4\pm1)$ &$(4.1\pm0.04)$&$\sim 4.3$\\
&$\mid\!\!\Gamma_{5}^{(3)}\rangle$ &$(9\pm 1)$& $(3.7\pm 0.4)$& $(4\pm1)$ &$(3.4\pm0.7)$&$\sim 3.5$\\
{\multirow{7}{*}{ $n=4$}} &$\mid\!\!\Gamma_{1}^{(4)}\rangle$ &$(6.3\pm 0.6)$& $(1.4\pm 0.1)$& $(2.8\pm0.8)$ &$(3.1\pm0.4)$&$\sim4.7$\\
    & $\mid\!\!\Gamma_{2}^{(4)}\rangle$  &$(4.3\pm 0.4)$& $(0.80\pm 0.06)$& $(2.2\pm0.4)$ &$(2.3\pm0.3)$&$\sim3.5$\\
     &$\mid\!\!\Gamma_{3}^{(4)}\rangle$ &$(5.9\pm 0.6)$& $(0.98\pm 0.09)$& $(1.7\pm 0.4)$ &$(2.1\pm0.3)$&$\sim2.6$\\
     &$\mid\!\!\Gamma_{4}^{(4)}\rangle$  &$(5.9\pm 0.7)$& $(0.89\pm 0.08)$& $(1.5\pm 0.4)$ &$(1.7\pm0.3)$&$\sim2.0$\\
      & $\mid\!\!\Gamma_{5}^{(4)}\rangle$  &$(5.9\pm 0.6)$& $(0.85\pm 0.08)$& $(1.4\pm0.4)$ &$(1.4\pm0.2)$&$\sim1.5$\\
     &    $\mid\!\!\Gamma_{6}^{(4)}\rangle$&$(5.1\pm 0.5)$& $(0.65\pm 0.04)$& $(1.1\pm0.2)$ &$(1.2\pm0.2)$&$\sim1.2$\\
     &$\mid\!\!\Gamma_{7}^{(4)}\rangle$  & $(6.0\pm 0.6)$ & $(0.72\pm 0.06)$ &$(1.1\pm0.3)$ &$(1.1\pm0.2)$&$\sim1.1$\\ 

     {\multirow{1}{*}{ $n=5$}}&     $\mid\!\!\Gamma_{9}^{(5)}\rangle$ &$(4.9\pm 0.4)$& $(0.18\pm 0.01)$& $(0.34\pm0.05)$ &$(0.34\pm0.03)$&$\sim 0.34$\\
     &     $\mid\!\!\Gamma_{8}^{(5)}$ &$(4.9\pm 0.4)$& $(0.19\pm 0.01)$& $(0.34\pm0.05)$ &$(0.38\pm0.03)$&$\sim 0.40$\\
     &     $\mid\!\!\Gamma_{7}^{(5)}\rangle$ &$(5.3\pm 0.3)$& $(0.21\pm 0.01)$& $(0.32\pm0.08)$ &$(0.41\pm0.02)$&$\sim 0.44$\\
     &$\mid\!\!\Gamma_{6}^{(5)}\rangle$  & $(4.9\pm 0.4)$ & $(0.22\pm 0.01)$ &$(0.4\pm0.1)$ &$(0.53\pm0.04)$&$\sim 0.56$\\ 
     &$\mid\!\!\Gamma_{5}^{(5)}\rangle$  & $(5.0\pm 0.3)$ & $(0.24\pm 0.01)$ &$(0.6\pm0.2)$ &$(0.63\pm0.05)$&$\sim 0.74$\\ 
     &$\mid\!\!\Gamma_{4}^{(5)}\rangle$  &$(4.9\pm0.3)$  &$(0.27\pm 0.2)$  & $(0.5\pm0.2)$ &$(0.77\pm0.05)$ &$\sim 0.99$\\
      &$\mid\!\!\Gamma_{3}^{(5)}\rangle$  & $(4.8\pm 0.3)$ & $(0.30\pm 0.02)$ &$(0.62\pm0.09)$ &$(0.89\pm0.06)$&$\sim 1.3$\\ 
      &$\mid\!\!\Gamma_{2}^{(5)}\rangle$  & $(4.7\pm 0.4)$ & $(0.33\pm 0.02)$ &$(0.7\pm0.3)$ &$(1.09\pm0.09)$&$\sim 1.8$\\ 
      &$\mid\!\!\Gamma_{1}^{(5)}\rangle$  & $(4.8\pm 0.3)$ & $(0.41\pm 0.03)$ &$(1.0\pm0.3)$ &$(1.33\pm0.08)$&$\sim 2.3$ \\
     \bottomrule
    \end{tabular}
    \footnotetext{Table of the experimental results obtained from the quantum processor ibmq$\_$quito. The displayed data are $P_{exp}(\mid\!\psi\rangle)$ and $W_{exp}(\mid\!\psi\rangle)$ of $\mid\!\!\Gamma^{(n)}_{t}\rangle$ for $n=3,4,5$ and $t=1,\dots, 2n-1$, estimated via randomized measurement with the error given by the standard error of the mean over the $N_{U}$ local Clifford operators. The fifth column contains the ratio between the corrected $W_{exp}(\psi)$, according to the correction monitored by $\epsilon$ (cfr. Sec. \ref{Sec: noisemodel}), and $P_{exp}$, while the sixth column contains the value of the ratio between $W(\psi_{p})$ and $P(\psi_{p})$ obtained from the noise model; the error on $W(\psi_{p})/P_{p}(\psi)$ comes from the error of the estimated $p$ from $P_{exp}(\mid\!\psi\rangle)$, see Sec. \ref{Sec: noisemodel}. The last column reports the theoretical value of $W(\mid\!\psi\rangle)$ achieved by the pure $\mid\!\!\Gamma^{(n)}_{t}\rangle$-states. The values for $P_{exp}(\mid\!\psi\rangle)$ are in $10^{-1}$ units, while all the value for $W_{exp}, W^{corr}_{exp}/P_{exp}, W(\psi_{p})/P(\psi_{p})$ and $ W_{th}$ are in $10^{-2}$ units.}
    \label{tableexp: quito}
\end{table}
\begin{table}[H]
    \centering
    \caption{ Experimental results of ibmq$\_$casablanca for n=3,4.5.}
    \begin{tabular}{ccccccc}
    \toprule
    \multicolumn{1}{c}{\# qubits} &$\mid\!\psi\rangle$   & $P_{exp}$ & $W_{exp} $ & $W_{exp}^{corr}/P_{exp}$ & $W_{p}/P_{p}$&$W_{th} $\\
    \midrule
{\multirow{6}{*}{ $n=3$}}&     $\mid\!\!\Gamma_{1}^{(3)}\rangle$ &$(8.3\pm 1.0)$& $(5.8\pm 0.8)$& $(7\pm2)$ &$(7.9\pm1.0)$&$\sim 9.4$\\
&$\mid\!\!\Gamma_{2}^{(3)}\rangle$ &$(8.4\pm 1.0)$& $(5.1\pm 0.7)$& $(6\pm2)$ &$(6.1\pm0.7)$&$\sim 7.0$\\
&$\mid\!\!\Gamma_{3}^{(3)}\rangle$ &$(9.1\pm 0.9)$& $(4.2\pm 0.4)$& $(5\pm1)$ &$(5.0\pm0.5)$&$\sim 5.3$\\
&$\mid\!\!\Gamma_{4}^{(3)}\rangle$ &$(8.3\pm 0.9)$& $(3.7\pm 0.4)$& $(4.5\pm1.0)$ &$(4.0\pm0.4)$&$\sim4.3$\\
&$\mid\!\!\Gamma_{5}^{(3)}\rangle$ &$(7.4\pm 0.7)$& $(2.8\pm 0.3)$& $(3.7\pm0.6)$ &$(3.4\pm0.3)$&$\sim 3.5$\\
{\multirow{6}{*}{ $n=4$}}&     $\mid\!\!\Gamma_{1}^{{(4)}}\rangle$ &$(5.1\pm 0.4)$& $(1.2\pm 0.1)$& $(3\pm1)$ &$(2.8\pm0.3)$&$\sim 4.7$\\
    & $\mid\!\!\Gamma_{2}^{(4)}\rangle$  &$(6.2\pm 0.9)$& $(1.4\pm 0.2)$& $(2.6\pm0.8)$ &$(2.6\pm0.4)$&$\sim 3.5$\\
     &$\mid\!\!\Gamma_{3}^{(4)}\rangle$ &$(6.4\pm 0.8)$& $(1.1\pm 0.1)$& $(1.8\pm 0.6)$ &$(2.1\pm0.4)$&$\sim 2.6$\\
     &$\mid\!\!\Gamma_{4}^{(4)}\rangle$  &$(7.4\pm 0.8)$& $(1.2\pm 0.1)$& $(1.6\pm 0.5)$ &$(1.8\pm0.3)$&$\sim 2.0$\\
      & $\mid\!\!\Gamma_{5}^{(4)}\rangle$  &$(6.2\pm 0.6)$& $(0.85\pm 0.08)$& $(1.3\pm0.3)$ &$(1.4\pm0.2)$&$\sim 1.5$\\
     &    $\mid\!\!\Gamma_{6}^{(4)}\rangle$&$(6.0\pm 0.6)$& $(0.71\pm 0.05)$& $(1.1\pm0.6)$ &$(1.2\pm0.2)$&$\sim 1.2$\\
     &$\mid\!\!\Gamma_{7}^{(4)}\rangle$  & $(5.9\pm 0.5)$ & $(0.68\pm 0.05)$ &$(1.1\pm0.2)$ &$(1.1\pm0.1)$&$\sim 1.1$\\ 
     {\multirow{1}{*}{ $n=5$}}&     $\mid\!\!\Gamma_{9}^{(5)}\rangle$ &$(4.4\pm 0.3)$& $(0.17\pm 0.01)$& $(0.38\pm0.05)$ &$(0.36\pm0.03)$&$\sim 0.34$\\
     &     $\mid\!\!\Gamma_{8}^{(5)}\rangle$ &$(3.7\pm 0.4)$& $(0.154\pm 0.008)$& $(0.40\pm0.06)$ &$(0.42\pm0.03)$&$\sim 0.40$\\
     &     $\mid\!\!\Gamma_{7}^{(5)}\rangle$ &$(3.2\pm 0.2)$& $(0.150\pm 0.008)$& $(0.45\pm0.04)$ &$(0.50\pm0.03)$&$\sim 0.44$\\
     &$\mid\!\!\Gamma_{6}^{(5)}\rangle$  & $(3.5\pm 0.2)$ & $(0.18\pm 0.01)$ &$(0.50\pm0.05)$ &$(0.57\pm0.03)$&$\sim 0.56$\\ 
     &$\mid\!\!\Gamma_{5}^{(5)}\rangle$  & $(3.7\pm 0.2)$ & $(0.23\pm 0.03)$ &$(0.6\pm0.1)$ &$(0.65\pm0.03)$&$\sim 0.74$\\ 
     &$\mid\!\!\Gamma_{4}^{(5)}\rangle$  &$(3.2\pm0.2)$  &$(0.162\pm 0.07)$  & $(0.47\pm0.10)$ &$(0.76\pm0.05)$ &$\sim 0.99$\\
      &$\mid\!\!\Gamma_{3}^{(5)}\rangle$  & $(2.8\pm 0.2)$ & $(0.171\pm 0.009)$ &$(0.62\pm0.10)$ &$(0.76\pm0.06)$&$\sim 1.3$\\ 
      &$\mid\!\!\Gamma_{2}^{(5)}\rangle$  & $(2.6\pm 0.2)$ & $(0.19\pm 0.01)$ &$(0.7\pm0.2)$ &$(1.0\pm0.08)$&$\sim 1.8$\\ 
      &$\mid\!\!\Gamma_{1}^{(5)}\rangle$  & $(2.4\pm 0.1)$ & $(0.162\pm 0.007)$ &$(0.7\pm0.1)$ &$(1.09\pm0.05)$&$\sim 2.3$ \\
     \bottomrule
    \end{tabular}
    \footnotetext{ Table of the experimental results obtained from the quantum processor ibmq$\_$casablanca. The displayed data are $P_{exp}(\mid\!\psi\rangle)$ and $W_{exp}(\mid\!\psi\rangle)$ of $\mid\!\!\Gamma^{(n)}_{t}\rangle$ for $n=3,4$ and $t=1,\dots, 2n-1$, estimated via randomized measurement with the error given by the standard error of the mean over the $N_{U}$ local Clifford operators. The fifth column contains the ratio between the corrected $W_{exp}(\psi)$, according to the correction monitored by $\epsilon$ (cfr. Sec. \ref{Sec: noisemodel}), and $P_{exp}$, while the sixth column contains the value of the ratio between $W(\psi_{p})$ and $P(\psi_{p})$ obtained from the noise model; the error on $W(\psi_{p})/P_{p}(\psi)$ comes from the error of the estimated $p$ from $P_{exp}(\mid\!\psi\rangle)$, see Sec. \ref{Sec: noisemodel}. The last column reports the theoretical value of $W(\mid\!\psi\rangle)$ achieved by the pure $\mid\!\!\Gamma^{(n)}_{t}\rangle$-states. The values for $P_{exp}(\mid\!\psi\rangle)$ are in $10^{-1}$ units, while all the value for $W_{exp}, W^{corr}_{exp}/P_{exp}, W(\psi_{p})/P(\psi_{p})$ and $ W_{th}$ are in $10^{-2}$ units.}
    \label{tableexp: casablanca}
\end{table}
\begin{table}[H]
\centering
\caption{ Experimental value for $\!\mid\!\!0\rangle^{\otimes n}$.}
         \begin{tabular}{ccccccc}
      \toprule
          \multicolumn{1}{c}{\# qubits} & $N_U$ & $N_M$&ibmq$\_$ & $P_{exp}$&$W_{exp}$& $\varepsilon$ \\
    \midrule
\multirow{2}{*} {$n=3$} &\multirow{2}{*} {$104$} &  \multirow{2}{*} {$149$} &quito&$10\pm1$&$10\pm1$&$0.36\pm0.09$\\
 &&&casablanca&$8.1\pm0.1$&$8\pm 1$&$0.16\pm0.09$\\
 \multirow{2}{*} {$n=4$} &\multirow{2}{*} {$196$} &  \multirow{2}{*} {$220$} &quito&$10\pm1$&$4.5\pm0.5$&$0.4\pm0.1$\\
 &&&casablanca&$9.0\pm0.1$&$4.3\pm 0.6$&$0.3\pm0.2$\\
  \multirow{2}{*} {$n=5$} &\multirow{2}{*} {$322$} &  \multirow{2}{*} {$406$} &quito&$8.7\pm0.9$&$1.7\pm0.2$&$0.3\pm0.1$\\
 &&&casablanca&$9.2\pm 0.8$&$2.2\pm 0.2$&$0.2\pm 0.1$\\
 \bottomrule
   \end{tabular}
       \footnotetext{The table show the experimental value for $P_{exp}(\mid\!\!0\rangle^{\otimes n})$ and $W_{exp}(\mid\!\!0\rangle^{\otimes n})$ with $n=3,4,5$ for both ibmq$\_$quito and ibmq$\_$casablanca. The first two columns report the number of optimal resources for $\!\mid\!\!0\rangle^{\otimes n}$ computed numerically according to the methods in Sec. \ref{Sec: experimentalprotocol}. The last column contains the value of the shift-angle $\epsilon$ modeling the non-accuracy in the application of random local Clifford operators, see Sec. \ref{Sec: noisemodel}.All the values for $P_{exp}$ are $10^{-1}$ units and all the values for $W_{exp}$ are in $10^{-2}$ units.}
\end{table}

%